\documentclass{aa}             %activate for journal version

\input epsf
\usepackage{longtable} 

\begin{document}

\def\AD{AD\,Boo}
\def\VZ{VZ\,Hya}
\def\WZ{WZ\,Oph}
\def\kms{\ifmmode{\rm km\thinspace s^{-1}}\else km\thinspace s$^{-1}$\fi}
\def\feh{$[\mathrm{Fe/H}]$}
\def\meh{$[\mathrm{M/H}]$}
\def\afe{$[\mathrm{\alpha/Fe}]$}
\def\ione{\,{\sc i}}
\def\itwo{\,{\sc ii}}

\title{
Absolute dimensions of eclipsing binaries. XXVI.
\thanks{Based on observations carried out at 
the Str{\"o}mgren Automatic Telescope (SAT) 
and the 1.5m and 2.2m telescopes at ESO, La Silla, Chile
(62.L-0284, 63.H-0080, 71.D-0554);
the 1.5m Wyeth reflector at the Oak Ridge Observatory, Harvard, 
Massachusetts, USA;
the 1.5-m Tillinghast reflector and the Multiple Mirror Telescope 
at the F.\ L.\ Whipple Observatory, Mt.\ Hopkins, Arizona.
}
}
\subtitle{
Setting a new standard: Masses, radii, and abundances for the F-type systems AD\,Bootis, 
VZ\,Hydrae, and WZ\,Ophiuchi.     
\thanks{Tables \ref{tab:adboo_rv}, \ref{tab:vzhya_rv} and \ref{tab:wzoph_rv}
will be available on electronic form at the CDS via anonymous ftp to 
130.79.128.5 or via http://cdsweb.u-strasbg.fr/Abstract.html.}
}
\author{
J.V. Clausen       \inst{1} 
\and G. Torres \inst{2}
\and H. Bruntt \inst{1,3}
\and J. Andersen \inst{1,4}
\and B. Nordstr\"om \inst{1}
\and R.P. Stefanik \inst{2}
\and D. W. Latham \inst{2}
\and J. Southworth \inst{1,5}   
}
\offprints{J.V.~Clausen, \\ e-mail: jvc@astro.ku.dk}

\institute{
Niels Bohr Institute, Copenhagen University,
Juliane Maries Vej 30,
DK-2100 Copenhagen {\O}, Denmark
\and
Harvard-Smithsonian Center for Astrophysics,
60 Garden Street, Cambridge, MA 02138, USA
\and
School of Physics A28, University of Sydney, 
2006 NSW, Australia
\and
Nordic Optical Telescope Scientific Association, Apartado 474, ES-38\,700
Santa Cruz de La Palma, Spain
\and
Department of Physics, University of Warwick, Coventry, CV47AL, UK
}

\date{Received 28 February 2008 / Accepted 5 April 2008}
 
\titlerunning{AD\,Boo, VZ\,Hya, and WZ\,Oph}
\authorrunning{J.V. Clausen et al.}

\abstract
%context
{Accurate mass, radius, and abundance determinations from 
binaries provide important information on stellar evolution, 
fundamental to central fields in modern astrophysics and cosmology.}
%aims
{We aim to determine absolute dimensions and abundances for the three F-type
main-sequence detached eclipsing binaries \AD, \VZ, and \WZ\ 
and to perform a detailed comparison with results from recent stellar 
evo\-lu\-tio\-nary models.}
%methods
{$uvby$ light curves and $uvby\beta$ standard photometry were obtained with
the Str\"omgren Automatic Telescope at ESO, La Silla, 
radial velocity observations at CfA facilities, 
and supplementary high-resolution spectra with ESO's FEROS spectrograph.
State-of-the-art methods were applied for the analyses:
the EBOP and Wilson-Devinney binary models, two-dimensional cross-correlation
and disentangling, and the VWA abundance analysis tool.}
%results
{Masses and radii that are precise to 0.5--0.7\% and 0.4--0.9\%, respectively, 
have been established for the components, which span the ranges of
1.1 to 1.4 $M_{\sun}$ and 1.1 to 1.6 $R_{\sun}$. 
The \feh\ abundances are from $-0.27$ to $+0.10$, with uncertainties
between 0.07 and 0.15 dex. We find indications of a slight $\alpha$-element
overabundance of \afe$\sim+0.1$ for \WZ.
The secondary component of \AD\ and both components of \WZ\ 
appear to be slightly active.
Yale-Yonsai and Victoria-Regina evolutionary models fit the components of
\AD\ and \VZ\ almost equally well, assuming coeval formation, at ages of about 
1.75/1.50 Gyr (\AD) and 1.25/1.00 Gyr (\VZ). 
BaSTI models, however, predict somewhat different ages for the primary
and secondary components.
For \WZ, the models from all three grids are significantly hotter than 
observed.  A low He content, decreased envelope convection 
coupled with surface activity, and/or higher interstellar absorption 
would remove the discrepancy, but its cause has not been definitively 
identified. 
}
%conclusions
{We have demonstrated the power of testing and comparing recent
stellar evolutionary models
using eclipsing binaries, provided their abundances are known. 
The strongest limitations and challenges are set by $T_{\rm eff}$ and
interstellar absorption determinations, and by their effects on and
correlation with abundance results. 
}

\maketitle

\section{Introduction}
\label{sec:intro}

Accurate stellar mass and radius data from eclipsing binary systems are 
valuable empirical test data for stellar evolution models, basically because 
they are free of any scale dependent calibrations. 
Additional tests become possible when the effective temperatures for
the components can also be determined, most often from well-calibrated photometry. 
The binaries then also serve as primary distance indicators, e.g.
for nearby stars, stellar clusters, and Local Group galaxies, and they define
the empirical mass-luminosity relations over a broad mass range.
However, the most stringent model tests require that spectroscopic
element abundances are available as well, minimising the number of
free parameters in the comparison with theory (e.g. Andersen \cite{ja91}).
Unfortunately, the number of binary components with complete and accurate 
data is still small.

In our efforts to provide such complete data for a larger sample of eclipsing 
binary systems, F-G type main-sequence stars have been given high priority for 
two reasons. First, their spectra offer favourable conditions for accurate 
radial velocity determinations and abundance analyses. Second, such stars 
cover the mass range within which convective cores begin to develop and affect 
the observational determination of stellar ages, in particular due to
difficulties and shortcomings related to the physics of core overshoot. 
Moreover, these ages cover the range of interest in studies of the evolution 
of the Galactic disk (e.g. Holmberg et al. \cite{holmberg07}). 

In this paper we present new analyses of the three detached F-type 
double-lined eclipsing binaries \object{\AD}, \object{\VZ}, and \object{\WZ}. 
From earlier studies they are known to have masses in the range 1.15--1.41  
$M_{\sun}$ and radii corresponding to the lower half of the main-sequence 
band, but better accuracy and - especially - abundance data are needed for 
meaningful tests of state-of-the art stellar mo\-dels.
We have therefore obtained new radial velocity observations, light curves, 
standard photometry, and high-resolution spectra covering a wide 
wavelength range, enabling us to determine all relevant parameters of the 
systems with high precision.

\begin{figure}
\epsfxsize=90mm
\epsfbox{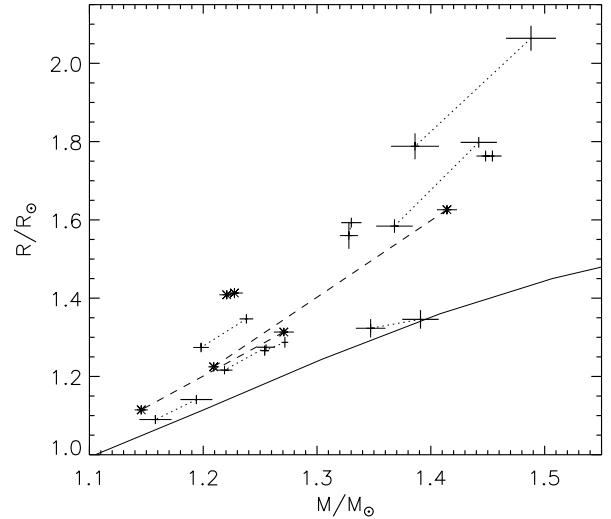}
\caption[]{\label{fig:debs}
Main-sequence eclipsing binaries with both components in the 
1.1--1.5 $M_{\sun}$ range having masses and radii accurate to 
2\% or better.
The nine systems shown are, in order of in\-crea\-sing mass of the 
primary component: 
\object{EW\,Ori}, \object{UX\,Men}, \object{HS\,Hya}, \object{V505\,Per},
\object{IT\,Cas}, \object{V1143\,Cyg}, \object{CD\,Tau}, \object{DM\,Vir}, 
and \object{BW\,Aqr}.
\WZ, \VZ, and \AD, are shown with star symbols.
The full drawn line is the $Y^2$ 0.3 Gyr isochrone for $Z=0.0181$
(Demarque et al. \cite{yale04}). 
}
\end{figure}

As pointed out by Popper (\cite{dmp98a}), and also seen in Fig.~\ref{fig:debs} 
\footnote{{\scriptsize\tt http://www.astro.keele.ac.uk/$\sim$jkt/}},
\AD\ is of particular interest for having components with the greatest 
differences in mass and radius among well-studied binaries in its mass range. 
Hence, it may provide stringent tests of stellar models, but the masses 
obtained in two recent analyses of the system disagree by slightly more than
their stated errors (Lacy \cite{lacy97}; Popper \cite{dmp98a}). Moreover, 
in both studies the radii were based on reanalyses of the relatively old 
$B,V$ light curves (normal points) by Zhai et al. (\cite{zhai82}), 
which have an accuracy of only about 0.02 mag and show out-of-eclipse 
variations by about 0.075 mag. 

Previous determinations of the dimensions of \VZ\ and \WZ\ were also based on 
old material (Popper \cite{dmp65}) and therefore had an accuracy of only 2--4\%;
see Popper (\cite{dmp80}). As in the case of \AD, the components of \VZ\ are 
fairly different, whereas those of \WZ\ are nearly identical.

The paper is structured as follows: First, Sect.~\ref{sec:obsmet} describes 
our observations and analysis techniques adopted. The spectroscopic and 
photometric analyses of the indi\-vi\-dual systems are then discussed in detail 
in Sect.~\ref{sec:specphot}. The resulting new, accurate absolute dimensions 
and abundances are presented in Sect.~\ref{sec:absdim} and compared with 
three recent grids of stellar evolutionary models in Sect.~\ref{sec:dis}.

Throughout the paper, the component eclipsed at the deeper eclipse at phase 
0.0 is referred to as the primary $(p)$ component, the other stars as the 
secondary $(s)$.

\section{Observations and analysis techniques}
\label{sec:obsmet}

\subsection{Spectroscopic observations}
\label{sec:obs_spec}
The radial velocity observations for the three binaries were conducted at the 
Harvard-Smithsonian Center for Astrophysics (CfA) with nearly identical echelle
spectrographs on the 1.5-m Wyeth reflector at the Oak Ridge Observatory 
(Harvard, Massachusetts), the 1.5-m Tillinghast reflector at the 
F.\ L.\ Whipple Observatory (Mt.\ Hopkins, Arizona), and the Multiple Mirror 
Telescope (also on Mt.\ Hopkins) prior to its conversion to a monolithic
mirror. A single echelle order 45\,\AA\ wide was recorded with
intensified photon-counting Reticon detectors, at a central wavelength
of approximately 5187\,\AA, which includes the Mg~I~b triplet. 
The resolving power of these instruments is
$\lambda/\Delta\lambda \approx 35\,000$, and the signal-to-noise
(S/N) ratios achieved per resolution element
of 8.5~\kms\ range from 10 to 43 for \AD, 15 to 39 for \VZ, and 10 to 25 for \WZ.
As seen in Figures~\ref{fig:adboo_rv}, \ref{fig:vzhya_rv}, and
\ref{fig:wzoph_rv}, the observations cover the orbits well.

For detailed abundance studies, we have obtained high-resolution spectra
at higher S/N ratios with the FEROS fiber echelle spectrograph at the ESO 
1.52-m and 2.2-m telescopes at La Silla, Chile (Kaufer et al. \cite{feros99}, 
\cite{feros00}).
The spectrograph, which resides in a temperature-controlled room,
covers without interruption the spectral region from the Balmer jump
to 8700\,\AA, at a constant velocity resolution of 2.7 \kms\
per pixel ($\lambda/\Delta\lambda=48\,000$). 
An ob\-ser\-ving log is given in Table~\ref{tab:feros}.

\begin{figure*}
\epsfxsize=190mm
\epsfbox{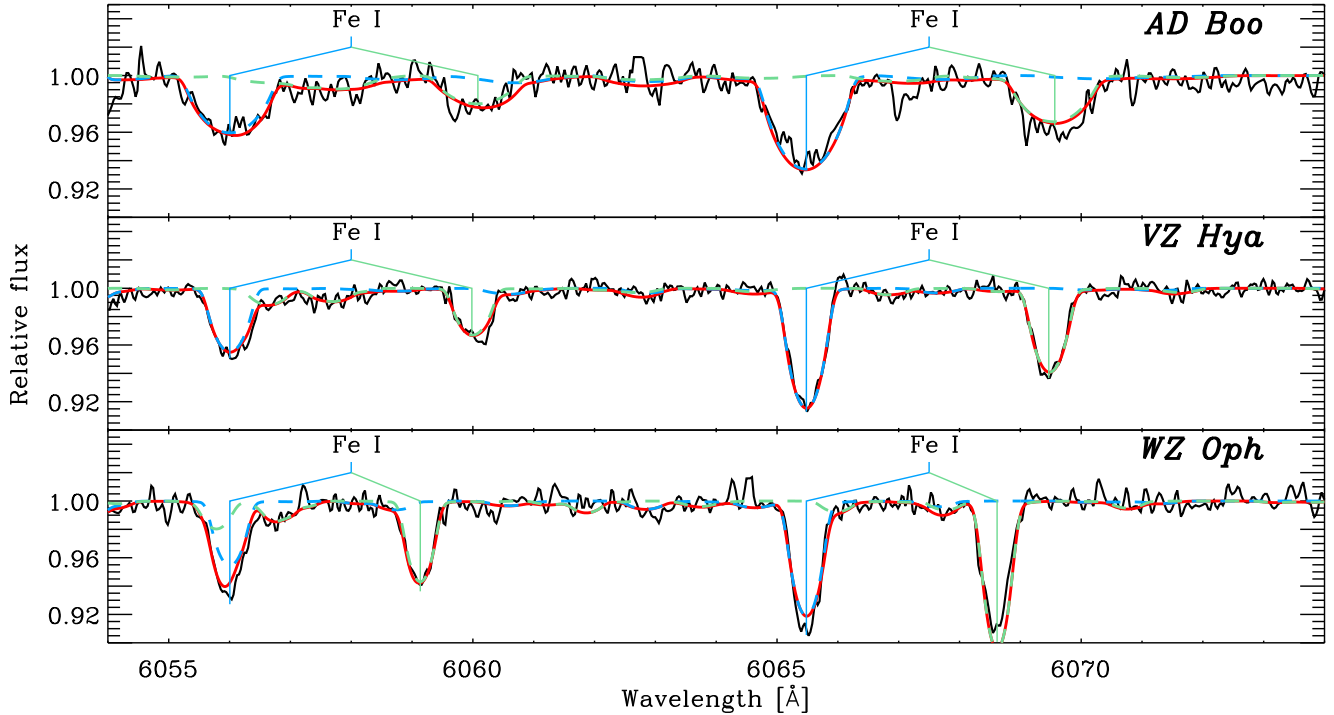}
\caption[]{\label{fig:feros}
FEROS spectra (black line) of \AD\ (phase 0.17), \VZ\ (phase 0.28), 
and \WZ\ (phase 0.33) in the region of the Fe\,I lines at 6056.0 and 
6065.5\,\AA, shifted to laboratory wavelengths for the primary components. 
Synthetic spectra have been included: 
primary components (dashed blue);
secondary components (dashed green);
combined (continuous red line).
}
\end{figure*}

A modified version of the MIDAS FEROS package, prepared by H. Hensberge,
was used for the basic reduction of the FEROS spectra
\footnote{{\scriptsize\tt http://www.ls.eso.org/lasilla/sciops/2p2/E2p2M/FEROS/DRS}}
\footnote{{\scriptsize\tt http://www.ls.eso.org/lasilla/sciops/2p2/E1p5M/FEROS/Reports}}.
Compared to the standard package, background removal, definition
and extraction of the individual orders, and wavelength calibration
are significantly improved.
The observations were reduced night by night using calibration exposures
(ThAr and flat field) obtained during the afternoon/morning.
Standard (rather than optimal) extraction was applied, and no order
merging was attempted. Typically, a standard error of the wavelength
calibration of 0.002--0.003\,\AA\ was obtained.
Subsequently, 
dedicated IDL\footnote{{\scriptsize\tt http://www.ittvis.com/idl/index.asp}} 
programs were applied to remove cosmic ray events and other defects, 
and for normalization of the individual orders. For each order, only 
the central part with acceptable signal-to-noise ratios was kept for further 
ana\-ly\-sis. 
A short region of sample FEROS spectra is shown in Fig.~\ref{fig:feros}.

\begin{table}
\caption[]{\label{tab:feros}
Log of the FEROS observations of \AD, \VZ, and \WZ.
HJD refers to mid-exposure.
The exposure time $t_{\rm exp}$ is given in seconds. The signal-to-noise ratio
per resolution element (S/N) of individual exposures was measured around 
6070\,\AA.
Observers: H = Heidelberg/Copenhagen guaranteed time; JVC = J.V.~Clausen;
JS = J.~Southworth. 
}
\begin{center}
\begin{tabular}{ccrrr}
\hline
\hline\noalign{\smallskip}
HJD$-$2\,400\,000               & observer & phase    &$t_{\rm exp}$ & S/N  \\
\noalign{\smallskip}
\hline
\noalign{\smallskip}
\AD               &    &             &      &        \\ 
51207.8746        &JVC &  0.6844     & 2280 &  125   \\   %  f3576  o27 125
51208.8792        &JVC &  0.1700     & 2400 &  200   \\   %  f3609  o27 200
\noalign{\smallskip}
\VZ               &    &             &      &        \\ 
51207.6144        &JVC &  0.2192     & 1800 &  165   \\   %  f3558  o27 165
51207.7489        &JVC &  0.2655     & 1800 &  160   \\   %  f3568  o27 160
51210.7054        &JVC &  0.2835     & 2400 &  200   \\   %  f3672  o27 200
51211.7137        &JVC &  0.6307     & 2400 &  240   \\   %  f3726  o27 240
51563.8019        &JVC &  0.8606     & 2400 &  160   \\   %  f3504  o27 160
\noalign{\smallskip}
\WZ               &    &             &      &        \\   %         o27
51390.6069        &H   &  0.3318     & 2700 &  210   \\   %  f8704  210
52776.7735        &JS  &  0.6727     &  300 &   50    \\  %  f0076   50
52776.8016        &JS  &  0.6794     &  300 &   45    \\  %  f0078   45
52776.8338        &JS  &  0.6871     &  300 &   55    \\  %  f0084   55
52776.8736        &JS  &  0.6966     &  300 &   50     \\ %  f0093   50
52776.9054        &JS  &  0.7042     &  300 &   45     \\ %  f0100   45
52777.6218        &JS  &  0.8754     &  300 &   55     \\ %  f0244   55
52777.7127        &JS  &  0.8972     &  300 &   45     \\ %  f0256   45
52777.7649        &JS  &  0.9096     &  300 &   75     \\ %  f0262   75
52777.8076        &JS  &  0.9198     &  300 &   70     \\ %  f0267   70
52777.8569        &JS  &  0.9316     &  300 &   55     \\ %  f0273   55
52778.7325        &JS  &  0.1409     &  300 &   30     \\ %  f0464   30
52778.7710        &JS  &  0.1501     &  420 &   50     \\ %  f0468   50
52778.8148        &JS  &  0.1606     &  300 &   35     \\ %  f0473   35

\hline
\end{tabular}
\end{center}
\end{table}

\subsection{Radial velocity determinations}
\label{sec:rv}

Radial velocities were obtained from the CfA spectra using the two-dimensional
cross-correlation algorithm TODCOR (Zucker \& Mazeh \cite{Zucker:94}). 
Templates were
selected from a library of synthetic spectra based on model
atmospheres by R.\ L.\ Kurucz (see also Nordstr\"om et al. \cite{Nordstrom:94};
Latham et al. \cite{Latham:02}), calculated over a wide range in effective 
temperature ($T_{\rm eff}$), rotational velocity ($v \sin i$), surface gravity
($\log g$), and metallicity ([M/H]). The templates producing the best
match to each component of the binary were identified by running
extensive grids of cross-correlations and selecting the combination
giving the highest correlation value averaged over all exposures,
weighted according to the strength of each spectrum
(e.g. Torres et al. \cite{Torres:02}). These grids were in general run in 
temperature and rotational velocity only, since those are the parameters 
that affect the velocities the most.  Surface gravities were fixed according 
to preliminary estimates. If needed, grids for different assumed
heavy element abundances were compared, leading indirectly to a rough
estimate of \meh\ for the sy\-stem, which was in turn compared with the
results from the detailed abundance analysis.

The stability of the zero-point of our velocity system was monitored
by means of exposures of the dusk and dawn sky, and small run-to-run
corrections were applied as described by Latham (\cite{Latham:92}).  In
addition to the radial velocities we derived the light ratio 
at the mean wavelength of our observations (5187\,\AA),
following Zucker \& Mazeh (\cite{Zucker:94}). 

Because of the narrow wavelength coverage of the CfA spectra, the
radial velocities were checked for systematic errors that might occur
as a result of lines of the stars mo\-ving in and out of the spectral
window with orbital phase (Latham et al. \cite{Latham:96}). Numerical 
simulations were carried out as described in more detail by Torres et al.
(\cite{Torres:97}), and corrections based on them were applied to the raw 
ve\-lo\-ci\-ties. 
These corrections are included in the measurements reported in
Tables~\ref{tab:adboo_rv}, \ref{tab:vzhya_rv}, and \ref{tab:wzoph_rv}.

\subsection{Photometric observations}
\label{sec:obs_phot}
$uvby$ light curves for \AD, \VZ, and \WZ\ were obtained with 
the Str{\"o}mgren Automatic Telescope (SAT) at ESO, La Silla during several
campaigns between March 1988 and March 1997. They contain 652, 1180, and 697
points in each band, respectively, and cover all phases well. The accuracy per 
point is 4--5 mmag in $vby$ and 6--7 mmag in $u$, far better than 
previous photometry of these systems. The light curves are kept in the stable $uvby$
instrumental system of the SAT in order not to introduce additional
scatter from transformation to the standard $uvby$ system. 

Improved linear ephemerides have been calculated for all three systems from
the new and numerous published times of minima. No signs of period changes 
were detected, and the orbits were assumed to be circular since there is no 
evidence for eccentricity, either in previous data or in our own.

In addition to the light curves, homogeneous standard $uvby\beta$ indices
have been established from SAT observations on dedicated nights in which the
binaries (and comparison stars) were observed together with a large sample
of $uvby$ and $\beta$ standard stars.
 
We refer to Clausen et al. (\cite{jvcetal08}, hereafter CVG08) for further 
details on the photometry and the new ephemerides.

\subsection{Photometric analyses}
\label{sec:phel}

The choice of binary model for the light curve analysis depends strongly
on the relative radii of the components and the irradiation between
them. In our case, the sy\-stems are well-detached with relative radii
between 0.10 and 0.17, the components are little deformed (oblateness
between 0.001 and 0.007), and reflection effects are small. Therefore,
the simple Nelson-Davis-Etzel model (Nelson \& Davis \cite{nd72}, 
Etzel \cite{e81}, Martynov \cite{m73}), which represents the deformed stars 
as biaxial ellipsoids and applies a simple bolometric reflection model, 
is expected to be perfectly adequate. We have adopted the corresponding 
EBOP code (Popper \& Etzel  \cite{pe81}), supplemented by an extended version 
JKTEBOP\footnote{{\scriptsize\tt http://www.astro.keele.ac.uk/$\sim$jkt/}} 
including Monte Carlo simulations (see e.g. Southworth et al. \cite{sms04a}, 
\cite{sms04b}), for the analyses of the three systems and the assignment 
of realistic errors for the photometric elements. 

Since the accuracy of the light curves is high, we have also
checked for model-dependent systematic errors in the photometric elements
by performing parallel analyses with two more advanced binary models,
WINK (Wood  \cite{dw72}) and the widely used Wilson-Devinney (WD) model
(Wilson \& Devinney \cite{rewd71}; Wilson \cite{rew93}, \cite{rew94}). 
WINK uses tri-axial ellipsoids, whereas WD is based on Roche geo\-me\-try. 
Our WINK version is modified and extended as described by Vaz 
(\cite{lpv84}, \cite{lpv86}), Vaz \& Nordlund (\cite{lpvan85}), and 
Nordlund \& Vaz (\cite{anlpv90}). Our WD code is a revised and extended
version of the 1994 code by Wilson; see Vaz et al. (\cite{lpv95}) and
Casey et al. (\cite{casey98}).
In both codes we use the predictions of ATLAS9 model atmospheres
(Kurucz \cite{ku93}) to connect the flux ratios of the two components 
in different passbands.

In their publicly available versions, EBOP, WINK, and WD carry out a
least-squares optimization of the parameters by means of the classical
differential corrections method. In JKTEBOP, this has been replaced by 
the Levenberg-Marquardt minimization algorithm (MRQMIN: Press et al. 
\cite{press92}).

Throughout this paper we have analysed the $uvby$ light curves independently, 
and for each band identical weights have been assigned to all observations.
A photometric scale factor (the magnitude at quadrature) was always included
as an adjustable parameter, and the phase of primary eclipse was allowed to
shift from 0.0.
The mass ratio between the components was kept at the spectroscopic value,
and synchronous rotation was assumed. 
Gravity darkening coefficients/exponents corresponding to convective 
atmospheres were applied. 
The bolometric reflection albedo coefficients were fixed at 0.5 in the WINK 
and WD analyses, again due to convection; for EBOP the simple bolometric 
reflection model was used.

Linear limb-darkening coefficients were either assigned from theoretical 
calculations (van Hamme \cite{vh93}; D\'\i az-Cordov\'es et al. \cite{dc95}; 
Claret \cite{c00}) according to the effective temperatures, surface gravities, 
and abundances, or included as adjustable parameters. 
For all components of the three systems, the linear coefficients by 
Claret (\cite{c00}) are about 0.10 higher than those by van Hamme (\cite{vh93}) 
and about 0.03 higher than those by D\'\i az-Cordov\'es et al. (\cite{dc95}). 
In general, a change in the coefficients by 0.10 causes systematic differences 
between the optimum theo\-re\-tical light curves of just a few mmag through the 
eclipses, and the overall quality of the fits to the observations is nearly 
identical. If adjusted, the coefficients for the components of 
\VZ\ and \WZ\ turn out closest to the values by van Hamme; for \AD\ the 
coefficients either become unrealistic or convergence is not achieved at all.
In addition, a few WD analyses with the non-linear law by Klinglesmith \& 
Sobieski (\cite{ks70}) and coefficients by van Hamme (\cite{vh93}) were 
included. No improvements in the light curve fits were seen. The influence on 
the other photometric elements of adopting different limb-darkening 
coefficients/laws are discussed below for the individual sy\-stems.
In order to include uncertainties of the limb-darkening
coefficients in the uncertainties of the final photometric elements,
random variations within +/- 0.05 have been introduced in the
JKTEBOP Monte Carlo simulations mentioned above.

In tables and text on photometric solutions we use the following symbols: 
$i$ orbital inclination; 
$r$ relative radius;
$k = r_s/r_p$;
$u$ linear limb darkening coefficient;
$y$ gravity darkening coefficient;
$J$ central surface brightness;
$L$ luminosity;
$\Omega$ surface potential;
$T_{\rm eff}$ effective temperature. 

\subsection{Effective temperatures and interstellar reddening}
\label{sec:teff}

\begin{table*}
\caption[]{\label{tab:teff}
Effective temperatures (K) of the `average' components of \AD, \VZ, and \WZ.
The $V_0$ magnitudes and the $(b-y)_0$ and $c_0$ indices are based on the 
out-of-eclipse $uvby$ standard indices from CVG08.
The 2MASS observations ($J,H,K_s$) were obtained at phases 0.216 (\AD), 0.181 (\VZ), and
0.495 (\WZ), respectively. For \WZ, the $V$ magnitude was calculated
from the value outside eclipses and the $y$ eclipse depth at phase 0.495. 
$A_V$ is the adopted visual interstellar absorption.
References are: 
A96, Alonso et al. (\cite{alonso96}); 
H07, Holmberg et al. (\cite{holmberg07});
RM05, Ram\'\i rez \& Mel\'endez (\cite{rm05}); 
M06, Masana et al. (\cite{masana06}).
The results from A96 are based on their $uvby$ calibration,
those from RM05 on their $uvby$, $(V-J)$, $(V-H)$, and $(V-K_s)$ calibrations (in that order); 
the calibration by M06 is for $(V-K_s)$. 
}
\scriptsize{
\begin{center}
\begin{tabular}{lccccccccccc} \hline
\hline\noalign{\smallskip}
System & $A_V$& [Fe/H] & $(b-y)_0$ &$c_0$&$(V-J)_0$&$(V-H)_0$&$(V-K_s)_0$& A96 & H07 &RM05 & M06  \\
\noalign{\smallskip}
\hline
\noalign{\smallskip}
AD Boo & 0.107&$+0.10$ &0.297&0.426&0.915&1.141&1.182 & 6456 & 6448 &6524/6370/6197/6296 &  6333 \\
VZ Hya & 0.086&$-0.20$ &0.280&0.385&0.817&1.036&1.096 & 6511 & 6517 &6528/6707/6463/6480 &  6513 \\
WZ Oph & 0.141&$-0.27$ &0.333&0.369&0.979&1.176&1.263 & 6115 & 6136 &6097/6190/6191/6176 &  6210 \\
\noalign{\smallskip}            
\hline
\end{tabular}            
\end{center}            
}
\end{table*}

Although significantly improved (semi)empirical temperature scales and
calibrations have become available during the last decade, the
determination of effective temperatures of binary components
remains a difficult task; see e.g. Popper (\cite{dmp80}), Andersen
(\cite{ja91}), and Clausen (\cite{jvc04}) for general remarks. 
For comparison with theoretical mo\-dels, luminosity calculations,
and distance determinations, it is on the other hand important to 
diminish both systematic and random errors as much as possible.
In the analyses of \AD, \VZ, and \WZ\ we have chosen to follow
three independent approaches: 

First, as part of the light curve analyses we calculate the individual 
$uvby$ indices for the components from the combined standard $uvby$ indices 
(CVG08) and the luminosity ratios between the components in the four bands 
(based on mean geometric elements). Interstellar reddening is determined 
from the intrinsic colour calibration by Olsen (\cite{eho88}), which has 
a precision of 0.009 mag, and the results are compared to those from other 
sources.
Effective temperatures can then be obtained from a series of (semi)empirical
calibrations such as those by Alonso et al. (\cite{alonso96}), 
Ram\'\i rez \& Mel\'endez (\cite{rm05}), and Holmberg et al. 
(\cite{holmberg07}).
Also, as an important check on these calibrations, effective tem\-pe\-ra\-ture 
differences between the components can be derived from the empirical flux 
calibration by Popper (\cite{dmp80}) and the $y$ surface flux ratios. They are
obtained to high precision in the light curve analyses, since they
depend mainly on the observed depths of primary and secondary eclipses.
We have applied both the BC scale used by Popper (\cite{dmp80}) and that by
Flower (\cite{flower96}), which in the temperature range of \AD, \VZ, and \WZ,
lead to nearly identical temperature differences with errors of only
about 25~K (excluding uncertainties in Popper's flux scale). 

Second, we attempt to obtain spectroscopic temperatures from the
high-resolution and high-S/N spectra, either combined or disentangled, 
as part of the abundance ana\-ly\-ses; see e.g. Santos et al. (\cite{santos04}) 
and Bruntt et al. (\cite{bruntt04}) for details. And thirdly, as mentioned in
Sect.~\ref{sec:rv}, temperatures are estimated by identifying the templates 
producing the best match to each component in the TODCOR analyses, and 
interpolating in $\log g$ or metallicity as necessary.
As discussed below for the individual systems, the two latter methods however 
tend to be problematic.

Finally, the photometric indices of the combined light of a system may provide a
first estimate of the effective temperature of its "average" component, and
if the components are nearly identical, as is the case for \WZ, a direct 
determination is possible. In Table~\ref{tab:teff} we present such 
temperatures derived from the combined $uvby$ indices and 
2MASS\footnote{{\scriptsize\tt http://www.ipac.caltec.edu/2mass}} 
$JHK_s$ observations. 
As seen, the temperatures from the different $uvby$ and $VJHK_s$ calibrations 
agree well for \WZ. The spread for \AD\ and \VZ\ pre\-su\-mably reflects the 
fact that their combined flux distributions are distorted compared to a single 
star, because the components differ by 400--600~K in effective temperature.

\subsection{Abundance analyses}
\label{sec:abund}

For meaningful comparisons between theoretical stellar models and 
empirical absolute dimensions --- as determined from detached double-lined
eclipsing binaries --- the che\-mi\-cal composition, and in particular the
metal abundance, must be known for the binary. However, detailed
abundance studies have only been done for a few systems. 
As part of the analyses of \AD, \VZ, and \WZ, we have used the new
spectroscopic and photometric data to establish their metal abundance. 
Several methods were applied/attempted: 
\begin{enumerate}
\item Spectroscopic analyses of observed or disentangled FEROS spectra
\item Spectroscopic analyses of disentangled CfA spectra
\item Template matching as part of the TODCOR analyses
\item Photometric $uvby$--[Fe/H] calibrations
\end{enumerate}

\subsubsection{FEROS spectra}
\label{sec:abund_feros}
The FEROS spectra (Table~\ref{tab:feros}) of \WZ\ were disentangled in order 
to extract the individual component spectra. 
We applied the disentangling method introduced by Simon \& Sturm 
(\cite{ss94}) and a revised version of the corresponding original code 
developed by E. Sturm. 
It assumes a constant light level, so only spectra taken outside of eclipse can
be included. We note that all three systems are in fact constant to within a 
few percent outside of eclipse. The orbital elements were fixed at the results 
based on the TODCOR analyses of the CfA spectra.
For \AD, with only two FEROS spectra available, disentangling is not possible.
Disentangling of the five \VZ\ spectra
was attempted, but the quality of the resulting component spectra was not 
acceptable; more than five spectra and/or a better phase distribution
seem to be needed. For both sy\-stems the observed double-lined spectra were 
therefore used directly for the abundance analyses.

The versatile semi-automatic IDL tool 
VWA\footnote{{\scriptsize\tt http://www.hans.bruntt.dk/vwa/}}, 
de\-ve\-lo\-ped by Bruntt, was applied for the abundance analyses. 
We refer to Bruntt et al.
(\cite{bruntt02}, \cite{bruntt04}, \cite{bruntt08}) for further details. 
For our particular study, VWA has been extended to allow the analysis of 
double-lined spectra in addition to normal spectra from single stars --- 
or disentangled binary component spectra.
We have adopted the effective tem\-pe\-ra\-tu\-res, surface gravities, and
rotational velocities determined for the binary components in this 
study (Table~\ref{tab:absdim}), but for comparison analyses for higher 
or lower temperatures were also done.
Microturbulence velocities were derived from the calibration by
Edvardsson et al. (\cite{be93}), which has a scatter of about 0.3 \kms,
but have been tuned, if needed, as part of the analyses.

In each spectrum, VWA selects the least blended lines from atomic
line lists extracted from the Vienna Atomic Line Database (VALD; 
Kupka et al. \cite{kupka99}). For each selected line, the
synthetic spectrum is then calculated, and the input abundance is
changed iteratively until the equi\-va\-lent width of the observed and
synthetic lines match. The synthetic spectra are generated with the
SYNTH software (Valenti \& Piskunov \cite{vp96}). Atmosphere models were 
either interpolated from the grid of modified ATLAS9 models by Heiter et al. 
(\cite{heiter02}) or calculated from the MARCS code. We have applied an 
improved version of the original model by Gustafsson et al. (\cite{gben75}); 
see J{\o}rgensen et al. (\cite{ugj92}) for further details. 

For the binary components, we calculate all abundances relative to the Sun,
line by line. We have used either an atlas spectrum of the solar flux 
(Wallace et al. \cite{whl98}) or a FEROS sky spectrum. In general, 
this approach gives better agreement between the abundances determined 
from the individual lines of a given element/ion.

If we adopt the VALD line data and the solar photosphere abundances by
Grevesse \& Sauval (\cite{gs98}), we obtain for the Sun the Fe abundances 
listed in Table~\ref{tab:sun_feh} for the solar atlas and FEROS spectra. 
Only lines with equivalent widths between 10 and 80 m{\AA} measured
in both spectra were included. 
It is gratifying to see that the results from the FEROS spectrum 
agree very well with those from the atlas spectrum.
For both spectra, Fe\,I and Fe\,II lines give slightly different 
$[\mathrm{Fe/H}]$ results.
They agree a bit better for the MARCS models than for ATLAS9 models, 
and the MARCS models also reproduce the solar result slightly better.

Although we will focus on Fe, we have also analysed other elements, and
results from the solar atlas spectrum are given in Table~\ref{tab:sun_abund}.
For most of the ions with many measured lines we obtain good
agreement with Grevesse \& Sauval (\cite{gs98}), but there are exceptions
such as Si\,I, Cr\,II, and Mn\,I. This may indicate errors in the VALD line 
data (a problem with Mn was found by Bruntt et al. \cite{bruntt04}) 
which should, however, be eliminated to a large extent in the binary 
abundance analyses performed relative to the Sun.  
Comparison with the most recent solar photosphere abundances by
Grevesse et al. (\cite{gas07}), which for many elements are derived
using updated line data, 3D models and departures from LTE, 
is less obvious, since we apply 1D, LTE atmosphere codes. 

\begin{table}
\caption[]{\label{tab:sun_feh}
Iron abundance (\feh) for the Sun determined from Solar atlas
spectrum (Wallance et al. \cite{whl98}) and FEROS spectra,
respectively. 131 Fe\,I and 20 Fe\,II lines, measured in both spectra, were
used. 
Results from ATLAS9 models 
and MARCS models are compared. 
}
\begin{center}
\begin{tabular}{lcc} \hline
\hline\noalign{\smallskip}
                     &  Fe\,I                      & Fe\,II               \\ 
\noalign{\smallskip}
\hline
\noalign{\smallskip}
ATLAS9:              &                   &          \\
Solar atlas          & $+0.11 \pm 0.10$            & $-0.02 \pm 0.07$      \\ 
FEROS                & $+0.11 \pm 0.10$            & $-0.04 \pm 0.07$      \\ 
\noalign{\smallskip}  
MARCS:               &                                                &    \\
Solar atlas          & $+0.07 \pm 0.09$            & $+0.00 \pm 0.10$      \\ 
FEROS                & $+0.07 \pm 0.10$            & $-0.01 \pm 0.09$      \\ 
\hline
\end{tabular}            
\end{center}            
\end{table}                       

\begin{table}
\caption[]{\label{tab:sun_abund}
Abundances for the Sun determined 
from Solar atlas spectrum (Wallace et al. \cite{whl98}) and
ATLAS9 models (Heiter et al. \cite{heiter02}).
N is the number of lines used; only ions with three or more measured lines
are listed. 
GS1998 is the logarithmic element abundances (H = 12.00) 
from Grevesse \& Sauval (\cite{gs98}). 
}
\begin{center}
\begin{tabular}{lrrrr} \hline
\hline\noalign{\smallskip}
Ion    &[El./H]  &  rms  &   N &  GS1998 \\
\hline\noalign{\smallskip}
  C  I & $ 0.21$ &  0.12 &   7 & $ 8.52 \pm  0.06$ \\                 % 8.39 0.05
  O  I & $ 0.00$ &  0.16 &   4 & $ 8.83 \pm  0.06$ \\                 % 8.66 0.05
 Na  I & $ 0.03$ &  0.07 &   9 & $ 6.33 \pm  0.03$ \\                 % 6.17 0.04
 Mg  I & $-0.08$ &  0.23 &  14 & $ 7.58 \pm  0.05$ \\                 % 7.53 0.05
 Al  I & $-0.17$ &  0.11 &   4 & $ 6.47 \pm  0.07$ \\                 % 6.37 0.06
 Si  I & $-0.17$ &  0.29 &  38 & $ 7.55 \pm  0.05$ \\ %ex 4.9-6       % 7.51 0.04
 Si II & $ 0.12$ &  0.33 &   3 & $ 7.55 \pm  0.05$ \\                 
  S  I & $ 0.11$ &  0.21 &   5 & $ 7.33 \pm  0.11$ \\                 % 7.14 0.05
 Ca  I & $ 0.06$ &  0.19 &  25 & $ 6.36 \pm  0.02$ \\                 % 6.31 0.04
 Sc  I & $-0.11$ &  0.14 &   8 & $ 3.17 \pm  0.10$ \\                 % 3.17 0.10
 Sc II & $ 0.06$ &  0.09 &  15 & $ 3.17 \pm  0.10$ \\
 Ti  I & $-0.02$ &  0.07 &  34 & $ 5.02 \pm  0.06$ \\                 % 4.90 0.06
 Ti II & $ 0.06$ &  0.12 &  21 & $ 5.02 \pm  0.06$ \\
  V  I & $ 0.00$ &  0.13 &  52 & $ 4.00 \pm  0.02$ \\                 % 4.00 0.02
  V II & $ 0.32$ &  0.28 &   4 & $ 4.00 \pm  0.02$ \\
 Cr  I & $ 0.07$ &  0.12 &  36 & $ 5.67 \pm  0.03$ \\                 % 5.64 0.10
 Cr II & $ 0.16$ &  0.15 &  22 & $ 5.67 \pm  0.03$ \\
 Mn  I & $ 0.19$ &  0.22 &  39 & $ 5.39 \pm  0.03$ \\ % ex 2-6        % 5.39 0.03
 Fe  I & $ 0.11$ &  0.10 & 238 & $ 7.50 \pm  0.05$ \\                 % 7.45 0.05
 Fe II & $-0.02$ &  0.08 &  30 & $ 7.50 \pm  0.05$ \\
 Co  I & $ 0.03$ &  0.18 &  49 & $ 4.92 \pm  0.04$ \\                 % 4.92 0.08
 Ni  I & $ 0.03$ &  0.10 &  93 & $ 6.25 \pm  0.04$ \\                 % 6.23 0.04
 Cu  I & $-0.17$ &  0.27 &   4 & $ 4.21 \pm  0.04$ \\                 % 4.21 0.04
 Zn  I & $ 0.09$ &  0.13 &   4 & $ 4.60 \pm  0.08$ \\                 % 4.60 0.03
  Y II & $ 0.08$ &  0.17 &  14 & $ 2.24 \pm  0.03$ \\                 % 2.21 0.02
 Zr  I & $-0.23$ &  0.07 &   3 & $ 2.60 \pm  0.02$ \\                 % 2.58 0.02
 Ba II & $ 0.23$ &  0.08 &   3 & $ 2.13 \pm  0.05$ \\                 % 2.17 0.07
 Ce II & $ 0.11$ &  0.21 &   8 & $ 1.58 \pm  0.09$ \\                 % 1.70 0.10
 Nd II & $ 0.13$ &  0.19 &   5 & $ 1.50 \pm  0.06$ \\                 % 1.45 0.05
 Sm II & $ 0.05$ &  0.23 &   3 & $ 1.01 \pm  0.06$ \\                 % 1.00 0.03
\noalign{\smallskip}
\hline
\end{tabular}
\end{center}
\end{table}

\subsubsection{Disentangled CfA spectra}
\label{sec:abund_cfa}
As seen in Tables~\ref{tab:adboo_orb}, \ref{tab:vzhya_orb}, and 
\ref{tab:wzoph_orb}, a large number of CfA spectra are available for all 
three systems. We have selected those collected outside eclipses with 
sufficient S/N ratios for disentangling.
The actual numbers of spectra used are 113 (\AD), 29 (\VZ), and 40 (\WZ).
After careful normalization of the individual spectra, the disentangled 
component spectra were constructed by adopting the orbital elements listed 
in the tables mentioned above. The luminosity ratios between the components, 
which cannot be determined from the disentangling procedure, were calculated 
by interpolation between the $b$ and $y$ luminosity ratios of the photometric 
solutions. The disentangled component spectra were then compared to the 
CfA library of synthetic spectra mentioned in Sect.~\ref{sec:rv}.

In general we find that the short wavelength range co\-ve\-red (45\,\AA), 
and the correspondingly small number of lines available, limits the success 
of this approach. Most importantly, it was difficult to break the correlation 
between metal abundance and effective temperature. Results for the individual 
systems will, however, be presented below.

\subsubsection{TODCOR template matching}
\label{sec;abund_todcor}
As mentioned in Sect.~\ref{sec:rv}, rough estimates of the metal abundance
for a system can be obtained directly from the observed composite spectra
by running extensive grids of two-dimensional cross-correlations (TODCOR)
using templates of different metallicity. Results from this exercise will
also be discussed below for the individual systems.

\subsubsection{$uvby$ calibration}
\label{sec:abund_uvby}
Finally, we have derived metal abundances from the $uvby$ indices for the
individual binary components (Table~\ref{tab:absdim}) and the recent 
calibrations by Holmberg et al. (\cite{holmberg07}), which are
based on a carefully selected sample of spectroscopic
\feh\ results for several hundred stars. 
The calibration adopted for $0.30 < b-y < 0.46$ has a
dispersion of only 0.07 dex; that for bluer stars has a dispersion
of 0.10 dex.
Taking into account the uncertainties of the $uvby$ indices and the
interstellar reddening, an overall precision of $0.10$ to $0.15$ dex can 
be achieved per component.

\section{Spectroscopic and photometric analyses}
\label{sec:specphot}

In this section we present in detail the spectroscopic and photometric
analyses of \AD, \VZ, and \WZ, including abundances, and compare our 
results with those obtained in previous studies. Additional references
to earlier work are given by CVG08.

\subsection{AD\,Boo}
\label{sec:adboo}

\AD\ = BD\,+25\degr2800 = BV135 ($P = 2\fd07$) was dis\-co\-ve\-red by 
Strohmeier et al.  (\cite{sm56}) to be an eclipsing binary, 
and Lacy (\cite{lacy85}) found it to be double-lined. It has most recently 
been studied by Lacy (\cite{lacy97}) and Popper (\cite{dmp98a}). 

\subsubsection{Radial velocities and spectroscopic orbit for \AD}
\label{sec:adboo_spec_orb}

Radial velocities for the components of \AD\ were determined by TODCOR
analyses of 124 usable CfA spectra outside of eclipse, observed from 
January 1989 to June 1997. 
Surface gravities ($\log g$) were fixed at 4.0 (primary) and 4.5 (secondary),
and the heavy element abundance was assumed to be solar,
the closest values in our grid of synthetic spectra to the 
actual values determined in Sect.~\ref{sec:absdim}. 
We inferred temperatures of $6250\pm150$~K for 
the primary of \AD\ (the more massive star), and $5680\pm300$~K for 
the secondary, both significantly lower than the photometric values, 
because the templates have lower metallicity than measured for \AD; 
see Sects.~\ref{sec:adboo_abund} and \ref{sec:absdim}. 
The measured $v \sin i$ values are $38\pm2$~\kms\ 
and $37\pm5$~\kms, respectively, with the secondary being considerably 
more uncertain due to its much weaker lines.
The synchronous values are 39~\kms\ for the primary and 30~\kms\ for the 
secondary. Lacy (\cite{lacy97}) reported $38\pm1$ and $31\pm1$ \kms, 
whereas Popper (\cite{dmp98a}) did not give any values. 
Templates from our library with parameters closest to the measured values 
were used to derive the radial velocities. 

\begin{figure}
\epsfxsize=95mm
\epsfbox{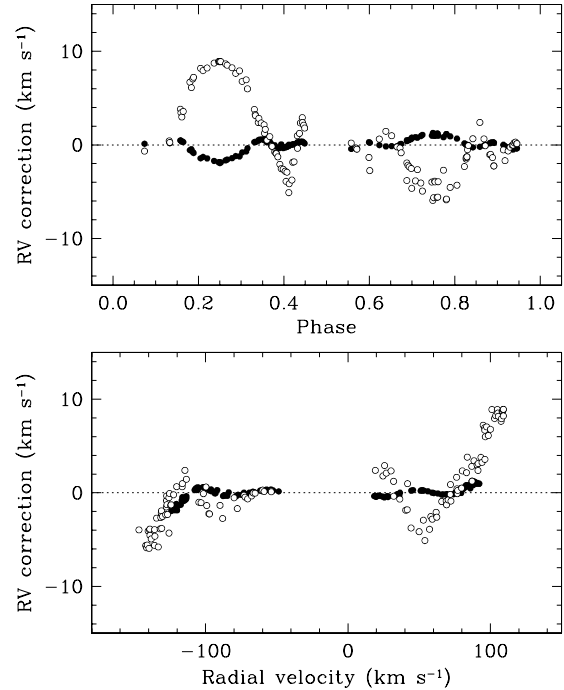}
\caption[]{\label{fig:adboo_cor}
Systematic errors in the raw TODCOR velocities of \AD\,
determined from simulations with synthetic binary spectra
(filled circles: primary; open circles: secondary).
The differences are plotted both as function of orbital phase
(upper panel) and radial velocity (lower panel), and
have been applied to the measured velocities as corrections.
Phase 0.0 corresponds to central primary eclipse.
}
\end{figure}

\begin{figure}
\epsfxsize=95mm
\epsfbox{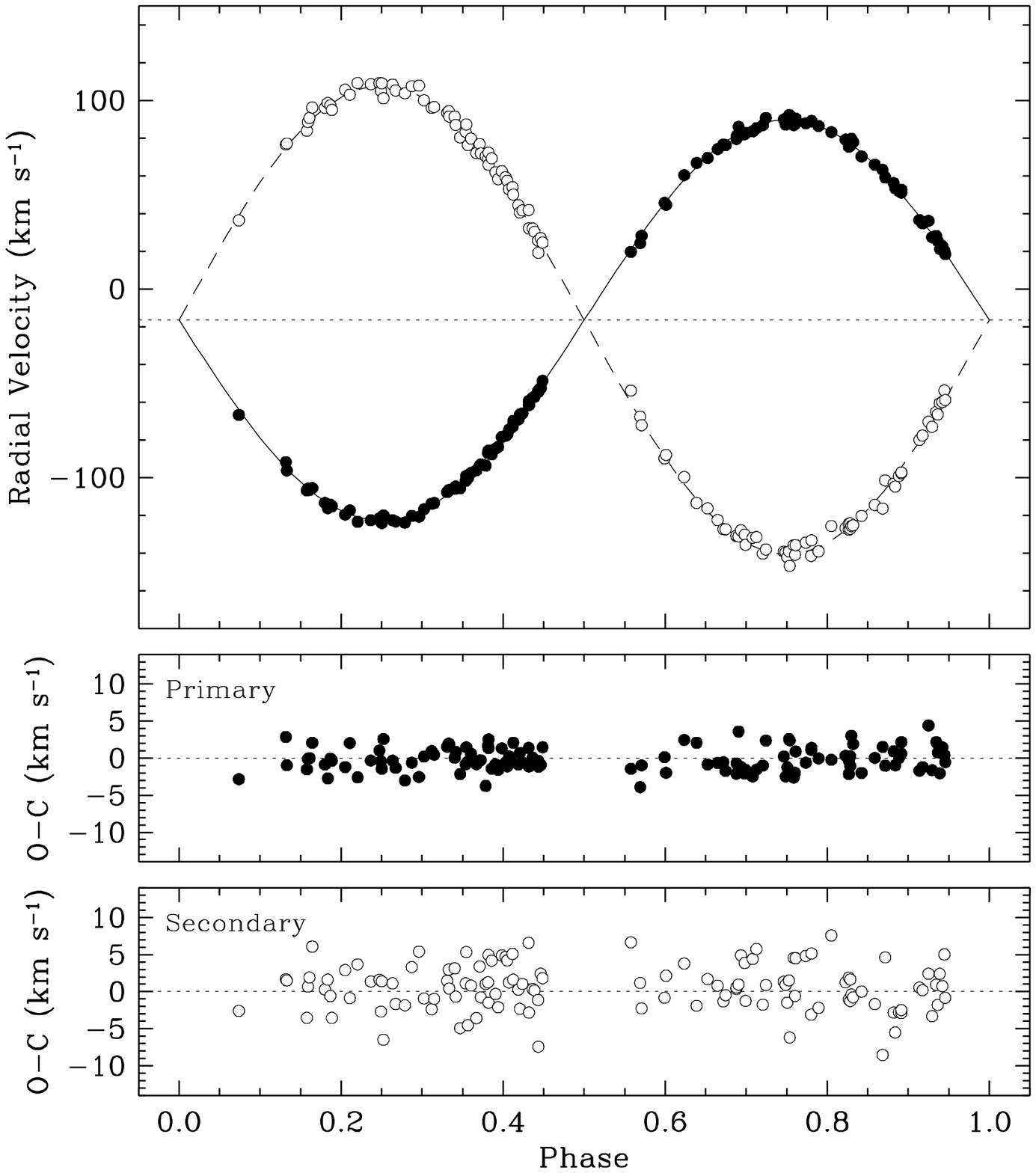}
\caption[]{\label{fig:adboo_rv}
Spectroscopic orbital solution for \AD\ (solid line: primary;
dashed line: secondary) and radial velocities (filled circles:
primary; open circles: secondary).
The dotted line (upper panel) represents the center-of-mass 
velocity of the system. 
Phase 0.0 corresponds to central primary eclipse.
}
\end{figure}

\begin{table}   
\caption[]{\label{tab:adboo_orb}
Spectroscopic orbital solution for AD\,Boo.
$T$ is the time of central primary eclipse.}
\begin{center}    
\begin{tabular}{lr} \hline   
\hline\noalign{\smallskip}    
Parameter            & \multicolumn{1}{c}{Value} \\ 
\noalign{\smallskip}
\hline
\noalign{\smallskip}    
Adjusted quantities:            &   \\ 
$K_p$~(\kms)                     &$106.13 \pm 0.19 $  \\
$K_s$~(\kms)                     &$124.11 \pm 0.36 $  \\
$\gamma$~(\kms)                  &$-16.52 \pm 0.13 $  \\
\noalign{\smallskip}  
Adopted quantities:             &     \\
$P$~(days)                      &  2.06880704 \\
$T$~(HJD$-$2\,400\,000)         &  49311.43169 \\
$e$                             &  0.00          \\ 
\noalign{\smallskip}  
Derived quantities:             &      \\
$M_p \sin^3i~\mathrm{(M_{\sun})}$       & $ 1.4103 \pm 0.0089 $ \\
$M_s \sin^3i~\mathrm{(M_{\sun}})$       & $ 1.2061 \pm 0.0056 $ \\
$a_p \sin i$~($10^6$~km)              & $ 3.0193 \pm 0.0053 $ \\
$a_s \sin i$~($10^6$~km)              & $ 3.5307 \pm 0.0103 $ \\
$a \sin i~\mathrm{(R_{\sun})}$          & $ 9.411  \pm 0.017  $ \\
\noalign{\smallskip}  
Other quantities pertaining to the fit:  &      \\
$N_{obs}$                      &   124 \\
Time span (days)               &  3074 \\
$\sigma_p$~(\kms)              &  1.63 \\
$\sigma_s$~(\kms)              &  3.15 \\
\noalign{\smallskip}  
\hline
\end{tabular}            
\end{center}            
\end{table}

In the case of \AD\ we found that the radial ve\-lo\-ci\-ty corrections
for systematics described in Sect.~\ref{sec:rv}, which are shown in 
Fig.~\ref{fig:adboo_cor}, were quite significant, reaching up 
to 10~\kms\ for the secondary star and up to 2~\kms\ for the primary. 
The effect on the absolute masses is large, $\sim$6\% for the 
primary and $\sim$4\% for the secondary. The final ve\-lo\-ci\-ti\-es 
including corrections are listed in Table~\ref{tab:adboo_rv}. 
The light ratio derived from the spectra, including also corrections for
systematics analogous to those derived for the velocities, is
$\ell_{\rm s}/\ell_{\rm p} = 0.37\pm0.02$ at the mean wavelength of 
our observations (5187\,\AA).

The spectroscopic orbital solution for \AD\ is given in 
Table~\ref{tab:adboo_orb}, and the observations and computed orbit 
are shown graphically in Figure~\ref{fig:adboo_rv} along with the
residuals for each star.
For comparison, we have also disentangled 113 CfA spectra 
outside of eclipse and obtain velocity semiamplitudes of $K_p = 105.95$ 
and $K_s = 124.50$, in good agreement with the results based on TODCOR 
velocities. 

Our velocity semiamplitudes agree well with those by Popper (\cite{dmp98a}),
($K_p, K_s$) = ($105.85\pm0.35$, $124.4\pm0.4 $),
whereas Lacy (\cite{lacy97}) obtained
($K_p, K_s$) = ($107.2\pm0.5 $, $124.6\pm0.6 $).

\subsubsection{Photometric elements for \AD}
\label{sec:adboo_phel}

\begin{figure}
\epsfxsize=85mm
\epsfbox{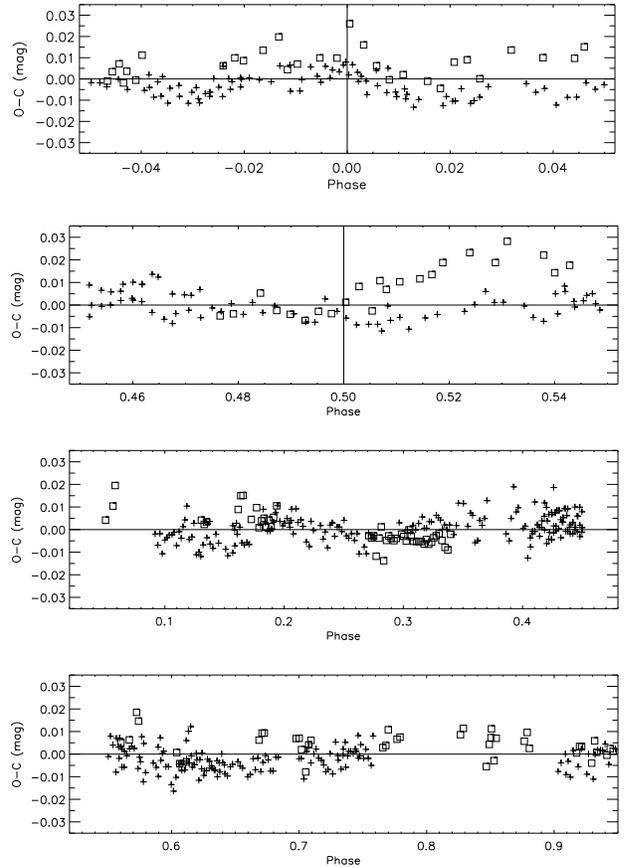}
\caption[]{\label{fig:adboo_res_vh_v}
($O\!-\!C$) residuals of the AD\,Boo $v$-band observations from the 
theoretical light curve computed for the photometric elements given 
in Table~\ref{tab:adboo_ebop_vh}. 1989 observations are shown with
square symbols. 
}
\end{figure}

The $uvby$ light curves of \AD\ contain 652 observations in 
each band and were observed on 42 nights during five periods between 
March 1988 and March 1992 (CVG08).
The accuracy per point is about 0.005 mag ($vby$) and 0.007 mag ($u$), but 
comparison of the data from the five periods reveals
that at some phases, systematic differences in the light level 
of 0.01--0.03 mag exist, increasing from $y$ to $u$. In particular, observations
during the 1989 eclipses are systematically fainter. The effect is shown
in Fig.~\ref{fig:adboo_res_vh_v} and indicates surface activity
(spots) at a rather low level on one or both components, the
cooler secondary being the most likely candidate. 
Zhai et al. (\cite{zhai82}) also noticed scatter in their $B$ and $V$ light 
curves, which were obtained from April to June 1981, but at a much
higher level of about 0.075 mag compared to their observing error of
about 0.02 mag. 

There is very little information on activity in the literature, 
and \AD\ does not appear to be an X-ray source.  
The presence of chromospheric activity in terms of the Rossby
number\footnote{Defined as the ratio of the rotation period to the
convective turnover time.} was studied by Hall (\cite{hall94}). The
secondary in \AD\ is perhaps cool enough to display significant
activity (see above) despite having a mass as high as 1.21 $M_{\sun}$,
which would imply a relatively thin convective envelope. An estimate of
the color index of $B-V = 0.58$ based on the temperature allows the
convective turnover time to be inferred, and this together with the
rotational period (assuming synchronous rotation) leads to a Rossby
number of approximately 0.20, which is in the range where other stars
tend to show spot activity (see Hall \cite{hall94}).
Furthermore, our two high-resolution FEROS spectra taken in January 1999 
(see Table ~\ref{tab:feros}) show in fact a weak emission feature
in the Ca~II~H and K lines at the position of the secondary
component, supporting also some activity. Nothing is seen in the
$\mathrm{H_{\alpha}}$ line.

EBOP and WD solutions, based on linear limb-darkening coefficients
by van Hamme (\cite{vh93}) are presented in Tables~\ref{tab:adboo_ebop_vh}
and \ref{tab:adboo_wd_vh}, respectively, and $O\!-\!C$ residuals of the $v$ 
observations from the theoretical light curve computed for the photometric 
elements given in Table~\ref{tab:adboo_ebop_vh} are shown in 
Fig.~\ref{fig:adboo_res_vh_v}.
As seen, differences between the photometric elements obtained from the
two models are small, and the mean values of $r_p + r_s$ and $k$ agree
within their uncertainties. WINK solutions, which are not included, agree
well with those from EBOP and WD. Slightly different orbital inclinations are 
obtained, however, for the four bands from all three codes. This could be due 
to either the syste\-matic light level differences mentioned above (spots), 
which increase from $y$ to $u$, to a small amount of (red) third light, or to 
a slightly incorrect theoretical wavelength dependence of the limb-darkening 
coefficients. 

\begin{table}
\caption[]{\label{tab:adboo_ebop_vh}
Photometric solutions for AD\,Boo (all data) from the EBOP code
adopting linear limb darkening coefficients by van Hamme (\cite{vh93}).
The errors quoted for the free parameters are the $formal$ errors determined 
from the iterative least squares solution procedure.
}
\begin{center}
\begin{tabular}{lrrrr} \hline
\hline\noalign{\smallskip}
                     &     $y$    &       $b$  &       $v$  &   $u$\\                   
\noalign{\smallskip}
\hline
\noalign{\smallskip}
$i$ \, (\degr)       &  87.58     &   87.69    &   87.83    &  87.89\vspace{-0.8mm}\\   
                     & $\pm12$    &  $\pm10$   &  $\pm10$   & $\pm15$\\                 

$r_p$                &  0.1711    &   0.1715   &   0.1709   &  0.1713\vspace{-0.8mm}\\  
                     &  $\pm 5$   &   $\pm 4$  &   $\pm 4$  &  $\pm 7$\\                

$r_s$                &  0.1291    &   0.1292   &   0.1289   &  0.1296\\                 
                                                                                        
$k$                  &   0.754    &   0.753    &   0.754    &   0.756\vspace{-0.8mm}\\  
                     &  $\pm 5$   &   $\pm4$   &   $\pm3$   &  $\pm 5$ \\               

$r_p + r_s$          &  0.3002    &   0.3007   &   0.2998   &  0.3009\\                 
                                                                                        
$u_p$                &  0.52      &   0.61     &   0.68     &  0.65\\  

$u_s$                &  0.56      &   0.65     &   0.73     &  0.75\\ 

$y_p$                &  0.34      &   0.39     &   0.45     &  0.53\\

$y_s$                &  0.36      &   0.42     &   0.48     &  0.57\\

$J_s/J_p$            &  0.7314    &    0.6888  &   0.6440   &  0.6775\vspace{-0.8mm}\\
                     &  $\pm33$   &   $\pm 30$ &   $\pm31$  &  $\pm48$\\

$L_s/L_p$            &  0.4079    &    0.3828  &   0.3566   &  0.3693 \\

$\sigma$ \, (mag.)   &  0.0064    &    0.0063  &   0.0067   &  0.0102\\

\noalign{\smallskip}            
\hline
\end{tabular}            
\end{center}            
\end{table}                       

\begin{table}
\caption[]{\label{tab:adboo_wd_vh}
Photometric solutions for AD Boo (all data) from the WD code
adopting linear limb darkening coefficients by van Hamme (\cite{vh93});
see Table~\ref{tab:adboo_ebop_vh}.
Gravity darkening exponents of 0.33 and bolometric albedo
coefficients of 0.5 were applied, as appropriate for convective envelopes.
$T_{\rm {eff},p}$ was assumed to be 6575~K.
}
\begin{center}
\begin{tabular}{lrrrr} \hline
\hline\noalign{\smallskip}
                     &     $y$    &       $b$  &       $v$  &   $u$\\                   
\noalign{\smallskip}
\hline
\noalign{\smallskip}
$i$ \, (\degr)       &  87.47     &   87.78    &   87.79    &  87.82\vspace{-0.8mm}\\           
                     & $\pm 5$    &  $\pm 3$   &  $\pm 3$   & $\pm 6$\\                 

$\Omega_p$           &   6.743    &    6.711   &    6.717   &   6.702\vspace{-0.8mm}\\  
                     &  $\pm11$   &   $\pm 8$  &   $\pm 9$  &  $\pm14$\\                

$\Omega_s$           &   7.646    &    7.696   &    7.675   &   7.636\vspace{-0.8mm}\\  
                     &  $\pm16$   &   $\pm10$  &   $\pm 9$  &  $\pm18$\\                

$r_p$                &  0.1705    &   0.1713   &   0.1711   &  0.1716\\                 
                                                                                        
$r_s$                &  0.1303    &   0.1293   &   0.1297   &  0.1305\\                 
                                                                                        
$k$                  &  0.764     &   0.755    &   0.758    &  0.761\\                  
                                                                                        
$r_p + r_s$          &  0.3008    &   0.3006   &   0.3008   &  0.3021\\                 
                                                                                        
$T_{\rm {eff},s}$    &  6078      &   6052     &   5999     &  5946\vspace{-0.8mm}\\
                     &$\pm 4$     & $\pm 4$    &$\pm  4$    &$\pm 4$\\

$L_s/L_p$            &  0.4205    &    0.3868  &   0.3615   &  0.3761 \\

$\sigma$ \, (mag.)   &  0.0064    &    0.0062  &   0.0066   &  0.0101\\

\noalign{\smallskip}            
\hline
\end{tabular}            
\end{center}            
\end{table}                       

\begin{table}
\caption[]{\label{tab:adboo_ebop_vh-1989}
Photometric solutions for AD Boo from the EBOP code,
excluding observations from 1989.
Linear limb darkening coefficients from van Hamme (\cite{vh93})
were adopted; see Table~\ref{tab:adboo_ebop_vh}.
}
\begin{center}
\begin{tabular}{lrrrr} \hline
\hline\noalign{\smallskip}
                     &     $y$    &       $b$  &       $v$  &   $u$\\                   
\noalign{\smallskip}
\hline
\noalign{\smallskip}
$i$ \, (\degr)       &  87.59     &   87.67    &   87.83    &  87.93\vspace{-0.8mm}\\   
                     & $\pm13$    &  $\pm11$   &  $\pm11$   & $\pm17$\\                 

$r_p$                &  0.1703    &   0.1707   &   0.1698   &  0.1698\vspace{-0.8mm}\\  
                     &  $\pm 5$   &   $\pm 5$  &   $\pm 4$  &  $\pm 7$\\                

$r_s$                &  0.1284    &   0.1284   &   0.1279   &  0.1279\\                 
                                                                                        
$k$                  &   0.754    &   0.752    &   0.753    &   0.753\vspace{-0.8mm}\\  
                     &  $\pm 6$   &   $\pm4$   &   $\pm3$   &  $\pm 6$ \\               

$r_p + r_s$          &  0.2987    &   0.2991   &   0.2977   &  0.2977\\                 
                                                                                        
$J_s/J_p$            &  0.7339    &    0.6872  &   0.6408   &  0.6707\vspace{-0.8mm}\\
                     &  $\pm38$   &   $\pm 35$ &   $\pm35$  &  $\pm57$\\

$L_s/L_p$            &  0.4090    &    0.3809  &   0.3538   &  0.3627 \\

$\sigma$ \, (mag.)   &  0.0059    &    0.0055  &   0.0056   &  0.0091\\

\noalign{\smallskip}            
\hline
\end{tabular}            
\end{center}            
\end{table}

\begin{figure}
\epsfxsize=85mm
\epsfbox{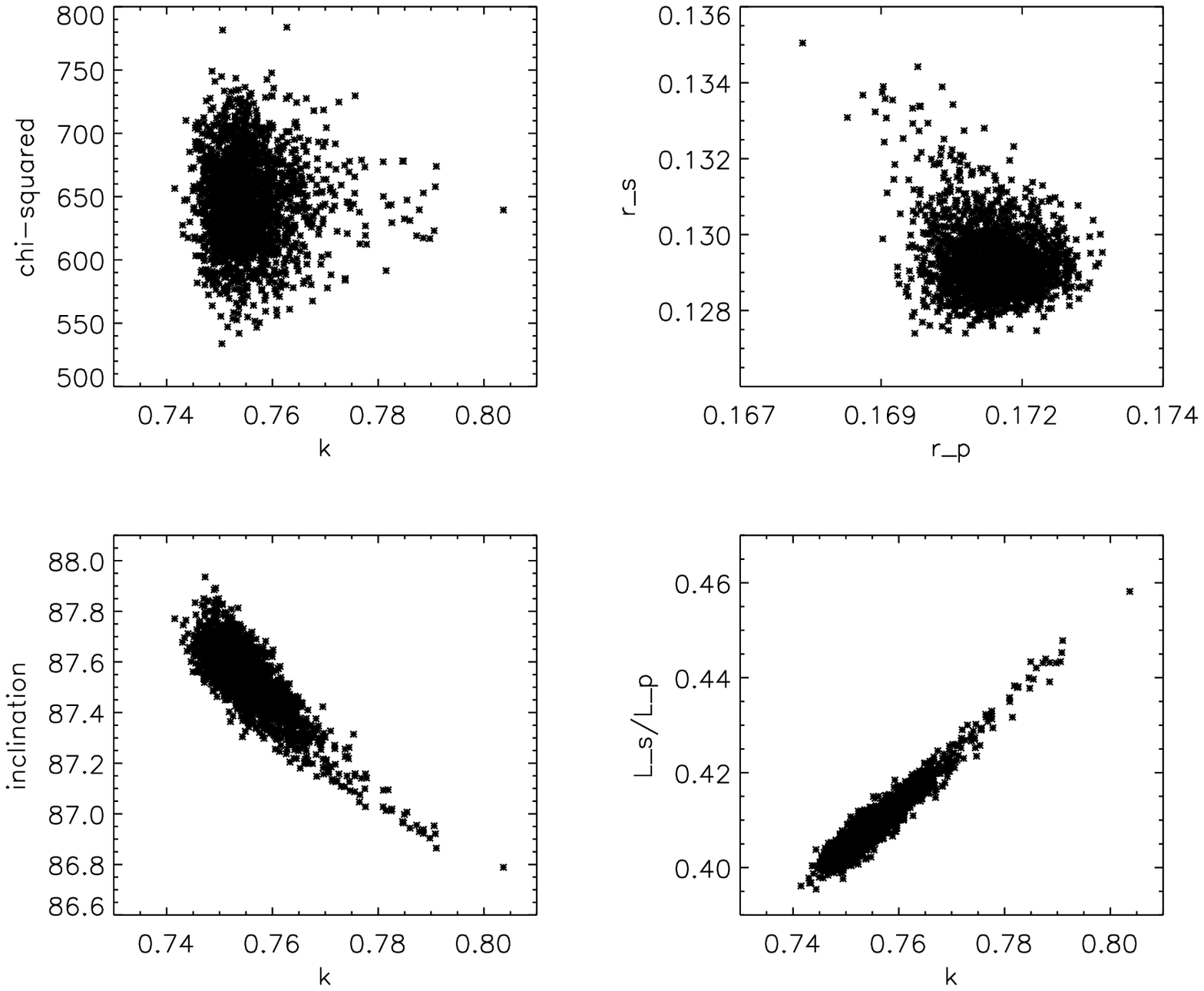}
\caption[]{\label{fig:adboo_mc_y}
Best fitting parameter values for the 10\,000 synthetic \AD\ $y$-band light 
curves created for the Monte Carlo analysis.
}
\end{figure}

\begin{figure}
\epsfxsize=85mm
\epsfbox{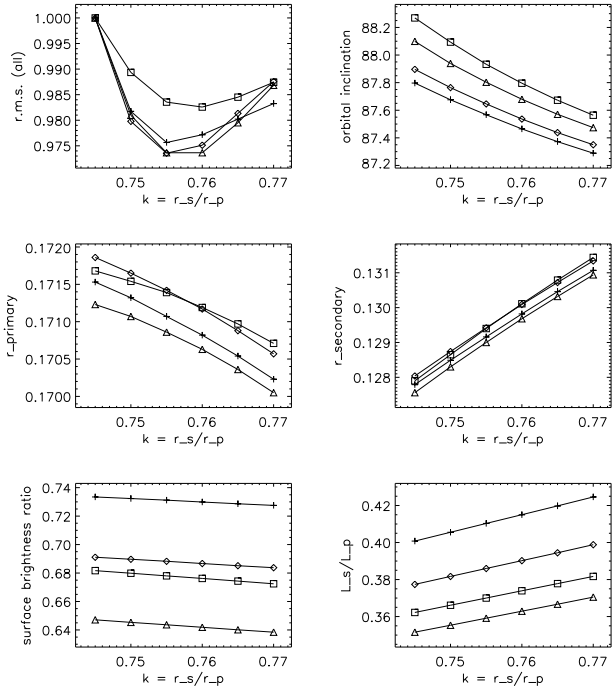}
\caption[]{\label{fig:adboo_ebop_vh}
EBOP solutions for \AD\ for a range of fixed $k$ va\-lues.
Linear limb-darkening coefficients by van Hamme (\cite{vh93}) 
were adopted. The upper left panel shows normalized rms
errors of the fit to the observations. Symbols are:
cross $y$; diamond $b$; triangle $v$; square $u$.
}
\end{figure}
 
Excluding the 1989 observations (see Fig~\ref{fig:adboo_res_vh_v}), we obtain 
the EBOP solutions listed in Table~\ref{tab:adboo_ebop_vh-1989}. The relative 
radii change by about 0.7\% in opposite directions and their sum by about 
$-$0.6\%, whereas the orbital inclination is unchanged in all four bands.
Including 2\% of third light changes $i$ by about $0\fdg3$, which is the 
maximum difference between the four bands, but we have not found any direct 
support for even such a small amount of third light in any of the analyses.
If we adopt linear limb-darkening coefficients from Claret (\cite{c00}), 
which are 0.04--0.07 higher, the orbital inclination decreases by $0\fdg16$, 
and the relative radius increases by about 0.0013 for the primary component 
but is unchanged for the secondary. If adjusted, the linear limb-darkening 
coefficients either become unrealistic and the orbital inclinations from 
the four bands disagree much more, or convergence fails.
Finally, WD solutions including non-linear limb darkening 
(see Sect.~\ref{sec:phel}) give photome\-tric elements very close to those 
presented in Table~\ref{tab:adboo_wd_vh}.
We are therefore unable to point at a clear explanation for the small $i$ 
discrepancies, which are, however, of no consequence for the determination 
of the absolute properties.

In order to assign realistic errors to the photome\-tric e\-le\-ments, we have 
supplemented these analyses with 10\,000 JKTEBOP Monte Carlo simulations in 
each band; see Fig.~\ref{fig:adboo_mc_y}. Furthermore, as shown in 
Fig.~\ref{fig:adboo_ebop_vh}, we have compared EBOP solutions for a range of 
fixed $k=r_s/r_p$ va\-lu\-es near 0.75.
The adopted photometric elements listed in Table~\ref{tab:adboo_phel}
are the weighted mean values of the EBOP solutions adopting the linear 
limb-darkening coefficients by van Hamme. Errors are based on the Monte 
Carlo simulations and comparison between the $uvby$ solutions, including 
those where the 1989 observations were excluded.
We find that secondary eclipse is in fact total, whereas at phase 0.0 about 
61\% of the $y$ light from the primary component is eclipsed. 
The spectroscopic light ratio of $\ell_{\rm s}/\ell_{\rm p} = 0.37\pm0.02$
at 5187\,\AA\ agrees quite well with that from the light curve solutions (0.40).

For comparison, Lacy (\cite{lacy97}), who applied the EBOP code 
to analyse the $B$ and $V$ normal points by Zhai et al. (\cite{zhai82}), 
obtained
$i = 87\fdg8 \pm 0\fdg2$, $r_p = 0.1702 \pm 0.0011$, $k = 0.750 \pm 0.010$, 
$J_s/J_p = 0.719 \pm 0.005$ and
$L_s/L_p = 0.403 \pm 0.015$ (both for the $V$ band).

\begin{table}            
\caption[]{\label{tab:adboo_phel}
Adopted photometric elements for AD\,Boo.
The individual flux and luminosity ratios are based
on the mean stellar and orbital parameters.
}
\begin{center}             
\begin{tabular}{ll}             
\noalign{\smallskip}             
\hline             
\noalign{\smallskip}             
$i$              & $87{\fdg}73 \pm 0{\fdg}15$ \\
$r_p$            & $0.1712 \pm 0.0015$ \\       
$r_s$            & $0.1291 \pm 0.0010$ \\       
$r_p + r_s$      & $0.3003 \pm 0.0020$ \\       
$k$              & $0.754\pm 0.006$ \\          
\noalign{\smallskip}             
\end{tabular}             
\begin{tabular}{lrrrr}             
\noalign{\smallskip}             
                 & $y$    & $b$    & $v$   & $u$  \\           
\noalign{\smallskip}             
$J_s/J_p$        & 0.734  & 0.690  & 0.642 & 0.678\\   
                 & $\pm11$&$\pm11$&$\pm10$ &$\pm11$\\   
$L_s/L_p$        & 0.410  & 0.384  & 0.356 & 0.369\vspace{-0.8mm} \\  
                 & $\pm 6$&$\pm 4$&$\pm 3$ &$\pm 5$\\   
\noalign{\smallskip}             
\hline             
\end{tabular}             
\end{center}            
\end{table}

\subsubsection{Abundances for \AD}
\label{sec:adboo_abund}

Two high S/N FEROS spectra are available for abundance analyses of \AD; 
see Table~\ref{tab:feros}.
We have concentrated on the one with the highest S/N, which gives significantly 
more accurate abundances. Due to the relatively high rotational velocities 
and low luminosity ratio, the number of suitable unblended lines is far 
lower than for \VZ\ and \WZ, especially for the secondary component. 
ATLAS9 models were used, and the effective temperatures, surface gravities and 
rotational velocities listed in Table~\ref{tab:absdim} were adopted, together 
with microturbulence velocities of 1.5 \kms\ for both components. 
The temperatures, which are derived from the $uvby$ indices of the components,
could not be constrained further from the spectroscopic analysis.

As seen in Table~\ref{tab:adboo_abund}, the \feh\ values from the Fe\,I lines 
(both components) and the few Fe\,II lines (primary) agree well, and
identical results are obtained if MARCS models are used. 
Changing model temperatures by the uncertainty of the effective temperatures 
($\pm 120$~K) modifies \feh\ from the Fe\,I lines by about $\pm 0.10$ dex, 
whereas the Fe\,II result is changed by only $\pm 0.04$ dex. 
If 0.5 \kms\ higher microturbulence velocities are adopted,
\feh\ decreases by about 0.10 dex for both neutral and ionized lines.
Taking these contributions into account, we obtain an average metal\-li\-ci\-ty
of \feh\,$=+0.07\pm0.14$ for \AD. Abundances for Ca and Ni are based on 
very few lines; although higher than for Fe, they agree within the 
uncertainties.

Metallicities have also been derived by comparing the disentangled CfA spectra 
to the extensive library of synthetic spectra especially prepared for the 
45\,\AA\ region centered at 5187\,\AA.
Surface gravities were fixed as above, and synchronous rotation of the 
components was adopted.
The best fits are obtained for \meh\,$=+0.13$ and $T_{\rm eff} = 6450$~K 
(primary), and \meh\,$=+0.30$ and $T_{\rm eff} = 6200$~K (secondary), 
but a strong correlation between metallicities and temperatures exists, and 
the determinations are therefore quite uncertain. We note that
the temperature difference between the components becomes smaller than 
established from the light curve analyses. Forcing the metallicities of the 
components to be identical, we derive \meh\,$=+0.25$ and temperatures of 
6530 and 6160~K for the components, in good agreement with the photometric
temperatures. Estimated uncertainties of metallicities and temperatures 
are (at least) 0.1 dex and 100~K.

Grids of TODCOR analyses of the  CfA spectra for a range of adopted template 
metallicities yield a metallicity estimate of \meh\,$=+0.33$, assuming 
temperatures of 6575 and 6145~K, and forcing the results for the components 
to be identical.

Finally, the calibrations by Holmberg et al. (\cite{holmberg07}) and the 
de-reddened $uvby$ indices (Table~\ref{tab:absdim}) yield 
\feh\,$=+0.28\pm0.15$ for the primary component ("blue" calibration) and 
\feh\,$=+0.10\pm0.13$ for the secondary.

Within the fairly large errors, the results from the different methods are
in reasonable agreement. Giving higher weight to the analysis of the FEROS 
spectra, we adopt \feh\,$=+0.10\pm0.15$ for \AD. This result should, 
however, be checked and preferably improved via analyses of disentangled 
component spectra based on a sufficiently large number of high-dispersion 
spectra.

\begin{table}
\caption[]{\label{tab:adboo_abund}
Abundances ($[\mathrm{El./H}]$) for the primary and se\-con\-da\-ry 
components of \AD.
N is the number of lines used per ion.}
\begin{center}
\begin{tabular}{lllrllr} \hline
\hline\noalign{\smallskip}
             &\multicolumn{3}{c}{Primary} & \multicolumn{3}{c}{Secondary} \\
Ion          &[El./H]&  rms&  N&[El./H]& rms & N  \\
\hline\noalign{\smallskip}
Ca I         &$+0.22$& 0.06&  4&$+0.20$& 0.23&  3 \\   
Fe I         &$+0.08$& 0.15& 33&$+0.06$& 0.17& 14 \\  
Fe II        &$+0.09$& 0.21&  4&       &     &    \\
Ni I         &$+0.22$& 0.11&  3&$+0.16$&     &  1 \\
\noalign{\smallskip}
\hline
\end{tabular}            
\end{center}            
\end{table}

\subsection{\VZ}
\label{sec:vzhya}

\VZ\ =  HD72257 = HIP41834 ($P = 2\fd90$) was dis\-co\-ve\-red by O'Connell 
(\cite{oconnell32}) to be an eclipsing binary, and spectroscopic elements 
were obtained from 75 {\AA}/mm spectra by Struve (\cite{struve45}).
The currently most reliable absolute dimensions of the components of 
\VZ\ are those given by Popper (\cite{dmp80}) in his critical review on 
stellar masses. 

\subsubsection{Radial velocities and spectroscopic orbit for \VZ}
\label{sec:vzhya_spec_orb}

\begin{figure}
\epsfxsize=095mm
\epsfbox{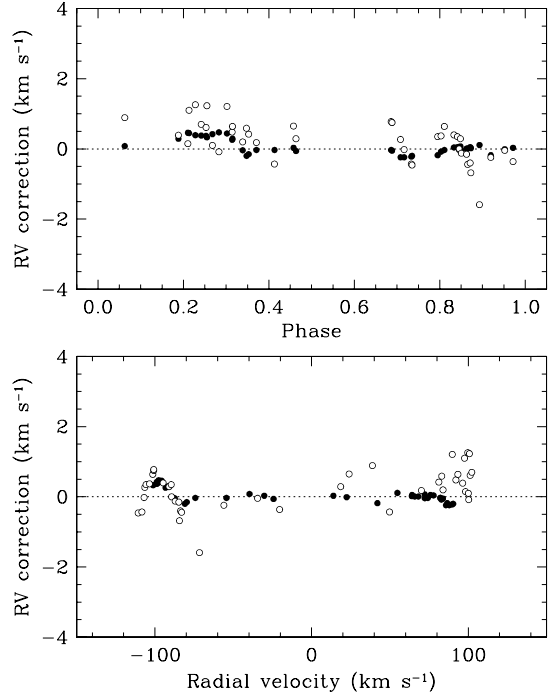}
\caption[]{\label{fig:vzhya_cor}
Systematic errors in the raw TODCOR velocities of \VZ, 
determined from simulations with synthetic binary spectra 
(filled circles: primary; open circles: secondary).
The differences are plotted both as function of orbital phase
(upper panel) and radial velocity (lower panel), and
have been applied to the measured velocities as corrections.
Phase 0.0 corresponds to central primary eclipse.
}
\end{figure}

\begin{figure}
\epsfxsize=095mm
\epsfbox{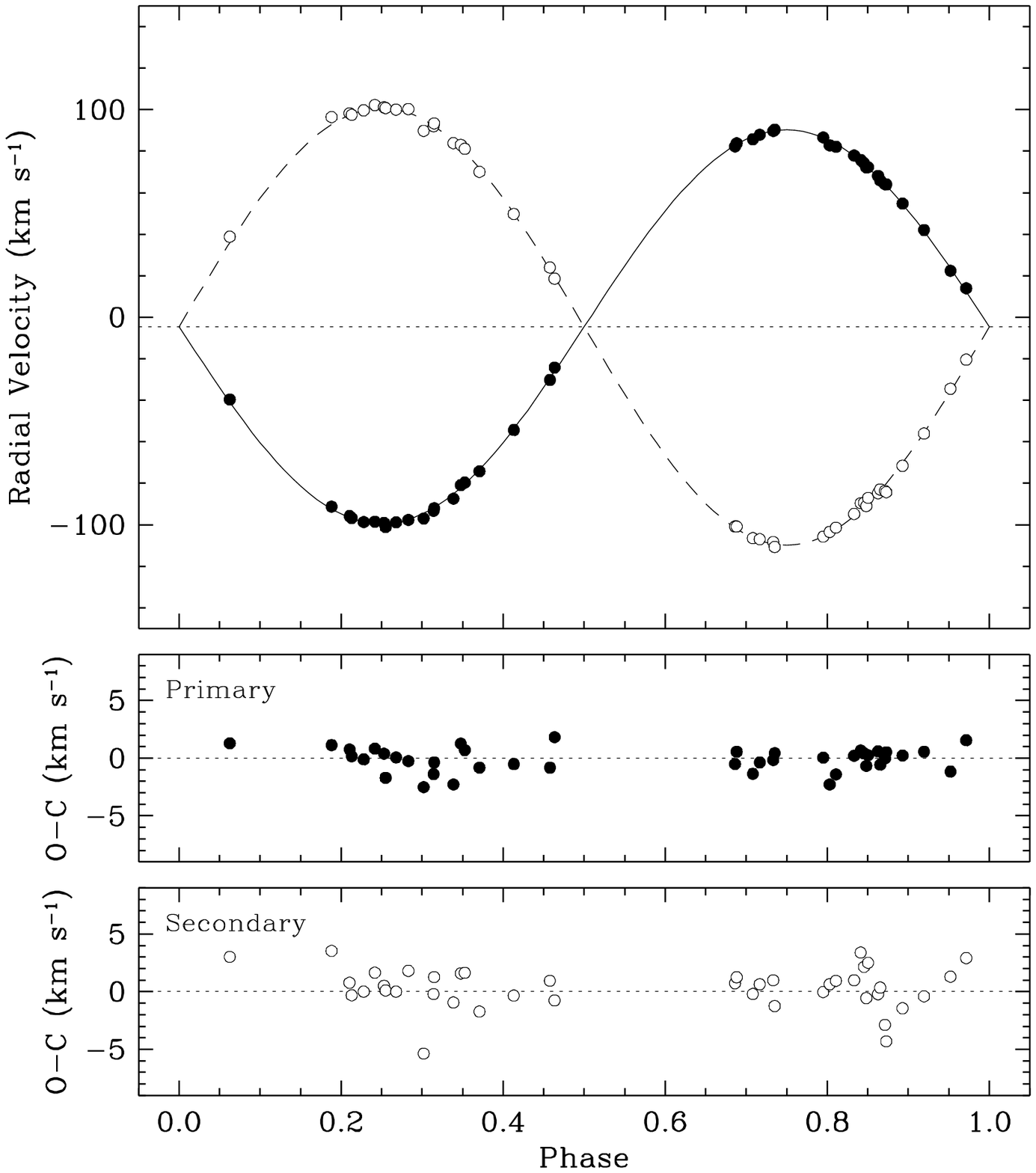}
\caption[]{\label{fig:vzhya_rv}
Spectroscopic orbital solution for \VZ\ (solid line: primary;
dashed line: secondary) and radial velocities (filled circles:
primary; open circles: secondary).
The dotted line (upper panel) represents the center-of-mass velocity of the
system. Phase 0.0 corresponds to central primary eclipse.
}
\end{figure}

\begin{table}   
\caption[]{\label{tab:vzhya_orb}
Spectroscopic orbital solution for VZ\,Hya.
$T$ is the time of central primary eclipse.}
\begin{center}    
\begin{tabular}{lr} \hline   
\hline\noalign{\smallskip}    
Parameter            & \multicolumn{1}{c}{Value} \\ 
\noalign{\smallskip}
\hline
\noalign{\smallskip}    
Adjusted quantities:            &   \\ 
$K_p$~(\kms)                    &$ 94.92 \pm 0.19 $  \\
$K_s$~(\kms)                    &$105.31 \pm 0.34 $  \\
$\gamma$~(\kms)                  &$ -4.57 \pm 0.14 $  \\
\noalign{\smallskip}  
Adopted quantities:             &     \\
$P$~(days)                      &  2.90430023 \\
$T$~(HJD$-$2\,400\,000)         &  48273.63450 \\
$e$                             &  0.00          \\ 
\noalign{\smallskip}  
Derived quantities:             &      \\
$M_p \sin^3 i$~(M$_{\sun}$)     & $ 1.2705 \pm 0.0090 $ \\
$M_s \sin^3 i$~(M$_{\sun}$)     & $ 1.1451 \pm 0.0061 $ \\
$a_p \sin i$~(10$^6$~km)      & $ 3.7907 \pm 0.0079 $ \\
$a_s \sin i$~(10$^6$~km)      & $ 4.2057 \pm 0.0140 $ \\
$a \sin i$~(R$_{\sun}$)         & $ 11.489 \pm 0.023  $ \\
\noalign{\smallskip}  
Other quantities pertaining to the fit:  &      \\
$N_{obs}$                      &    42 \\
Time span (days)               &   799 \\
$\sigma_p$~(\kms)              &  1.04 \\
$\sigma_s$~(\kms)              &  1.85 \\
\noalign{\smallskip}  
\hline
\end{tabular}            
\end{center}            
\end{table}

Our TODCOR analysis of \VZ\ was based on 42 usable CfA spectra outside of 
eclipse, observed from January 1989 to April 1991.
Templates of solar metallicity, $\log g$ = 4.5 (both components), and 
temperatures of 6750~K (primary) and 6250~K (secondary) were used to derive 
the radial ve\-lo\-ci\-ties.
As seen in Fig.~\ref{fig:vzhya_cor}, TODCOR corrections for syste\-ma\-tics 
in the radial velocities are fairly small, typically under 1 \kms\ for 
both components. They make very little dif\-fe\-ren\-ce in the masses, 
changing them by only +0.4\% (primary) and $-$0.1\% (secondary), which 
is well within the errors. 
The scatter around the orbit is about the same with and without corrections
applied.
Interpolating between $\log g$ = 4.0 and 4.5 grids to the exact gravities 
given in Table~\ref{tab:absdim}, we obtain temperatures of
$6590 \pm 100$~K for the primary and $6150 \pm 150$~K for the secondary,
both in excellent agreement with the photometric values; 
see Sect.~\ref{sec:absdim}.
The corresponding light ratio is $\ell_{\rm s}/\ell_{\rm p} = 0.53 \pm 0.02$ 
at the mean wavelength of our observations (5187\,\AA). Finally,
the grids of cross-correlations give best $v \sin i$ values of
21 and 20 \kms, with errors of 2 \kms\ and 3 \kms, respectively; 
they are consistent with synchronous rotation. 

The spectroscopic orbital solution for \VZ\ is given in
Table~\ref{tab:vzhya_orb}, and the observations and computed orbit
are shown graphically in Figure~\ref{fig:vzhya_rv} along with the
residuals for each star.
Our velocity semiamplitudes agree well with those by Popper (\cite{dmp65}),
($K_p, K_s$) = ($94.3 \pm 2.3$, $103.9 \pm 1.8$) but are
significantly more accurate.

For comparison, we have also disentangled the 29 best CfA spectra 
outside of eclipse and obtain velocity semiamplitudes of $K_p = 94.62$ 
and $K_s = 105.87$, in good agreement with the results based on TODCOR 
velocities. 

\subsubsection{Photometric elements for \VZ}
\label{sec:vzhya_phel}

The $uvby$ light curves of \VZ\ contain 1180 observations in 
each band and were observed on 44 nights during four periods between 
February 1989 and April 1992 (CVG08).
The average observational accuracy per point is about 4 mmag ($vby$) and 
6 mmag ($u$), and throughout all phases the points scatter at
these levels, indicating that the components of \VZ\ are constant
within the precision of our data.
Furthermore, we see no signs of emission in Ca\,II\,H and K, or in
$\mathrm{H_{\alpha}}$, in the FEROS spectra.

\begin{figure}
\epsfxsize=85mm
\epsfbox{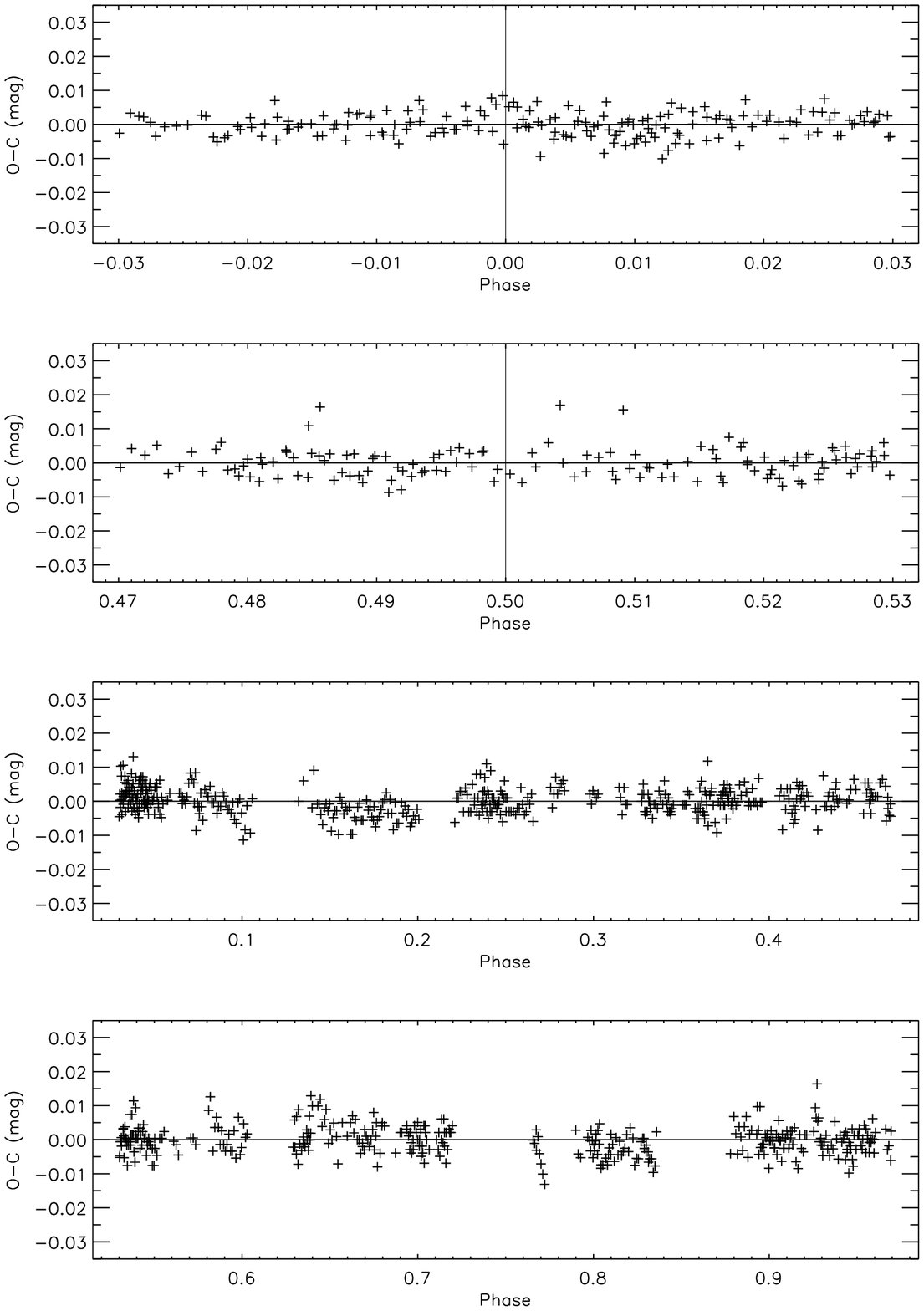}
\caption[]{\label{fig:vzhya_res_vh_b}
($O\!-\!C$) residuals of the VZ\,Hya $b$-band observations from the theoretical
light curve computed for the photometric elements given in 
Table~\ref{tab:vzhya_ebop_vh}.
}
\end{figure}

EBOP and WD solutions based on linear limb-darkening coefficients
by van Hamme (\cite{vh93}) are shown in Tables~\ref{tab:vzhya_ebop_vh}
and \ref{tab:vzhya_wd_vh}, respectively. As seen, the results from the
different bands agree very well, and the two models lead to practically 
identical photometric elements. The same holds for WINK solutions not 
included here.
If we adopt instead linear limb-darkening coefficients by Claret (\cite{c00}), 
which are 0.05--0.08 higher, the orbital inclination decreases by $0\fdg2$, 
and the relative radius is unchanged for the primary component but increases 
by 0.0009 for the secondary. If adjusted, the linear limb-darkening 
coefficients by van Hamme are reproduced to within $\pm 0.05$ and the 
solutions are close to the results in Tables~\ref{tab:vzhya_ebop_vh} 
and \ref{tab:vzhya_wd_vh}. 
WD solutions in\-clu\-ding non-linear limb darkening (see Sect.~\ref{sec:phel})
give photometric elements very close to those presented in 
Table~\ref{tab:vzhya_wd_vh}.
$O\!-\!C$ residuals of the $b$ observations from the theoretical light curve
computed for the photometric elements given in Table~\ref{tab:vzhya_ebop_vh} 
are shown in Fig.~\ref{fig:vzhya_res_vh_b}.

\begin{table}
\caption[]{\label{tab:vzhya_ebop_vh}
Photometric solutions for VZ\,Hya from the EBOP code
adopting linear limb darkening coefficients by van Hamme (\cite{vh93}).
The errors quoted for the free parameters are the $formal$ errors determined 
from the iterative least squares solution procedure.
}
\begin{center}
\begin{tabular}{lrrrr} \hline
\hline\noalign{\smallskip}
                     &     $y$    &       $b$  &       $v$  &   $u$\\                   
\noalign{\smallskip}
\hline
\noalign{\smallskip}
$i$ \, (\degr)       &  88.83     &   88.91    &   88.88    &  88.89\vspace{-0.8mm}\\   
                     & $\pm 6$    &  $\pm 5$   &  $\pm 5$   & $\pm 8$\\                 

$r_p$                &  0.1142    &   0.1144   &   0.1144   &  0.1140\vspace{-0.8mm}\\  
                     &  $\pm 4$   &   $\pm 3$  &   $\pm 3$  &  $\pm 5$\\                

$r_s$                &  0.0969    &   0.0967   &   0.0969   &  0.0970\\                 
                                                                                        
$k$                  &   0.849    &   0.845    &   0.847    &   0.850\vspace{-0.8mm}\\  
                     &  $\pm 7$   &   $\pm4$   &   $\pm5$   &  $\pm 8$ \\               

$r_p + r_s$          &  0.2111    &   0.2111   &   0.2113   &  0.2110\\                 
                                                                                        
$u_p$                &  0.52      &   0.60     &   0.67     &  0.64\\                   % from vanHamme
$u_s$                &  0.54      &   0.63     &   0.71     &  0.71\\                   % from vanHamme    
$y_p$                &  0.34      &   0.39     &   0.44     &  0.52\\

$y_s$                &  0.35      &   0.41     &   0.47     &  0.55\\

$J_s/J_p$            &  0.7983    &    0.7665  &   0.7306   &  0.7649\vspace{-0.8mm}\\
                     &  $\pm20$   &   $\pm 16$ &   $\pm18$  &  $\pm29$\\

$L_s/L_p$            &  0.5701    &    0.5402  &   0.5143   &  0.5364 \\

$\sigma$ \, (mag.)   &  0.0046    &    0.0040  &   0.0044   &  0.0068\\

\noalign{\smallskip}            
\hline
\end{tabular}            
\end{center}            
\end{table}                       
          
\begin{table}
\caption[]{\label{tab:vzhya_wd_vh}
Photometric solutions for VZ\,Hya from the WD code
adopting linear limb darkening coefficients by van Hamme (\cite{vh93});
see Table~\ref{tab:vzhya_ebop_vh}.
Gravity darkening exponents of 0.33 and bolometric albedo
coefficients of 0.5 were adopted, as appropriate for convective envelopes.
$T_{{\rm eff},p}$ was assumed to be 6650 K.
}
\begin{center}
\begin{tabular}{lrrrr} \hline
\hline\noalign{\smallskip}
                     &     $y$    &       $b$  &       $v$  &   $u$\\                   
\hline\noalign{\smallskip}            
$i$ \, (\degr)       &  88.90     &   88.92    &   88.95    &  88.98\vspace{-0.8mm}\\           
                     & $\pm 2$    &  $\pm 1$   &  $\pm 2$   & $\pm 2$\\                 

$\Omega_p$           &   9.649    &    9.666   &    9.639   &   9.668\vspace{-0.8mm}\\  
                     &  $\pm 8$   &   $\pm 7$  &   $\pm 8$  &  $\pm11$\\                

$\Omega_s$           &  10.349    &   10.329   &   10.355   &  10.343\vspace{-0.8mm}\\  
                     &  $\pm 8$   &   $\pm 7$  &   $\pm 8$  &  $\pm11$\\                

$r_p$                &  0.1144    &   0.1142   &   0.1146   &  0.1142\\                 
                                                                                        
$r_s$                &  0.0969    &   0.0971   &   0.0968   &  0.0969\\                 
                                                                                        
$k$                  &  0.847     &   0.850    &   0.845    &  0.849\\                  
                                                                                        
$r_p + r_s$          &  0.2113    &   0.2113   &   0.2114   &  0.2111\\                 
                                                                                        
$T_{{\rm eff},s}$    &  6272      &   6252     &   6229     &  6182\vspace{-0.8mm}\\
                     &$\pm 2$     & $\pm 1$    &$\pm  2$    &$\pm 3$\\

$L_s/L_p$            &  0.5686    &    0.5464  &   0.5125   &  0.5347 \\

$\sigma$ \, (mag.)   &  0.0046    &    0.0039  &   0.0043   &  0.0068\\
\noalign{\smallskip}
\hline
\end{tabular}
\end{center}
\end{table}

\begin{figure}
\epsfxsize=85mm
\epsfbox{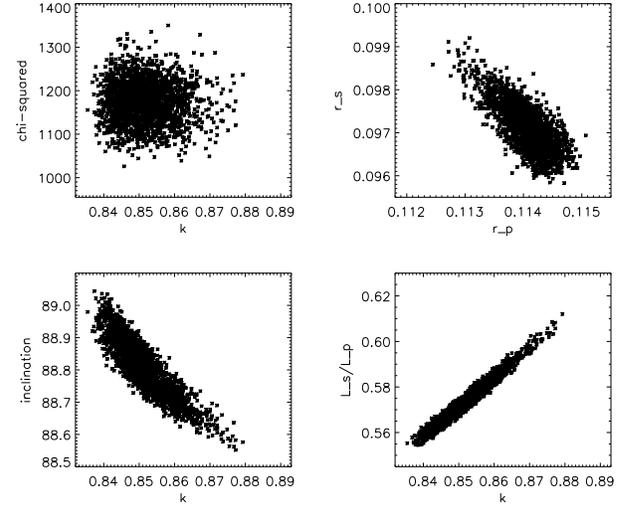}
\caption[]{\label{fig:vzhya_mc_y}
Best fitting parameter values for the 10\,000 synthetic \VZ\ $y$ light curves
created for the Monte Carlo analysis.
}
\end{figure}

\begin{figure}
\epsfxsize=85mm
\epsfbox{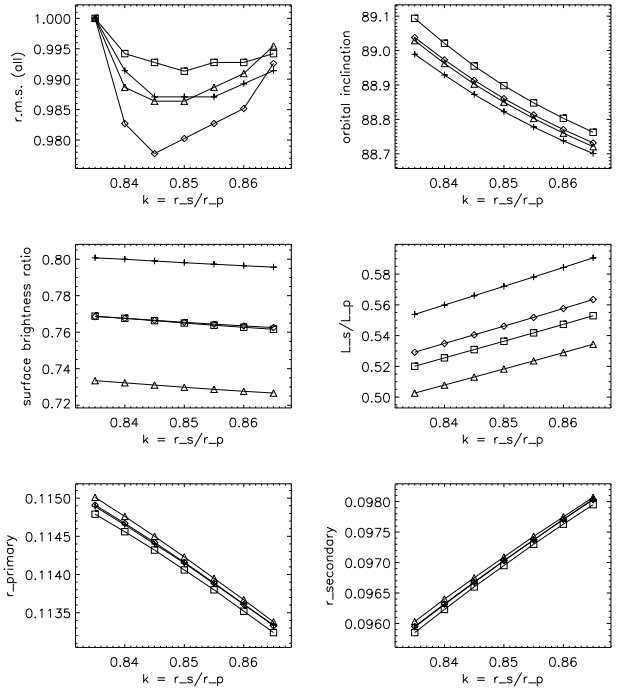}
\caption[]{\label{fig:vzhya_ebop_vh}
EBOP solutions for \VZ\ for a range of fixed $k$ va\-lues.
Linear limb-darkening coefficients by van Hamme (\cite{vh93}) 
were adopted. The upper left panel shows normalized rms
errors of the fit to the observations. Symbols are:
cross $y$; diamond $b$; triangle $v$; square $u$.
}
\end{figure}

As done for \AD, in order to assign realistic errors to the photometric 
elements, we have supplemented these analyses with 10\,000 JKTEBOP 
Monte Carlo simulations in each band, and EBOP solutions for a range of 
fixed $k$ values near 0.85; see Fig.~\ref{fig:vzhya_mc_y} and 
Fig.~\ref{fig:vzhya_ebop_vh}.
The adopted photometric elements listed in Table~\ref{tab:vzhya_phel}
are the weighted mean values of the EBOP solutions adopting the linear 
limb-darkening coefficients by van Hamme. Errors are based on the 
Monte Carlo simulations and a comparison between the $uvby$ solutions. 
Our results are significantly more accurate than obtained in previous 
studies of \VZ.
At phase 0.0, about 76\% of the $y$ light from the primary component is 
eclipsed, and at phase 0.5 about 98\% of the $y$ light of the secondary is 
blocked out, meaning that secondary eclipse is almost total. 
The spectroscopic light ratio of $\ell_{\rm s}/\ell_{\rm p} = 0.53 \pm 0.02$
at 5187\,\AA\ agrees quite well with that from the light curve solutions (0.56).

Previous analyses based on the photoelectric $UBV$ light curves by Walker 
(\cite{walker70}) gave 3--4\% larger relative radii for the primary component, 
whereas those for the secondary component agree well with our result 
(Walker \cite{walker70}, Wood \cite{wood71}, Cester et al. \cite{cester78}).
Another set of photoelectric $UBV$ light curves was published by Padalia \& 
Srivastava (\cite{ps75}), who found primary eclipse to be an occultation 
(larger star in front) rather than a transit (smaller star in front). 
This picture of \VZ\ was soon shown by Popper (\cite{dmp76}) to be wrong but 
was nevertheless defended once more by Padalia (\cite{p86}).

\begin{table}            
\caption[]{\label{tab:vzhya_phel}
Adopted photometric elements for VZ\,Hya.
The individual flux and luminosity ratios are based
on the mean stellar and orbital parameters.
}
\begin{center}             
\begin{tabular}{ll}             
\noalign{\smallskip}             
\hline             
\noalign{\smallskip}             
$i$              & $88{\fdg}88 \pm 0{\fdg}09$ \\
$r_p$            & $0.1143 \pm 0.0004$ \\       
$r_s$            & $0.0968 \pm 0.0006$ \\       
$r_p + r_s$      & $0.2111 \pm 0.0004$ \\       
$k$              & $0.847\pm 0.007$ \\          
\noalign{\smallskip}             
\end{tabular}             
\begin{tabular}{lrrrr}             
\noalign{\smallskip}             
                 & $y$    & $b$    & $v$   & $u$  \\           
\noalign{\smallskip}             
$J_s/J_p$        & 0.799  & 0.766  & 0.730 & 0.764\\   
                 & $\pm12$&$\pm12$&$\pm11$ &$\pm12$\\   
$L_s/L_p$        & 0.569  & 0.543  & 0.515 & 0.532\vspace{-0.8mm} \\  
                 & $\pm 9$&$\pm 6$&$\pm 6$ &$\pm 9$\\   
\noalign{\smallskip}             
\hline             
\end{tabular}             
\end{center}            
\end{table}

\subsubsection{Abundances for \VZ}
\label{sec:vzhya_abund}

\begin{table}
\caption[]{\label{tab:vzhya_abund}
Abundances ($[\mathrm{El./H}]$) for the primary and se\-con\-da\-ry 
components of \VZ.
N is the number of lines used per ion.}  
\begin{center}
\begin{tabular}{lllrllr} \hline
\hline\noalign{\smallskip}
             & \multicolumn{3}{c}{Primary} & \multicolumn{3}{c}{Secondary} \\
Ion          &[El./H]&  rms&  N&[El./H]& rms & N  \\
\hline\noalign{\smallskip}
Si I         &$-0.14$& 0.17&  7&$-0.18$& 0.16& 12 \\  
Ca I         &$-0.13$& 0.15&  8&$-0.17$& 0.16&  7 \\
Sc II        &$-0.36$& 0.14&  2&$-0.33$& 0.07&  3 \\
Ti I         &$-0.25$&     &  1&$-0.21$& 0.17&  4 \\
Ti II        &$-0.48$& 0.43&  3&$-0.30$& 0.09&  2 \\
Cr I         &$-0.23$& 0.16&  6&$-0.24$& 0.11&  4 \\
Cr II        &$-0.29$& 0.03&  4&$-0.43$& 0.06&  4 \\
Fe I         &$-0.21$& 0.16& 91&$-0.19$& 0.18& 69 \\
Fe II        &$-0.32$& 0.18& 12&$-0.14$& 0.26& 11 \\
Ni I         &$-0.20$& 0.23&  9&$-0.07$& 0.27& 16 \\
\noalign{\smallskip}
\hline
\end{tabular}            
\end{center}            
\end{table}

Five FEROS spectra are available for the detailed abundance analysis of \VZ; 
see Table~\ref{tab:feros}. As mentioned in Sect.~\ref{sec:abund}, disentangling 
of so few spectra was unsuccessful, and we have instead used the three most 
suitable observed spectra and analysed them as double-lined.
The results from the VWA analysis are presented in Table~\ref{tab:vzhya_abund}. 
ATLAS9 models were used, and the final effec\-ti\-ve tem\-pe\-ra\-tu\-res, 
surface gravities and rotational velocities listed in Table~\ref{tab:absdim} 
were adopted, together with microturbulence velocities of 2.0 and 1.6 \kms, 
respectively, calculated from the calibration by Edvardsson et al. 
(\cite{be93}). 
For this choice of parameters there is no dependence on equivalent width and 
excitation potential for abundances determined from individual lines, which 
on the other hand appears if temperatures are changed by about 150~K. 

As seen, the \feh\ values from the many Fe\,I lines agree well, whereas 
those from the much fewer Fe\,II lines are lower (primary) and higher 
(secondary). Similar results are obtained if MARCS models are used, and 
we have not been able to identify the cause of the discrepancies.
Changing model temperatures by the uncertainty of the effective temperatures 
($\pm 150$~K) modifies \feh\ from the Fe\,I lines by $\pm 0.10$ dex, 
whereas the Fe\,II results are changed by only $\pm 0.04$ dex. 
If 0.5 \kms\ higher microturbulence velocities are adopted,
\feh\ decreases by about 0.06 dex for both neutral and ionized lines.
Taking these contributions into account, we obtain a weighted average of 
\feh\,$=-0.20\pm0.12$ for \VZ.

Excluding ions with few measurements, we find that
within the uncertainties, identical abundances are derived from 
Cr\,I and Ni\,I lines. The $\alpha$-elements Si and Ca seem slightly 
enhanced, but the difference is barely significant.

For the metallicities derived by comparing the disentangled CfA spectra with 
the library of synthetic spectra, surface gravities were fixed as above, 
and synchronous rotation of the components was adopted.
The best fits are obtained for \meh\,$=+0.09$ and $T_{\rm eff} = 6630$~K 
(primary), and \meh\,$=-0.23$ and $T_{\rm eff} = 6130$~K (secondary), 
but as for \AD, a strong correlation between metallicities and temperatures 
exist, and the determinations are therefore quite uncertain. 
Forcing the metallicities of the components to be identical, we derive 
\meh\,$=-0.15$, and temperatures of 6595 and 6185~K for the components, 
in good agreement with the FEROS result and the photometric temperatures. 
Estimated uncertainties of these CfA metallicities and temperatures are 
(at least) 0.1 dex and 100~K, respectively.

Comparing TODCOR analyses of the observed CfA spectra for a range of adopted 
template metallicities yields a metallicity estimate of \meh\,$=+0.02$, 
assuming temperatures of 6650 and 6300~K and forcing the metal abundance of 
the components to be identical.

Finally, the calibrations by Holmberg et al. (\cite{holmberg07}) and the 
de-reddened $uvby$ indices in Table~\ref{tab:absdim} yield 
\feh\,$=+0.05\pm0.15$ for the primary component ("blue" calibration) and 
\feh\,$=-0.08\pm0.12$ for the secondary, marginally in agreement with 
the FEROS result, which we adopt for \VZ.

\subsection{\WZ}
\label{sec:wzoph}

\WZ\ = HD154676 = HIP83719 ($P = 4\fd18$) was discovered by 
J.\ H.\ Metcalf to be an eclipsing binary (Pickering \cite{p17}), 
and Sanford (\cite{s37}) obtained the first radial velocities.
The currently best modern analysis of \WZ, leading to masses and radii 
accurate to $3$--$4$\%, was done by Popper (\cite{dmp65}) from photoelectric 
light curves and 20 {\AA}/mm spectra.

For the spectroscopic and photometric analyses of \WZ\, we have adopted the
well established ephemeris from CVG08, where the zero point has been shifted 
by half a period compared to that by Popper (\cite{dmp65}).
We note that the primary component eclipsed at phase 0.0 according to our 
adopted ephemeris, and corresponding to the slightly deeper minimum for 
the $uvby$ light curves, is given by Popper (\cite{dmp65}) as component 2. 
Popper found that his component 2, i.e. our primary component, has slightly 
stronger spectral lines. This is consistent with our measurements of 
equivalent widths in the disentangled FEROS spectra for lines stronger than 
about 100 m{\AA}.

\subsubsection{Radial velocities and spectroscopic orbit for \WZ}
\label{sec:wzoph_spec_orb}

\begin{figure}
\epsfxsize=095mm
\epsfbox{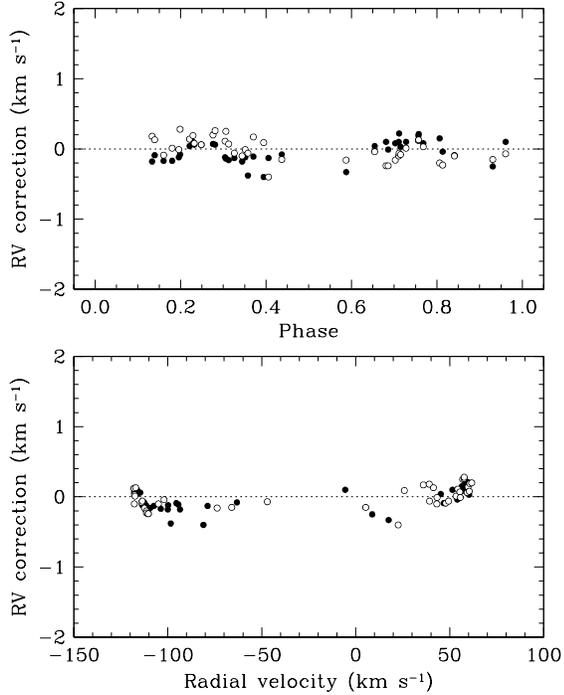}
\caption[]{\label{fig:wzoph_cor}
Systematic errors in the raw TODCOR velocities of \WZ, 
determined from simulations with synthetic binary spectra
(filled circles: primary; open circles: secondary).
The differences are plotted both as function of orbital phase
(upper panel) and radial velocity (lower panel), and
have been applied to the measured velocities as corrections.
Phase 0.0 corresponds to central primary eclipse.
}
\end{figure}

Radial velocities for the components of \WZ\ were determined with TODCOR
on the basis of 40 usable CfA spectra outside of eclipse, observed from 
February 1989 to September 1991.
Identical templates of solar composition with $\log g$ = 4.0 and a 
temperature of 6250~K were adopted for the two stars. 
The TODCOR corrections for systematics shown in Fig.~\ref{fig:wzoph_cor} are
less than 0.5 \kms\ at all phases for both components. This makes a
difference of only +0.3\% in the minimum masses, about half their
uncertainties, and the scatter of the orbit does not change
much with corrections. $v \sin i$ values of 18 and 17 \kms\ are
obtained for the primary and secondary components, respectively,
with estimated uncertainties of 1~\kms. They are consistent with 
synchronous rotation.

\begin{figure}
\epsfxsize=095mm
\epsfbox{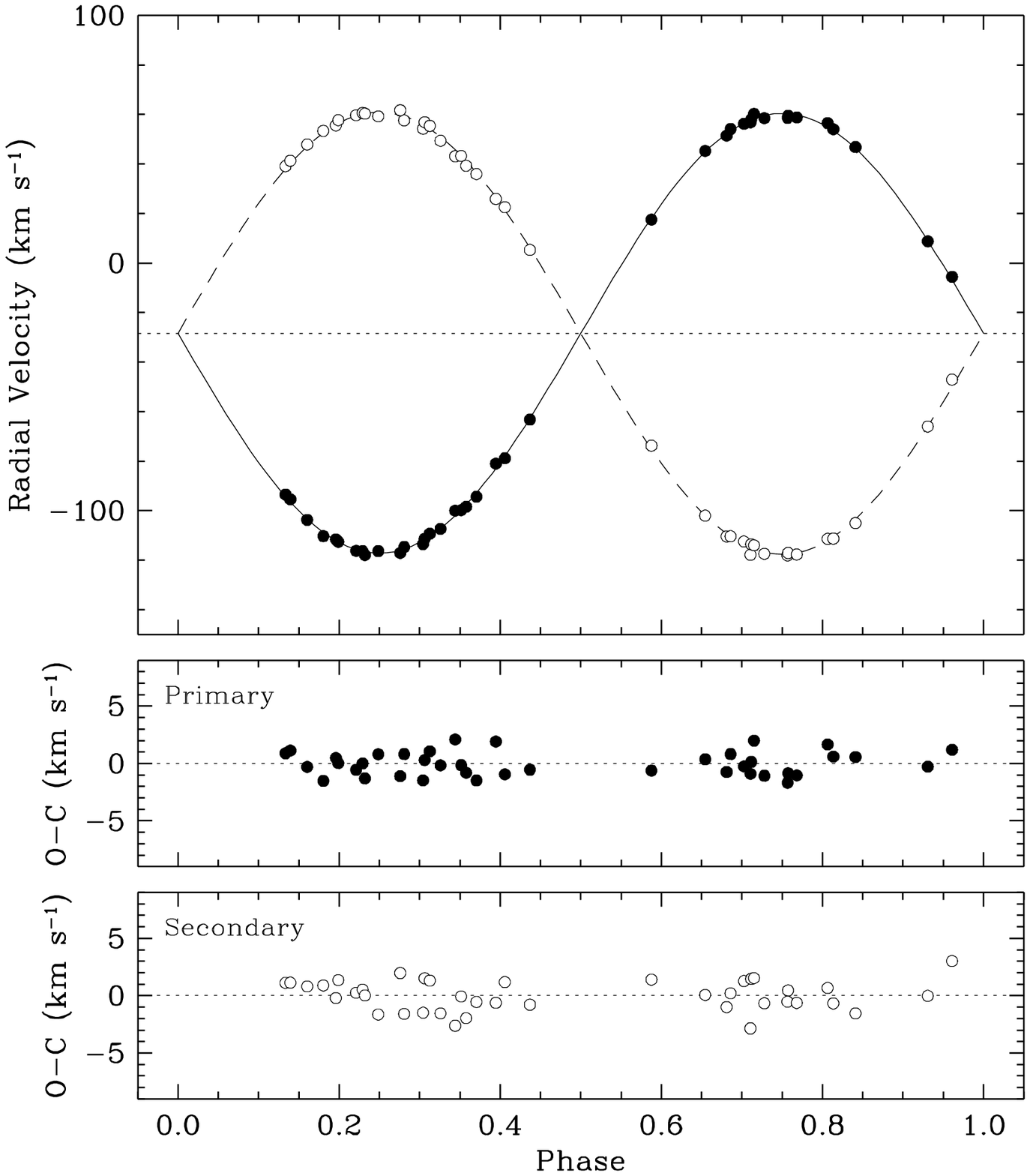}
\caption[]{\label{fig:wzoph_rv}
Spectroscopic orbital solution for \WZ\ (solid line: primary;
dashed line: secondary) and radial velocities (filled circles:
primary; open circles: secondary).
The dotted line (upper panel) represents the center-of-mass velocity of the
system. Phase 0.0 corresponds to central primary eclipse.
}
\end{figure}

\begin{table}   
\caption[]{\label{tab:wzoph_orb}
Spectroscopic orbital solution for WZ\,Oph.
$T$ is the time of central primary eclipse.}
\begin{center}    
\begin{tabular}{lr} \hline   
\hline\noalign{\smallskip}    
Parameter            & \multicolumn{1}{c}{Value} \\ 
\noalign{\smallskip}
\hline
\noalign{\smallskip}    
Adjusted quantities:            &   \\ 
$K_p$~(\kms)                    &$ 88.77 \pm 0.19 $  \\
$K_s$~(\kms)                    &$ 89.26 \pm 0.24 $  \\
$\gamma$~(\kms)                 &$-28.36 \pm 0.13 $  \\
\noalign{\smallskip}  
Adopted quantities:             &     \\
$P$~(days)                      &  4.18350681 \\
$T$~(HJD$-$2\,400\,000)         &  50353.78331 \\
$e$                             &  0.00          \\ 
\noalign{\smallskip}  
Derived quantities:             &      \\
$M_p \sin^3 i$~(M$_{\sun}$)                     & $1.2262 \pm 0.0073 $ \\
$M_s \sin^3 i$~(M$_{\sun}$)                     & $1.2194 \pm 0.0063 $ \\
$a_p \sin i$~(10$^6$~km)                      & $5.106 \pm 0.011 $ \\
$a_s \sin i$~(10$^6$~km)                      & $5.135 \pm 0.014 $ \\
$a \sin i$~(R$_{\sun}$)                         & $14.714 \pm 0.026 $ \\
\noalign{\smallskip}  
Other quantities pertaining to the fit:  &      \\
$N_{obs}$                      &    40 \\
Time span (days)               &   918 \\
$\sigma_p$~(\kms)              &  1.04 \\
$\sigma_s$~(\kms)              &  1.33 \\
\noalign{\smallskip}  
\hline
\end{tabular}            
\end{center}            
\end{table}

Our best spectroscopic estimates of the effective temperatures of the
components by interpolating to a $\log g$ value close to that listed
in Table~\ref{tab:absdim} are 6260~K for the primary and 6340~K
for the secondary, assuming solar metallicity. Interpolating to the
metallicity of \meh\,$=-0.27$ (see below) gives 6050~K and 6120~K. 
The lower values for the primary compared to the secondary disagree
formally with the trend from photometry, which gives a slightly hotter 
primary based on the slightly deeper primary eclipse. However, the 
difference is well within the spectroscopic error estimate of 150~K.
The light ratio is $\ell_{\rm s}/\ell_{\rm p} = 0.98 \pm 0.02$
at the mean wavelength of our observations (5187\,\AA).

The spectroscopic orbital solution for \WZ\ is given in
Table~\ref{tab:wzoph_orb} and the observations and computed orbit
are shown graphically in Figure~\ref{fig:wzoph_rv} along with the
residuals for each star.
For comparison, we have also disentangled the 40 CfA spectra 
outside of eclipse and obtain velocity semiamplitudes of $K_p = 88.62$ 
and $K_s =  88.92$, in good agreement with the results based on the 
TODCOR velocities. 

Our velocity semiamplitudes are larger than those by Popper (\cite{dmp65}),
($K_p, K_s$) = ($ 85.9  \pm 1.8  $, $ 87.0 \pm 2.0 $), and as in the case of
\VZ\ they are of much higher precision.

\subsubsection{Photometric elements for \WZ}
\label{sec:wzoph_phel}

The $uvby$ light curves of \WZ\ contain 697 observations in 
each band and were observed on 41 nights during six periods between 
March 1991 and March 1997 (CVG08).
The average observational accuracy per point is about 4 mmag ($vby$) and 
6 mmag ($u$), but throughout all phases the points scatter by 1--4
mmag more than this, highest in $u$. At a given phase, the observations from
different seasons do not differ systematically, and it is thus unclear
if the extra scatter in the light curves is due to slight variability of 
one or both stars (activity), or perhaps to an unknown 
instrumental/observational effect not seen in other systems. 
The Rossby number inferred for the mean component of \WZ\ does not clearly 
place it in the group of stars that show variability due to spots (Hall
\cite{hall94}), and is thus inconclusive. On the other hand, the high S/N 
FEROS spectrum taken at phase 0.33 (see Table~\ref{tab:feros}) indicates 
very weak emission from both components in the Ca II\,H and K lines, 
but not in $\mathrm{H_{\alpha}}$.  This suggests \WZ\ is in fact mildly active.

\begin{table}
\caption[]{\label{tab:wzoph_ebop_vh}
Photometric solutions for WZ\,Oph from the EBOP code
adopting linear limb darkening coefficients by van Hamme (\cite{vh93}).
The errors quoted for the free parameters are the $formal$ errors 
determined from the iterative least squares solution procedure.
}
\begin{center}
\begin{tabular}{lrrrr} \hline
\hline\noalign{\smallskip}
                     &     $y$    &       $b$  &       $v$  &   $u$\\                   
\hline\noalign{\smallskip}            
$i$ \, (\degr)       &  89.08     &   89.09    &   89.06    &  89.11\vspace{-0.8mm}\\   
                     & $\pm 2$    &  $\pm 2$   &  $\pm 2$   & $\pm 5$\\                 

$r_p$                &  0.0956    &   0.0950   &   0.0951   &  0.0949\vspace{-0.8mm}\\  
                     &  $\pm 8$   &   $\pm 7$  &   $\pm 7$  &  $\pm11$\\                

$r_s$                &  0.0963    &   0.0963   &   0.0960   &  0.0975\\                 
                                                                                        
$k$                  &   1.008    &   1.014    &   1.009    &   1.027\vspace{-0.8mm}\\  
                     &  $\pm16$   &   $\pm14$  &   $\pm14$  &  $\pm22$ \\               

$r_p + r_s$          &  0.1919    &   0.1913   &   0.1911   &  0.1924\\                 
                                                                                        
$u_p = u_s$          &  0.56      &   0.65     &   0.73     &  0.74\\                   

$y_p = y_s$          &  0.36      &   0.42     &   0.48     &  0.57\\

$J_s/J_p$            &  0.9878    &    0.9815  &   0.9831   &  0.9668\vspace{-0.8mm}\\
                     &  $\pm38$   &   $\pm 41$ &   $\pm46$  &  $\pm73$\\

$L_s/L_p$            &  1.0035    &    1.0102  &   1.0001   &  1.0209 \\

$\sigma$ \, (mag.)   &  0.0053    &    0.0057  &   0.0065   &  0.0104\\

\noalign{\smallskip}            
\hline
\end{tabular}            
\end{center}            
\end{table}                       
        
\begin{table}
\caption[]{\label{tab:wzoph_wd_vh}
Photometric solutions for WZ Oph from the WD code
adopting linear limb darkening coefficients by van Hamme (\cite{vh93});
see Table~\ref{tab:wzoph_ebop_vh}.
Gravity darkening exponents of 0.33 and bolometric albedo
coefficients of 0.5 were adopted, as appropriate for convective envelopes.
$T_{{\rm eff},p}$ was assumed to be 6165 K.
}
\begin{center}
\begin{tabular}{lrrrr} \hline
\hline\noalign{\smallskip}
                     &     $y$    &       $b$  &       $v$  &   $u$\\                   
\noalign{\smallskip}
\hline
\noalign{\smallskip}
$i$ \, (\degr)       &  89.11     &   89.11    &   89.08    &  89.14\vspace{-0.8mm}\\   
                     & $\pm 1$    &  $\pm 1$   &  $\pm 2$   & $\pm 3$\\                 

$\Omega_p$           &  11.454    &   11.510   &   11.554   &  11.444\vspace{-0.8mm}\\  
                     &  $\pm11$   &   $\pm15$  &   $\pm14$  &  $\pm32$\\                

$\Omega_s$           &  11.334    &   11.299   &   11.310   &  11.276\vspace{-0.8mm}\\  
                     &  $\pm12$   &   $\pm14$  &   $\pm17$  &  $\pm30$\\                

$r_p$                &  0.0957    &   0.0952   &   0.0948   &  0.0958\\                 
                                                                                        
$r_s$                &  0.0963    &   0.0967   &   0.0966   &  0.0969\\                 
                                                                                        
$k$                  &  1.007     &   1.016    &   1.019    &  1.012\\                  
                                                                                        
$r_p + r_s$          &  0.1920    &   0.1919   &   0.1914   &  0.1927\\                 
                                                                                        
$T_{{\rm eff},s}$    &  6149      &   6141     &   6141     &  6121\vspace{-0.8mm}\\
                     &$\pm 2$     & $\pm 2$    &$\pm  3$    &$\pm 5$\\

$L_s/L_p$            &  1.0023    &    1.0126  &   1.0180   &  0.9936 \\

$\sigma$ \, (mag.)   &  0.0053    &    0.0057  &   0.0065   &  0.0104\\

\noalign{\smallskip}            
\hline
\end{tabular}            
\end{center}            
\end{table}

\begin{figure}
\epsfxsize=85mm
\epsfbox{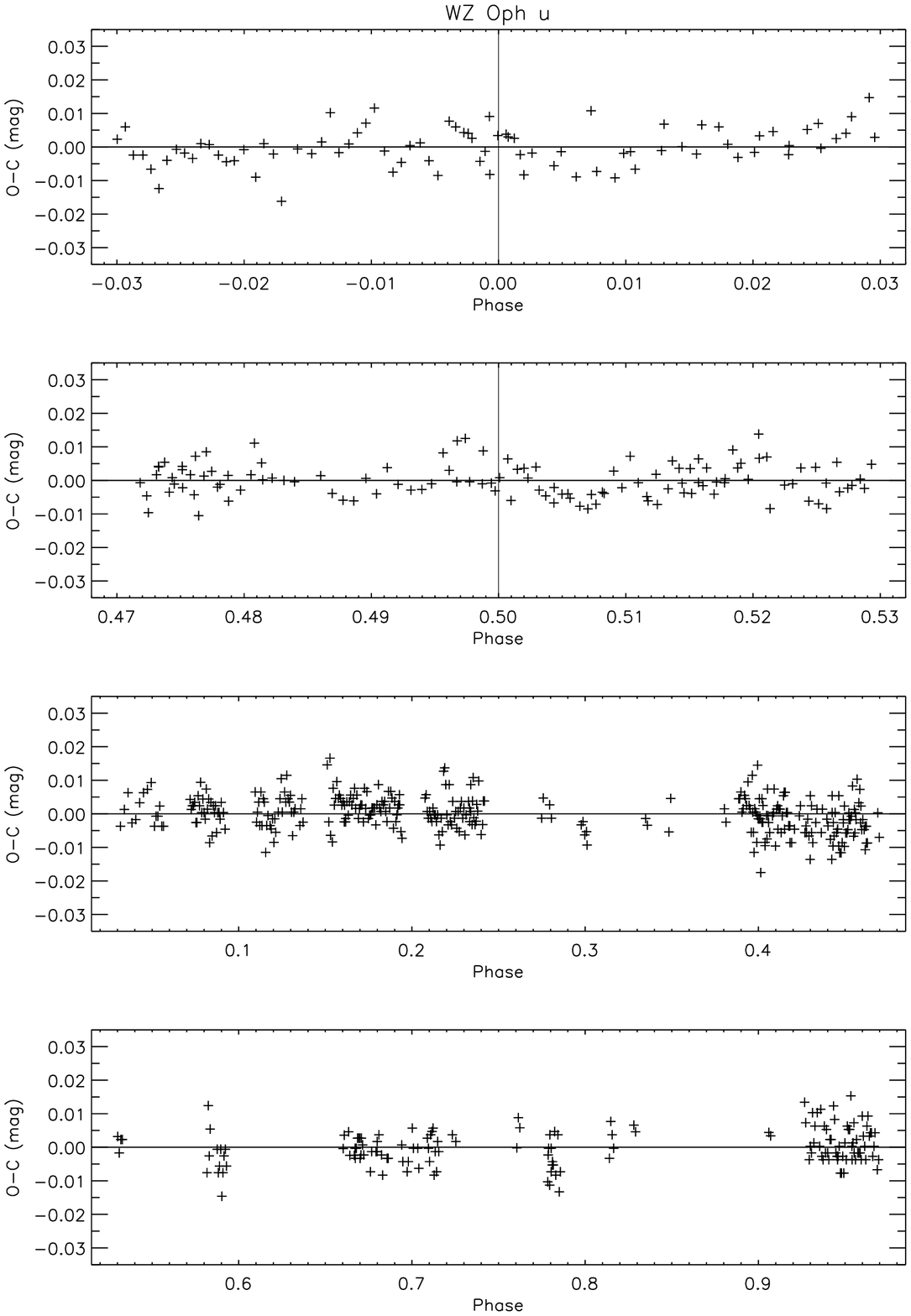}
\caption[]{\label{fig:wzoph_res_vh_y}
($O\!-\!C$) residuals of the WZ\,Oph $y$-band observations from the theoretical
light curve computed for the photometric elements given in 
Table~\ref{tab:wzoph_ebop_vh}.
}
\end{figure}

\begin{figure}
\epsfxsize=85mm
\epsfbox{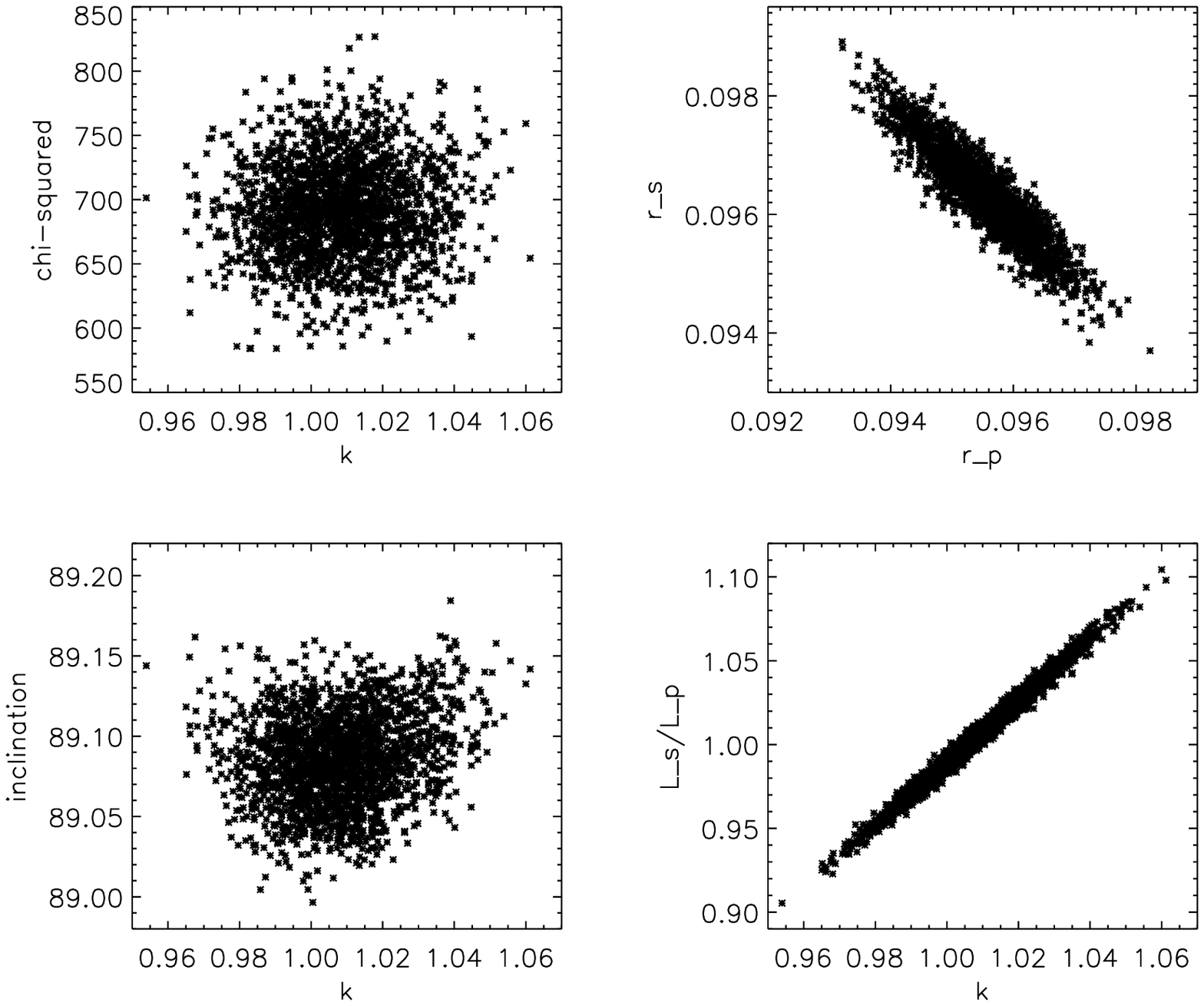}
\caption[]{\label{fig:wzoph_mc_y}
Best fitting parameter values for the 10\,000 synthetic \WZ\ $y$ light curves
created for the Monte Carlo analysis.
}
\end{figure}

\begin{figure}
\epsfxsize=90mm
\epsfbox{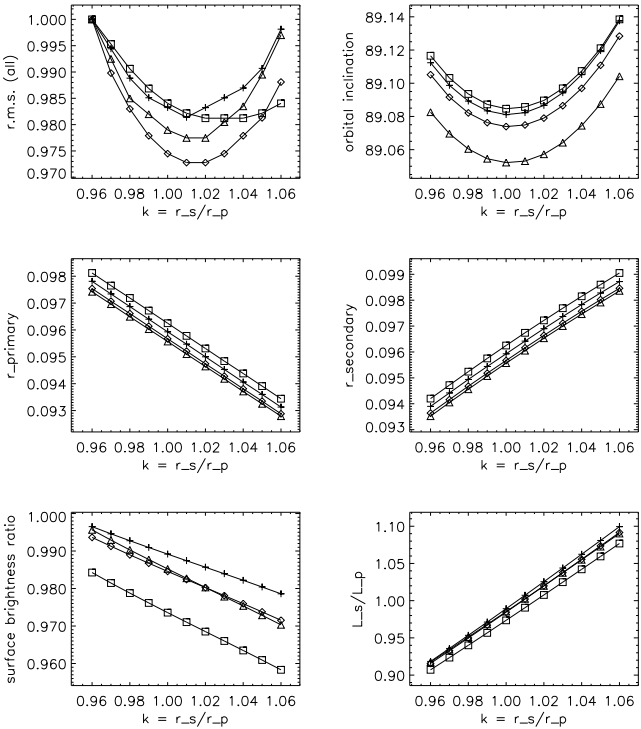}
\caption[]{\label{fig:wzoph_ebop_vh}
EBOP solutions for \WZ\ for a range of fixed $k$ va\-lues.
Linear limb-darkening coefficients by van Hamme (\cite{vh93}) 
were adopted. The upper left panel shows normalized rms
errors of the fit to the observations. Symbols are:
cross $y$; diamond $b$; triangle $v$; square $u$.
}
\end{figure}

EBOP and WD solutions, based on linear limb-darkening coefficients
by van Hamme (\cite{vh93}) are presented in Tables~\ref{tab:wzoph_ebop_vh}
and \ref{tab:wzoph_wd_vh}, respectively. As seen, the results from the
different bands agree well, and the two models lead to nearly identical 
photometric elements. WINK solutions not included here lead to changes in
the relative radii of $-0.0008$ (primary) and $+0.0008$ (secondary). 
Formally, the less massive secondary component
is slightly larger and cooler, but more luminous, than the primary for all
three models. In reality the two stars are, however, very nearly identical. 
If we adopt linear limb-darkening coefficients by Claret (\cite{c00}), 
which are 0.10 higher than those by van Hamme, the orbital inclination 
decreases by $0\fdg1$, and the relative radius increases by about 0.0004 
for the primary component but is unchanged for the secondary. If adjusted, 
the $vby$ linear limb-darkening coefficients by van Hamme
are reproduced to within $\pm 0.05$ and the solutions are close to the 
results in Table~\ref{tab:wzoph_ebop_vh}. The $u$ coefficients, however, 
become unrealistically low (0.60), and $i$ changes by $0\fdg2$. 
WD solutions including non-linear limb dar\-ke\-ning (see Sect.~\ref{sec:phel})
give photometric elements very close to those presented in 
Table~\ref{tab:wzoph_wd_vh} for the $uvb$ bands, but the $u$ solution now 
disagrees in having $k \approx 1.05$.
$O\!-\!C$ residu\-als of the $y$ observations from the theoretical light curve
computed for the photometric elements given in Table~\ref{tab:wzoph_ebop_vh} 
are shown in Fig.~\ref{fig:wzoph_res_vh_y}.

Realistic errors have been assigned to the photo\-me\-tric elements by 
performing 10\,000 JKTEBOP Monte Carlo simulations in each band, and EBOP 
solutions for a range of fixed $k$ values near 1.0; see 
Fig.~\ref{fig:wzoph_mc_y} and Fig.~\ref{fig:wzoph_ebop_vh}.
The adopted photometric elements listed in Table~\ref{tab:wzoph_phel}
are the weighted mean values of the EBOP solutions adopting the linear 
limb-darkening coefficients by van Hamme. Errors are based on the Monte 
Carlo simulations and comparison between the $uvby$ solutions.
Adopting the spectroscopic light ratio of 
$\ell_{\rm s}/\ell_{\rm p} = 0.98 \pm 0.02$ at 5187\,\AA\ would lead to 
$k = 0.996$ (formal error $\pm 0.011$), $r_p = 0.0960$, and 
$r_s = 0.0956$.
We find that about 92\% of the light of the components is shadowed out at both
central eclipses.

For comparison, Popper (\cite{dmp65}) obtained slightly smaller relative 
radii ($r_p = 0.0945 \pm 0.002$, $r_s = 0.092 \pm 0.001$), and an orbital 
inclination of $i = 89\fdg4 \pm 0\fdg2$ (all uncertainties given by Popper 
are estimated). From a WINK re-analysis of Popper's photoelectric light curve,
Cester et al. (\cite{cester78}) derived $r_p = 0.096 \pm 0.001$,  
$k = 0.971 \pm 0.017$, and $i = 89\fdg0 \pm 0\fdg1$.

\begin{table}            
\caption[]{\label{tab:wzoph_phel}
Adopted photometric elements for WZ\,Oph.
The individual flux and luminosity ratios are based
on the mean stellar and orbital parameters.
}
\begin{center}             
\begin{tabular}{ll}             
\noalign{\smallskip}             
\hline             
\noalign{\smallskip}             
$i$              & $89{\fdg}08 \pm 0{\fdg}10$ \\
$r_p$            & $0.0952 \pm 0.0008$ \\       
$r_s$            & $0.0964 \pm 0.0008$ \\       
$r_p + r_s$      & $0.1916 \pm 0.0004$ \\       
$k$              & $1.013\pm 0.015$ \\          
\noalign{\smallskip}             
\end{tabular}             
\begin{tabular}{lrrrr}             
\noalign{\smallskip}             
                 & $y$    & $b$    & $v$   & $u$  \\           
\noalign{\smallskip}             
$J_s/J_p$        & 0.987  & 0.981  & 0.982 & 0.972\\   
                 & $\pm15$&$\pm16$&$\pm16$ &$\pm17$\\   
$L_s/L_p$        & 1.013  & 1.007  & 1.008 & 0.997\vspace{-0.8mm}\\  
                 & $\pm30$&$\pm27$&$\pm26$ &$\pm37$\\   
\noalign{\smallskip}             
\hline             
\end{tabular}             
\end{center}            
\end{table}

\subsubsection{Abundances for \WZ}
\label{sec:wzoph_abund}

14 FEROS spectra are available of \WZ\ (see Table~\ref{tab:feros}),
which allowed us to obtain disentangled spectra of high quality for
the abundance studies. Results from the VWA analysis are presented in 
Table~\ref{tab:wzoph_abund}. ATLAS9 models were adopted, and the effective 
temperatures, surface gravities and rotational velocities listed in
Table~\ref{tab:absdim} were used, together with microturbulence velocities of
1.7 \kms\ for both components, calculated from the calibration by Edvardsson 
et al. (\cite{be93}). 
This choice of parameters is supported by the spectroscopic analyses, 
since we see no dependence on equivalent width and excitation potential 
for abundances determined from individual lines.
The precision of the temperatures ($\pm 100$~K) could not be improved from 
the abundance analysis.

As seen, a robust \feh\ is obtained with identical results for Fe\,I and 
Fe\,II lines from both components, also if MARCS models are used.  
Changing model temperatures by $\pm 100$~K modifies \feh\ from the Fe\,I 
lines by $\pm0.06$ dex, whereas the Fe\,II results are nearly unchanged. 
If 0.5 \kms\ higher microturbulence velocities are adopted,
\feh\ decreases by about 0.04 dex for both neutral and ionized lines.
Taking these contributions into account, we find \feh\,$=-0.27\pm0.07$ for \WZ.

\begin{table}
\caption[]{\label{tab:wzoph_abund}
Abundances ($[\mathrm{El./H}]$) for the primary and se\-con\-da\-ry
components of \WZ.
N is the number of lines used per ion.}  
\begin{center}
\begin{tabular}{lllrllr} \hline
\hline\noalign{\smallskip}
             &  \multicolumn{3}{c}{Primary} & \multicolumn{3}{c}{Secondary} \\
Ion          &[El./H]&  rms&  N&[El./H]& rms & N  \\
\hline\noalign{\smallskip}
Na I         &$-0.32$& 0.05&  4&$-0.35$& 0.05&  3 \\
Si I         &$-0.19$& 0.12& 20&$-0.18$& 0.09& 16 \\
Ca I         &$-0.21$& 0.13& 11&$-0.36$& 0.18&  9 \\
Sc II        &$-0.28$& 0.14&  5&$-0.30$& 0.07&  5 \\
Ti I         &$-0.32$& 0.23&  3&$-0.26$& 0.20&  3 \\
Ti II        &$-0.12$& 0.03&  4&$-0.17$& 0.10&  4 \\
V I          &$+0.03$& 0.25& 10&$+0.06$& 0.16&  3 \\
Cr I         &$-0.36$& 0.25& 11&$-0.39$& 0.25& 10 \\
Cr II        &$-0.17$& 0.17&  8&$-0.14$& 0.13&  6 \\
Mn I         &$-0.54$& 0.27&  9&$-0.64$& 0.22&  5 \\
Fe I         &$-0.27$& 0.12& 60&$-0.28$& 0.16& 74 \\
Fe II        &$-0.28$& 0.04&  9&$-0.27$& 0.08&  9 \\
Co I         &$+0.08$& 0.13&  4&$-0.18$& 0.31&  2 \\
Ni I         &$-0.25$& 0.23& 13&$-0.25$& 0.16& 15 \\
\noalign{\smallskip}
\hline
\end{tabular}            
\end{center}            
\end{table}

Excluding ions with only a few lines measured, we find that
within the uncertainties identical abundances are derived for Cr and Ni, 
noting that Cr\,II deviates, as seen for the Sun (Table~\ref{tab:sun_abund}), 
although our abundances are relative so that the effect should cancel out.
The $\alpha$-elements Si and Ca seem to be enhanced by about 0.1 dex, 
whereas Na and Ti (with only few lines measured) appear normal.
The Si result may perhaps be questioned, as in the case of Cr\,II above, 
since Si\,I gave a low abundance for the Sun.
For Mn, which together with Al is expected to anticorrelate with 
$\alpha$-elements, we actually derive a rather low abundance of 
$-0.54 \pm 0.13$, taking into account temperature and
microturbulence uncertainties as above. Unfortunately, problems with 
the Mn\,I line were also seen for the Sun. 
We tend, however, to conclude, that a slight $\alpha$-enhancement of about 
0.1 dex may be present. Finally, we derive an abundance for V close to solar.

\begin{table*}   
\caption[]{\label{tab:absdim}
Astrophysical data for \AD, \VZ, and \WZ.
$T_{eff\sun} = 5780$ K, 
$M_{bol\sun} = 4.74$, 
and bolometric corrections ($BC$) by Flower (\cite{flower96}) have been applied.
$v_{sync}$ is the equatorial velocity for synchronous rotation.
}
\begin{center}    
\begin{tabular}{lrrrrrr} \hline    
\noalign{\smallskip}    
\hline    
\noalign{\smallskip}    
                     &\multicolumn{2}{c} {AD Boo}            &\multicolumn{2}{c} {VZ Hya}            &\multicolumn{2}{c} {WZ Oph} \\
                     &    Primary        &    Secondary      &    Primary        &    Secondary      &    Primary        &    Secondary \\ 
\noalign{\smallskip}    
\hline    
\noalign{\smallskip}    
Absolute dimensions: &                   &                   &                   &                   &                   &
 \\ 
$M/M_{\sun}$         &$1.414 \pm 0.009$  &$1.209 \pm 0.006$  &$1.271 \pm 0.009$  &$1.146 \pm 0.006$  &$1.227 \pm 0.007$  &$1.220 \pm 0.006$
\\ 
$R/R_{\sun}$         &$1.612 \pm 0.014$  &$1.216 \pm 0.010$  &$1.314 \pm 0.005$  &$1.112 \pm 0.007$  &$1.401 \pm 0.012$  &$1.419 \pm 0.012$
\\ 
$\log g$ (cgs)       & $4.173 \pm 0.008$ & $4.351 \pm 0.007$ & $4.305 \pm 0.005$ & $4.405 \pm 0.006$ & $4.234 \pm 0.008$ & $4.221 \pm 0.008$
\\
$v \sin i$ (\kms)    & $38 \pm 2$        & $37 \pm 5$        & $21 \pm 2$        & $20 \pm 3$        & $18 \pm 1$        & $17 \pm 1$
\\ 
$v_{sync}$ (\kms)    & $39.4 \pm 0.4$    & $29.7 \pm 0.2$    & $22.9 \pm 0.1$    & $19.4 \pm 0.1$    & $16.9 \pm 0.1$    & $17.2 \pm 0.1$
\\ 
 & & \\ 
Photometric data:&                   &                       &                   &                   &                   &
\\ 
$V$\dag          & $9.752 \pm 0.008$ & $10.721 \pm 0.013$    &$9.442 \pm 0.008$  &   $10.054 \pm 0.012$ &$9.856 \pm 0.017$  &  $9.841 \pm 0.017$
\\  
$(b-y)$\dag      & $0.302 \pm 0.006$ & $ 0.371 \pm 0.007$    &$0.281 \pm 0.005$  &   $ 0.333 \pm 0.005$ &$0.362 \pm 0.005$  &  $0.370 \pm 0.005$
\\
$m_1$\dag        & $0.183 \pm 0.011$ & $ 0.198 \pm 0.011$    &$0.165 \pm 0.008$  &   $ 0.169 \pm 0.008$ &$0.149 \pm 0.008$  &  $0.141 \pm 0.008$
\\
$c_1$\dag        & $0.464 \pm 0.015$ & $ 0.340 \pm 0.016$    &$0.420 \pm 0.009$  &   $ 0.329 \pm 0.010$ &$0.369 \pm 0.011$  &  $0.382 \pm 0.012$
\\
$E(b-y)$    & \multicolumn{2}{c}   {$0.025 \pm 0.015$}       &\multicolumn{2}{c}   {$0.020 \pm 0.020$}  &\multicolumn{2}{c}   {$0.033 \pm 0.012$}
\\
$T_{\mbox{\scriptsize eff}}\,$ &  $6575 \pm 120$    & $6145 \pm 120$   &$6645 \pm 150$     &   $6290 \pm 150$ &  $6165 \pm 100$    &   $6115 \pm 100$
\\
$M_{\mbox{\scriptsize bol}}\,$ &  $3.14  \pm 0.08$  & $4.05  \pm 0.09$ &$6645 \pm 150$     &   $6290 \pm 150$ &  $3.73  \pm 0.07$  &   $3.74  \pm 0.07$ 
\\
$\log L/L_{\sun}$              & $0.64 \pm 0.03$    & $0.28 \pm 0.04$  &$0.48 \pm 0.04$ &    $0.24 \pm 0.04$  & $0.41 \pm 0.03$ &    $0.40 \pm 0.03$
\\
$BC$                           & $ 0.01$            & $-0.03$          &$ 0.02$         &    $-0.01$          & $-0.02$         &    $-0.03$
\\
$M_V$                          & $ 3.13 \pm 0.08$   & $ 4.08 \pm 0.09$ &$ 3.52 \pm 0.10$&   $ 4.15 \pm 0.10$  &$ 3.75 \pm 0.07$&   $ 3.77 \pm 0.07$
\\
$V_0-M_V$                      & $ 6.51  \pm 0.11 $ & $ 6.54  \pm 0.11$&$ 5.83  \pm 0.13 $& $ 5.82  \pm 0.14 $&$ 5.96 \pm 0.09 $& $5.94 \pm 0.09 $ 
\\
Distance \, (pc)               & $ 201   \pm  10  $ & $ 203   \pm  10 $&$ 147   \pm   9  $& $ 146   \pm   9  $ &$ 156   \pm   7  $& $154  \pm   7  $
\\
 & & & & & &\\ 
Abundance:                     &                    &                                     &                    &                  &\\              
\feh\                          & \multicolumn{2}{c}   {$+0.10 \pm 0.15$} & \multicolumn{2}{c}   {$-0.20 \pm 0.12$} & \multicolumn{2}{c}   {$-0.27 \pm 0.07$}
\\
\noalign{\smallskip}            
\hline
\noalign{\smallskip}            
\multicolumn{7}{l}{\dag\ Not corrected for interstellar absorption/reddening.}
\end{tabular}            
\end{center}            
\end{table*}

As for \AD\ and \VZ, metallicities have also been derived from the disentangled 
CfA spectra. Surface gravities were fixed as above, and synchronous rotation 
of the components was adopted.
The best fits are obtained for \meh\,$=-0.14$ and $T_{\rm eff} = 6280$~K 
(primary), and \meh\,$=-0.29$ and $T_{\rm eff} = 6250$~K (secondary), but as 
in the other two binaries there is a strong correlation between metallicity 
and temperature, and the determinations are therefore quite uncertain.
Forcing the metallicities of the components to be identical, we derive 
\meh\,$=-0.23$ and temperatures of 6230 and 6310~K for the components, in 
good agreement with the FEROS abundance result and in reasonable agreement 
with the photometric temperatures. Estimated uncertainties of metallicities 
and temperatures from this method are (at least) 0.1 dex and 100~K, 
respectively.

TODCOR analyses of the observed (composite) CfA spectra for a range of fixed 
template metallicities yield an estimate of \meh\,$=-0.27$, 
assuming temperatures of 6165 and 6115~K and constraining the results for 
the components to be identical.

Finally, the calibration by Holmberg et al. (\cite{holmberg07}) and the 
de-reddened $uvby$ indices yield \feh\,$=-0.17\pm0.11$ and
$-0.24\pm0.12$ for the primary and secondary components, respectively, 
also in good agreement with the FEROS result, which we adopt for \WZ. 

\section{Absolute dimensions}
\label{sec:absdim}

Absolute dimensions for \AD, \VZ, and \WZ\ are presented 
in Table~\ref{tab:absdim}, as calculated from the new
spectroscopic and photometric
elements given in Tables~\ref{tab:adboo_orb}, \ref{tab:adboo_phel},
\ref{tab:vzhya_orb}, \ref{tab:vzhya_phel}, \ref{tab:wzoph_orb},
and \ref{tab:wzoph_phel}.

As seen, masses and radii precise to 0.5--0.7\% and 0.4--0.9\%, respectively, 
have been established for the binary components. 
For \AD, the masses agree well with those determined by Popper (\cite{dmp98a})
but are significantly more accurate, whereas those by Lacy (\cite{lacy97})
are about 2\% higher. Both studies report radii in agreement with ours.
Popper (\cite{dmp65}, \cite{dmp80}) reached masses and radii accurate to about 
3\% for \VZ, and within this uncertainty his results agree with ours.
The less massive secondary component of \WZ\ appears slightly larger and
more luminous but cooler than the primary. However, within the uncertainties
the two components can be considered identical.
On average, we obtain masses and radii for \WZ\ which are about 8\% and 5\% 
larger, respectively, than those given by Popper (\cite{dmp65}), and they are
far more accurate.

The $uvby$ indices for the components of \AD, \VZ, and \WZ, included in 
Table~\ref{tab:absdim}, were calculated from the the combined indices of the 
systems at phase 0.25 (CVG08) and the luminosity ratios 
between their components (Tables~\ref{tab:adboo_phel}, \ref{tab:vzhya_phel}, 
\ref{tab:wzoph_phel}). The $E(b-y)$ interstellar reddenings, also given in
Table~\ref{tab:absdim}, were determined from $uvby\beta$ standard photometry
and other sources, as described in more detail in Sect.~\ref{sec:eby}.

The adopted effective temperatures listed in Table~\ref{tab:absdim} were
calculated from the calibration by Holmberg et al. (\cite{holmberg07})
assuming the final \feh\ abundances. The uncertainties ($100$--$150$ K) 
include those of the $uvby$ indices, $E(b-y)$, \feh, and the calibration 
itself. 
Within errors, temperatures from the $(b-y)$, $c_1$ calibration by Alonso et 
al.  (\cite{alonso96}) and the $b-y$ calibration by Ram\'\i rez \& 
Mel\'endez (\cite{rm05}) agree, but for \AD\ we note that both calibrations 
lead to a temperature difference between the components that is about 
120~K higher. A similar trend was noticed by Holmberg et al. 
(\cite{holmberg07}).
The empirical flux scale by Popper (\cite{dmp80})
and the $y$ flux ratios between the components (Tables~\ref{tab:adboo_phel},
\ref{tab:vzhya_phel}, \ref{tab:wzoph_phel}) yield well established temperature 
differences of 450~K (\AD), 335~K (\VZ), and 20~K (\WZ), supporting the 
temperature calibration we have chosen. Details on temperatures derived
as part of the TODCOR analyses of the observed CfA spectra (assuming the
final metallicities) and the abundance analyses of the disentangled CfA 
spectra are given in Sect.~\ref{sec:specphot}. In general, the results agree 
well with those included in Table~\ref{tab:absdim}, except for
\WZ\ where slightly higher temperatures were obtained from the analysis
of the disentangled spectra.

The distances to the binaries were calculated from the "classical" relation,
\begin{eqnarray}
\label{eq:m-M}
V_0-M_V & = & (V - A_V) - M_{\rm bol\,\odot} +\; 5\:{\rm log}(R/R_{\,\odot})\\
            &   & +\; 10\:{\rm log}(T_{\rm eff}/T_{\rm eff\,\odot}) + BC \nonumber
\end{eqnarray}
\noindent
adopting the solar values and bolometric corrections given in 
Table~\ref{tab:absdim}, and $A_V/E(b-y) = 4.27$.
As seen, the distances have been established to $5$--$6$ \%, accounting for all
error sources and including the use of other $BC$ scales
(e.g. Code et al. \cite{code76}, Bessell et al. \cite{bessell98}, 
Girardi et al. \cite{girardi02}).
For \AD\ and \VZ, the empirical $K$ surface brightness - $T_{\rm eff}$ relation
by Kervella et al. (\cite{kervella04}) leads to nearly identical and perhaps
even more precise distances; see Southworth et al. (\cite{southworth05}) for
details. $K$ photometry outside eclipses is unfortunately not available for \WZ.  
\VZ\ and \WZ\ are included in the Hipparcos ca\-ta\-log (ESA \cite{hip97}), 
which gives parallaxes of $5.03 \pm 1.34$ mas and $7.99 \pm 1.37$ mas, 
respectively, corresponding to distances of $199 \pm 53$ and $125 \pm 21$ pc.
Matching the shorter Hipparcos distance for \WZ\ would require
lower temperatures, whereas higher interstellar absorption would
imply higher temperatures and therefore a nearly unchanged distance.
\WZ\ therefore does not formally belong to the group of eclipsing binaries 
within 125 pc discussed by Popper (\cite{dmp98b}), which could be of use to
improve the radiative flux scale, provided more precise astrometric
parallaxes were available.

\subsection{Interstellar reddening and absorption}
\label{sec:eby}

The interstellar reddenings of \AD, \VZ, and \WZ\ have been determined as 
follows: 

For \AD, we adopt an interstellar reddening of $E(b-y) = 0.025 \pm 0.015$, 
which is derived from the
calibration by Olsen (\cite{eho88}) and the standard $uvby\beta$ indices at 
phase 0.0, where a significant fraction of the primary component is eclipsed.
Ideally, indices at phase 0.5, where the secondary component is totally 
eclipsed, should be used, but such $\beta$ observations are unfortunately 
not available. 
Combining instead theoretical $\beta$ values (Moon \& Dworetsky \cite{md85}) 
and the individual $uvby$ indices, reddenings of $E(b-y) \approx 0.01 \pm 0.02$
are obtained for both components. 
Alternatively, 2MASS $JHK_s$ photometry for stars within $1\degr$ of \AD\ on 
the sky and closer than 200 pc show a mean absorption of $A_V = 0.18 \pm 0.04$,
corresponding to $E(b-y) = 0.04 \pm 0.01$ (Knude, private communication). 
Finally, the model by Hakkila et al. (\cite{hak97}) yields $A_V = 0.10$ or
$E(b-y) = 0.02$, in the direction of and at the distance of \AD, and the 
Schlegel et al.  (\cite{sch98}) dust maps give the same amount of total 
extinction.

For \VZ, the calibration by Olsen and the standard $uvby\beta$ indices at 
phase 0.5, where the secondary is nearly totally eclipsed, lead to negligible
interstellar reddening, as do the indices at phase 0.0, where a significant
fraction of the primary component is eclipsed.
On the other hand, 2MASS $JHK_s$ photometry for stars within $30\arcmin$ of 
\VZ\ on the sky and closer than 150 pc show a mean absorption of 
$A_V = 0.17 \pm 0.06$, corresponding to $E(b-y) = 0.040 \pm 0.014$.
Furthermore, the model by Hakkila et al. also yields $A_V = 0.17$
in the direction of and at the distance of \VZ, whereas the Schlegel et al. 
dust maps give a total extinction of $A_V = 0.12$, corresponding to
$E(b-y) = 0.028$. 
To encompass the different results we have adopted $E(b-y) = 0.02 \pm 0.02$.

For \WZ\ we adopt an interstellar reddening of $E(b-y) = 0.033 \pm 0.012$, 
which is derived from the calibration by Olsen and the standard 
$uvby\beta$ indices outside of eclipse, which represent the nearly 
identical components well.
2MASS $JHK_s$ photometry for stars within $30\arcmin$ of \WZ\ on the
sky and closer than 180 pc show a mean absorption of $A_V = 0.10 \pm 0.04$, 
corresponding to $E(b-y) = 0.024 \pm 0.009$.
In the direction and at the distance of \WZ, the model of Hakkila et al. 
yields a very high absorption of $A_V = 0.52$, whereas the Schlegel et al. 
dust maps give of total extinction of $A_V = 0.28$ or $E(b-y) = 0.066$.

\section{Discussion}
\label{sec:dis}

In the following, we compare the new dimensions obtained for
\AD, \VZ, and \WZ\ with properties predicted from some of the latest stellar 
evolutionary mo\-dels. Below, we first briefly present the key ingredients 
of the selected grids, referring to the original papers for full descriptions.

The confrontation between observations and theory for these three systems 
can be constrained better than in most previous studies, first of all
because our \feh\ determinations and abundance results for
other heavy elements are probably among the most detailed done to date for 
binaries. This means that we can select models not only
with respect to mass, but also with respect to $Z$.
On the other hand, for a given mass and $Z$, the observable properties 
of models at a given age such as radius, effective temperature, and 
luminosity still depend on the adopted input physics, including the treatment 
of core and envelope convection, diffusion, etc., and on the assumed element 
mixture and $Y$,$Z$ relation as well. In order to constrain such free model 
parameters, accurate results from many binaries are required, supplemented 
by other results such as those from asteroseismology and cluster analysis. 
We will focus on the three binaries studied here; a larger sample will be
dealt with in forthcoming papers, cf. Fig.~\ref{fig:debs}.

From the binary perspective, masses and radii are the most direct
parameters available, free of any scale dependent calibrations. 
The $M-R$ plane is therefore well suited for isochrone tests,
especially when the binary components are significantly different.
In addition, the $T_{\rm eff}-R$ or $T_{\rm eff}-\mathrm{log}(g)$ planes
allow tests of model temperatures, which for a given mass and $Z$ 
depend, e.g., on the abundance mixture, $Y$, core overshoot treatment, 
and surface convection efficiency.
In our case, the binary temperatures are based on several calibrations
and methods, and furthermore the temperature difference between
the components is well constrained from the light curves; cf.
Sect.~\ref{sec:teff}. Adopting the usual assumption of coeval formation
from the same raw material, identical ages must be reached in these
planes, as well as in the $M-\mathrm{log}(L)$ plane.

\subsection{Summary of models}
\label{sec:models}

As a reference, we have adopted the $Y^2$ (Yonsei-Yale) evolutionary 
tracks and isochrones by Demarque et al. 
(\cite{yale04})\footnote{{\scriptsize\tt http://www.astro.yale.edu/demarque/yystar.html}}, 
which are based on up-to-date input physics and include an improved core 
overshoot treatment where $\Lambda_{OS} = \lambda_{ov}/H_p$
depends on mass, and which also takes into account 
the composition dependence of $M_{crit}^{conv}$
\footnote{Defined as "the mass above which stars continue to have a
substantial convective core even after the end of the pre-MS phase."}
The mixing length parameter in convective envelopes is calibrated using
the Sun, and is held fixed at $l/H_p = 1.7432$.
Helium diffusion has been included. 
All tracks are started at the pre-main-sequence birth line.
Tracks and isochrones are available for a broad range of masses, ages, and
heavy ele\-ment abundances. The enrichment law $Y = 0.23 + 2Z$ is adopted, 
together with the solar mixture by Grevesse et al. (\cite{gns96}), leading to
($X$,$Y$,$Z$)$_{\sun}$ = (0.71564,0.26624,0.01812).
Models with enhanced $\alpha$-element abundances are also
included; see Kim et al. (\cite{yale2}) for further details.
We have used the abundance, mass, and age interpolation routines 
provided by the $Y^2$ group. 

Another recent example of stellar evolution calculations is given by the 
extensive Victoria-Regina grids (VandenBerg et al., 
\cite{vr06})\footnote{{\scriptsize\tt http://www1.cadc-ccda.hia-iha.nrc-cnrc.gc.ca/cvo/
community/VictoriaReginaModels/}}, 
which adopt a somewhat different physically based description of the core 
overshoot with $F_{over}$ depending on mass and metallicity and 
ca\-li\-bra\-ted observationally via cluster CMD's. Diffusive processes 
are not treated.
All tracks have been computed from the pre-main-sequence stage.
The VRSS grids (scaled-solar) used here adopt $Y = 0.23544 + 2.2Z$ and the 
solar mixture by Grevesse \& Noels (\cite{gn93}), leading to
($X$,$Y$,$Z$)$_{\sun}$ = (0.7044,0.2768,0.0188).
The mixing length parameter in convective envelopes is calibrated using
the Sun and held at $\alpha_{MLT} = 1.90$.
For further details on the chemical abundances adopted for the 
VR0A, VR2A, and VR4A grids (different \afe) we refer the reader to 
VandenBerg et al.  (\cite{vdb00}).
We have used the isochrone interpolation routines provided by the 
Victoria-Regina group; track interpolations have not been done. 

A third source of modern evolutionary tracks and isochrones for specific 
masses and ages is the BaSTI database 
(Pietrinferni et al., \cite{basti04})\footnote{{\scriptsize\tt http://www.te.astro.it/BASTI/index.php}},
available for several compositions.
It is based on a recent version of the FRANEC evolutionary code 
(Degl'Innocenti et al., \cite{franec07}).
Models without and with core overshoot are available; in the latter
case the choice of $\lambda_{ov}$ depends on mass but not on metallicity.
$Y = 0.245 + 1.4Z$ and the solar mixture by Grevesse \& Noels 
(\cite{gn93}) are used, and the mixing length parameter in convective 
envelopes is calibrated using the Sun and held fixed at $l/H_p = 1.913$. 
Mass loss is included, but atomic diffusion is taken into 
account only in the standard solar model.
All tracks have been computed from the pre-main-sequence phase.

\begin{figure}
\epsfxsize=90mm
\epsfbox{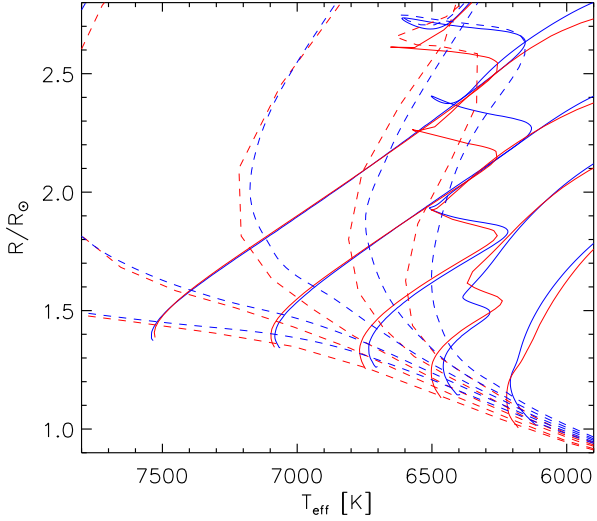}
\caption[]{\label{fig:yalevic}
Comparison between $Y^2$ (red) and Victoria-Regina VRSS (blue) models
for \feh\,$=-0.190$ and \afe\,$= 0.0$. 

$Y^2$: ($X$,$Y$,$Z$) = (0.73400,0.25400,0.01200). 
VRSS:  ($X$,$Y$,$Z$) = (0.72456,0.26294,0.01250). 
Tracks (solid lines) for 1.1, 1.2, 1,3. 1.4, and 1.5 $M_{\sun}$ and isochrones
(dashed) between 0.5 and 2.5 Gyr (step 0.5 Gyr) are shown. 
}
\end{figure}

\begin{figure}
\epsfxsize=90mm
\epsfbox{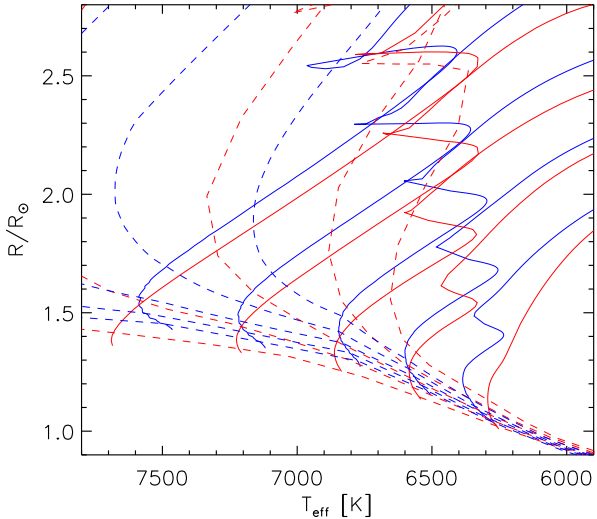}
\caption[]{\label{fig:yalebas}
Comparison between $Y^2$ (red) and BaSTI core overshoot (blue) models
for \feh\,$=-0.253$ and \afe\,$=0.0$.

$Y^2$: ($X$,$Y$,$Z$) = (0.73865,0.25090,0.01045).
BaSTI: ($X$,$Y$,$Z$) = (0.7310,0.2590,0.0100). 
Tracks (solid lines) for 1.1, 1.2, 1.3, 1.4, and 1.5 $M_{\sun}$ and isochrones
(dashed) between 0.5 and 2.5 Gyr (step 0.5 Gyr) are shown. 
}
\end{figure}

A comparison of the main sequence models between 1.1 and 1.5 $M_{\sun}$ for 
identical \feh\ values close to those of our stars reveals some differences 
between the three grids described above. This is exemplified in 
Figs.~\ref{fig:yalevic} and ~\ref{fig:yalebas}. 
In the lower main-sequence band, the $Y^2$ and the Victoria-Regina tracks 
agree quite well. The isochrones are slightly shifted, and for part of the
mass interval the effect increases in the upper part of the main-sequence 
band due to differences in the  treatment of core overshoot and the ramping 
procedure adopted. 
The BaSTI tracks differ somewhat in shape and location, and
consequently significant differences are seen in the isochrones as well.

Before focusing on the confrontation between binary observations and models,
we point out that we have not considered here models based on scaling of the 
most recent solar mixture (Grevesse et al. \cite{gas07}), which implies a 
significantly lower metallicity of $Z_{\sun} \approx 0.013$ and a different 
conversion between [Fe/H] and $Z$.
At present, solar models computed with these abundances disagree with
precise results inferred from helioseismology, and the cause(s) have
yet to be identified (e.g. Basu \& Antia \cite{basu07}).

\subsection{Comparison with the binary results}
\label{sec:obscomp}

\begin{figure}
\epsfxsize=90mm
\epsfbox{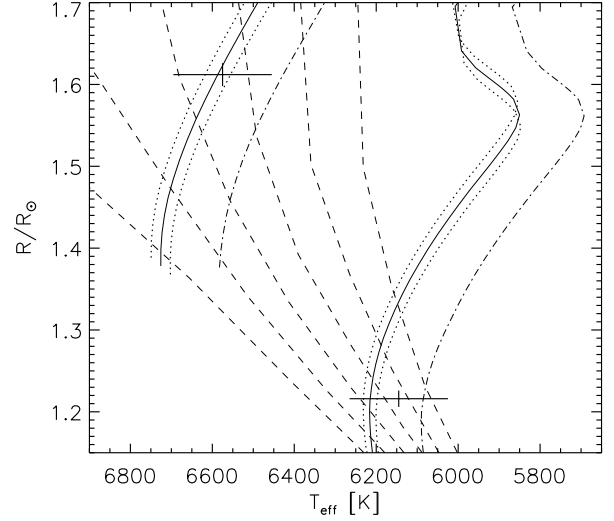}
\caption[]{\label{fig:adboo_tr}
AD Boo compared with $Y^2$ models for ($X$,$Y$,$Z$) = (0.7028,0.2748,0.0224), 
equivalent to the measured abundance \feh\,$=+0.10\pm0.15$
for \afe\,$=0.0$.
Tracks for the component masses (solid lines) and isochrones from 0.5
to 3.0 Gyr (dashed; step 0.5 Gyr) are shown.
The uncertainty in the location of the tracks coming from
the mass errors are indicated (dotted lines).
To illustrate the effect of the abundance uncertainty, tracks for 
\feh\,$=+0.25$ and \afe\,$=0.0$ (dash-dot), corresponding to 
($X$,$Y$,$Z$) = (0.6785,0.2910,0.0305), are included.
}
\end{figure}

\begin{figure}
\epsfxsize=90mm
\epsfbox{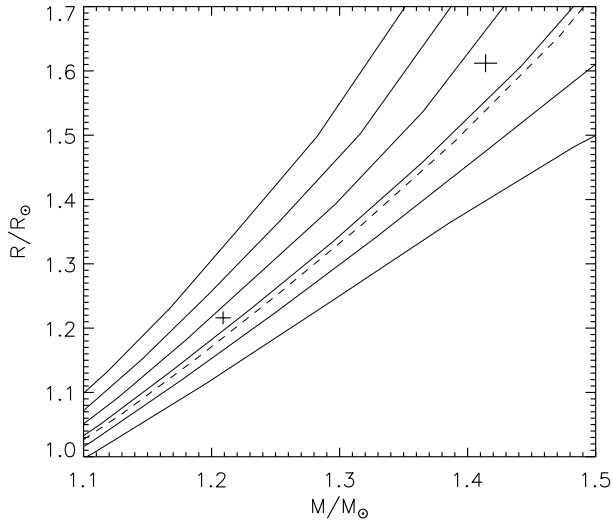}
\caption[]{\label{fig:adboo_mr}
AD Boo compared with $Y^2$ models for ($X$,$Y$,$Z$) = (0.7028,0.2748,0.0224), 
equivalent to \feh\,$=+0.10$ for \afe\,$=0.0$. 
Iso\-chro\-nes from 0.5 to 3.0 Gyr (step 0.5 Gyr) are shown.
The 1.5 Gyr isochrone for \feh\,$=+0.25$ and \afe\,$=0.0$ (dashed),
corresponding to ($X$,$Y$,$Z$) = (0.6785,0.2910,0.0305), is 
included for comparison.
}
\end{figure}

\begin{figure}
\epsfxsize=90mm
\epsfbox{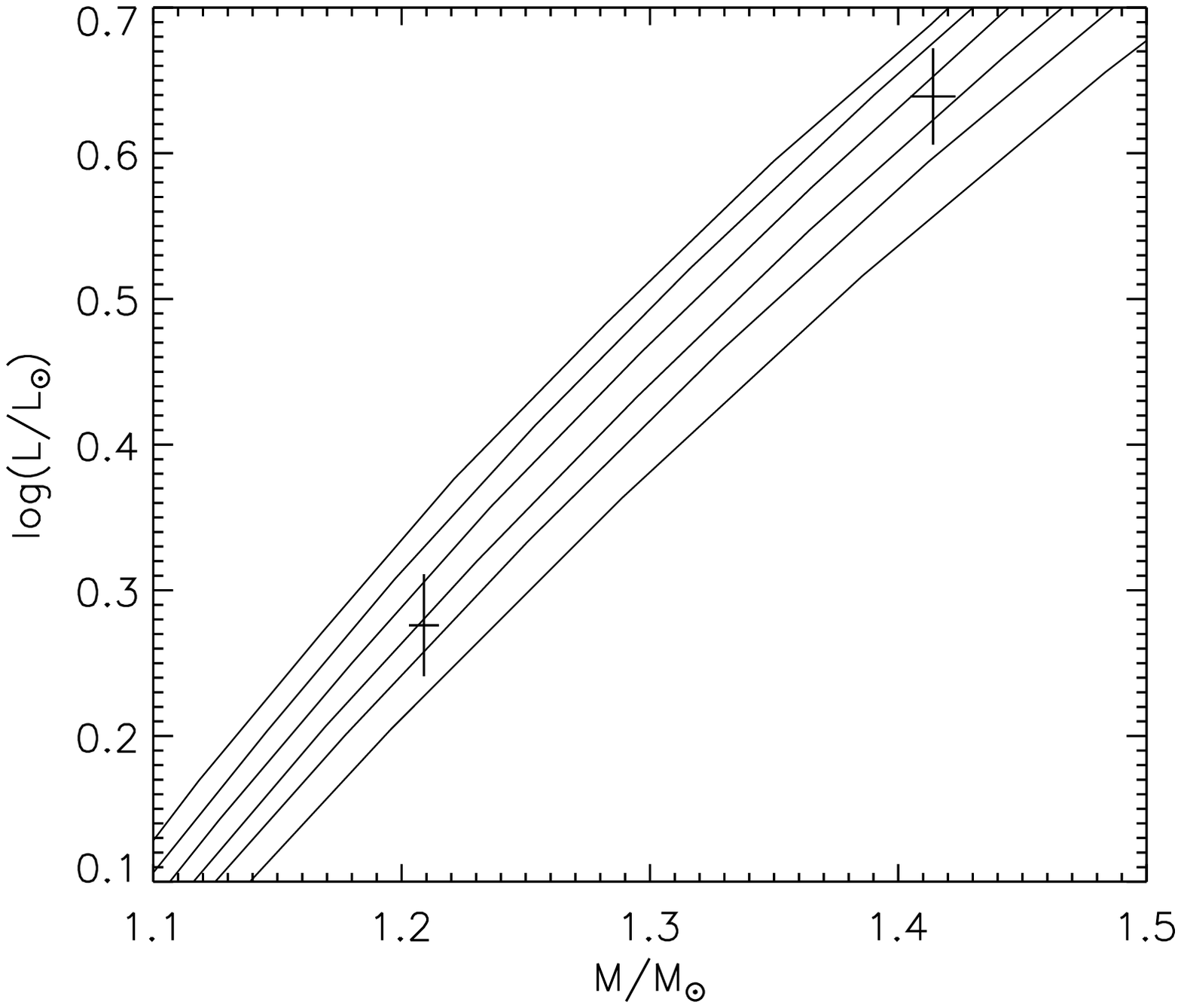}
\caption[]{\label{fig:adboo_ml}
AD Boo compared with $Y^2$ models for ($X$,$Y$,$Z$) = (0.7028,0.2748,0.0224), 
equivalent to \feh\,$=+0.10$ for \afe\,$=0.0$. 
Iso\-chro\-nes from 0.5 to 3.0 Gyr (step 0.5 Gyr) are shown.
}
\end{figure}

For \AD, comparisons between the absolute dimensions 
and $Y^2$ models calculated for the measured \feh\ are shown in 
Figs.~\ref{fig:adboo_tr}--\ref{fig:adboo_ml}.
As seen in Fig.~\ref{fig:adboo_mr}, an isochrone corresponding to 
$\sim$1.75 Gyr would fit both components very well, and for the primary 
component this age is also deduced from Figs.~\ref{fig:adboo_tr} and 
\ref{fig:adboo_ml}.
The abundance uncertainty of $\pm 0.15$ dex translates to an age uncertainty 
of about $\pm 0.15$ Gyr, comparable to that from mass and radius uncertainties.
The model tracks fit the components well but predict a $\sim100$~K lower 
temperature difference between the components than established from
both photometry and the empirical flux scale; see Sect.~\ref{sec:absdim}.
Consequently, the predicted model luminosity of the secondary is also 
marginally higher than observed (Fig.~\ref{fig:adboo_ml}). 
If real, the slight discrepancy may be related to the activity of 
the secondary component. 
The same picture is seen for Victoria-Regina models, which give
a slightly lower age of $\sim$1.5 Gyr.
On the other hand, the BaSTI models fit \AD\ less well. 
For the observed masses and radii, they predict quite different ages of 
about 2.25 and 1.75 Gyr for the primary and secondary, respectively. 
This is probably related to the different shape of the BaSTI tracks 
(Fig.~\ref{fig:yalebas}) in the lower main-sequence region. 

\begin{figure}
\epsfxsize=90mm
\epsfbox{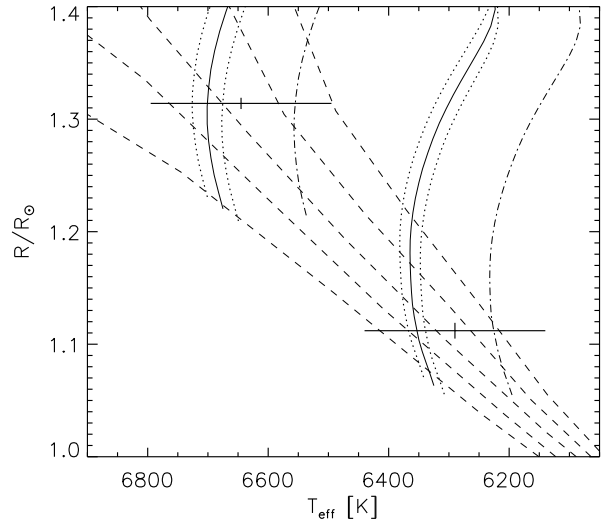}
\caption[]{\label{fig:vzhya_tr}
VZ Hya compared with $Y^2$ models for ($X$,$Y$,$Z$) = (0.73478,0.25348,0.01174) 
equivalent to the measured abundance \feh\,$=-0.20\pm0.12$ 
for \afe\,$=0.0$. 
Tracks for the component masses (solid lines) and isochrones from 0.5
to 2.5 Gyr (dashed; step 0.5 Gyr) are shown.
The uncertainty in the location of the tracks coming from
the mass errors are indicated (dotted lines).
To illustrate the effect of the abundance uncertainty,
tracks for \feh\,$=-0.08$ and \afe\,$=0.0$ (dash-dot),
corresponding to ($X$,$Y$,$Z$) = (0.72425,0.26050,0.01525), are 
included.
}
\end{figure}

\begin{figure}
\epsfxsize=90mm
\epsfbox{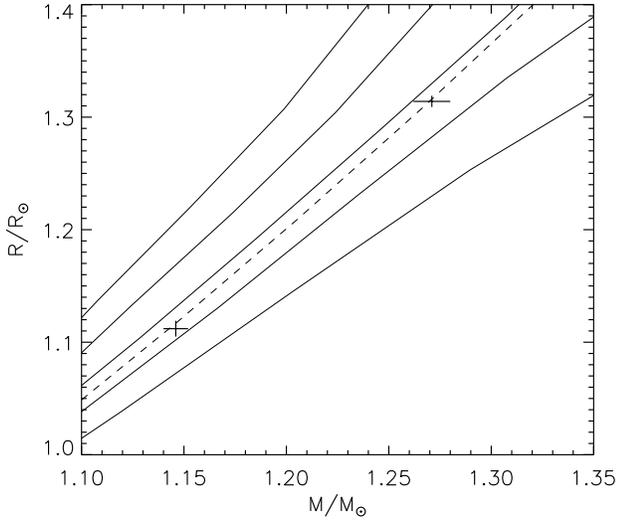}
\caption[]{\label{fig:vzhya_mr}
VZ Hya compared with $Y^2$ models for ($X$,$Y$,$Z$) = (0.73478,0.25348,0.01174) 
equivalent to \feh\,$=-0.20$. 
Iso\-chro\-nes from 0.5 to 2.5 Gyr (step 0.5 Gyr) are shown.
The 1.5 Gyr isochrone for \feh\,$=-0.08$ and \afe\,$=0.0$ (dashed),
corresponding to ($X$,$Y$,$Z$) = (0.72425,0.26050,0.01525), is 
included for comparison.
}
\end{figure}

\begin{figure}
\epsfxsize=90mm
\epsfbox{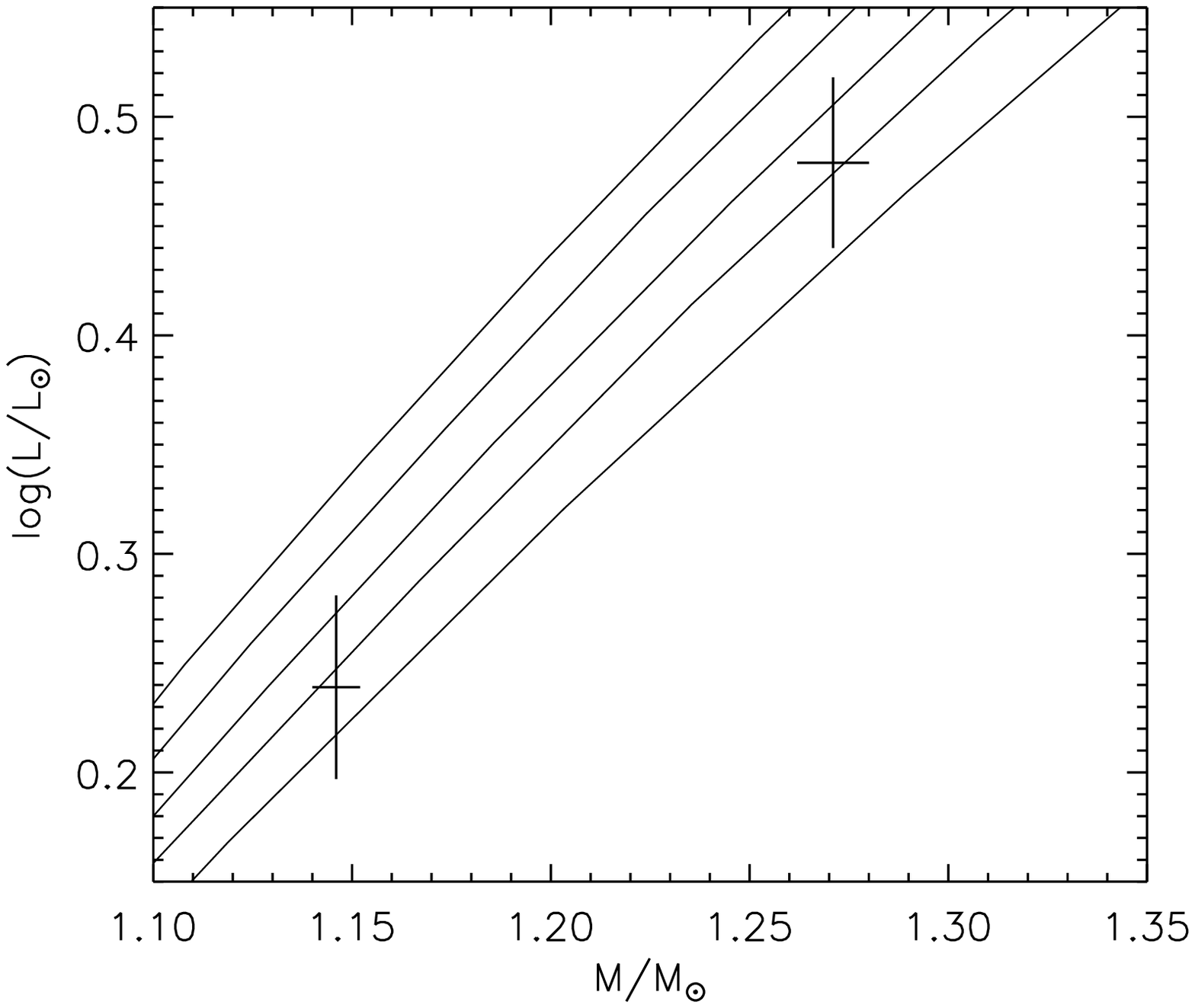}
\caption[]{\label{fig:vzhya_ml}
VZ Hya compared with $Y^2$ models for ($X$,$Y$,$Z$) = (0.73478,0.25348,0.01174) 
equivalent to \feh\,$=-0.20$. 
Iso\-chro\-nes from 0.5 to 2.5 Gyr (step 0.5 Gyr) are shown.
}
\end{figure}

Figs.~\ref{fig:vzhya_tr}--\ref{fig:vzhya_ml} compare the absolute
dimensions for \VZ\ with the $Y^2$ models calculated for the measured
\feh. 
As seen in Fig.~\ref{fig:vzhya_mr}, an isochrone for an age of 1.25 Gyr fits 
both components within errors, although a slightly steeper isochrone
slope would formally improve the fit still more. 
The abundance uncertainty of $\pm 0.12$ dex corresponds to about
$\pm 0.25$ Gyr in age, similar to the contribution from mass and radius
uncertainties. 
The model tracks are slightly hotter than observed, and
consequently the ages derived from Figs.~\ref{fig:vzhya_tr} and 
\ref{fig:vzhya_ml} become too high and low, respectively. 
Both tracks would fit well for a slightly higher \feh\ of $-0.15$, 
safely within the uncertainty of the measured value. 
Adopting \feh\,$=-0.15$, identical ages of 1.25 Gyr are obtained 
from all three diagrams.
Victoria-Regina tracks for the observed \feh\ fit the components
well, being slightly cooler than the $Y^2$ tracks.  For the observed
masses and radii, ages become lower, about 1.0 and 0.7 Gyr for primary 
and secondary, respectively; they agree within errors.
The BaSTI models yield about 1.25 and 0.75 Gyr, and in this case the 
difference is significant.

\begin{figure}
\epsfxsize=90mm
\epsfbox{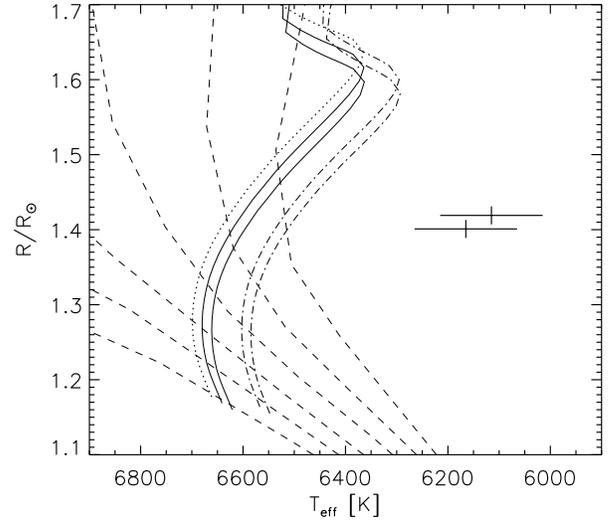}
\caption[]{\label{fig:wzoph_tr}
WZ Oph compared with $Y^2$ models for ($X$,$Y$,$Z$) = (0.73985,0.25010,0.01005),
equivalent to the measured abundance 
\feh\,$-0.27\pm0.07$ for \afe\,$=0.0$. 
Tracks for the component masses (solid lines) and isochrones from 0.5
to 3.0 Gyr (dashed; step 0.5 Gyr) are shown.
The uncertainty in the location of the primary track coming from
the mass error is indicated (dotted line).
For comparison, tracks for ($X$,$Y$,$Z$) = (0.7349,0.2534,0.0117) and 
\afe\,$=+0.1$, also equivalent to \feh\,$=-0.27$, are included
(dash-dot).
}
\end{figure}

Whereas the $Y^2$ models in general match the observed properties of 
\AD\ and \VZ\ well, the situation for \WZ\ is completely different,
as seen from Fig.~\ref{fig:wzoph_tr}.
Models for the measured \feh\ are much too hot, both for 
\afe\,$=0.0$ and $+0.1$ (cf. Sect.~\ref{sec:wzoph_abund}),
and the same difference is seen for Victoria-Regina and BaSTI
models.
In order to reproduce \WZ\ for the observed \feh, an 
unrealistically high \afe\ of about $+0.5$ would be required. 
Alternatively, the models would need to have a significantly lower helium 
content and/or an envelope mixing length parameter well below $l/H_p = 1.7432$.
The latter can not a priori be ruled out, since the
photometry and spectroscopy indicate the presence of some degree of surface 
activity on both components of \WZ; see Sect.~\ref{sec:wzoph_phel}.
It would therefore be of great interest to have available models for the
observed \feh\ and a range of $l/H_p$ values.
Such mo\-dels have been published by Baraffe et al. 
(\cite{baraffe98})\footnote{{\scriptsize\tt ftp.ens-lyon.fr}} 
but only for solar metallicity. They suggest
that $l/H_p \sim 1$ or even lower would be needed, although models
computed speci\-fi\-cal\-ly for this case would be
required in order to evaluate the combined effects on the stellar
radii and temperatures in detail. 
The possible connection between activity and convection, as
indicated from studies of solar-type binaries, 
has been discussed by Clausen et al. (\cite{granada99}),
Torres et al. (\cite{wt06}), and others.

\begin{figure}
\epsfxsize=90mm
\epsfbox{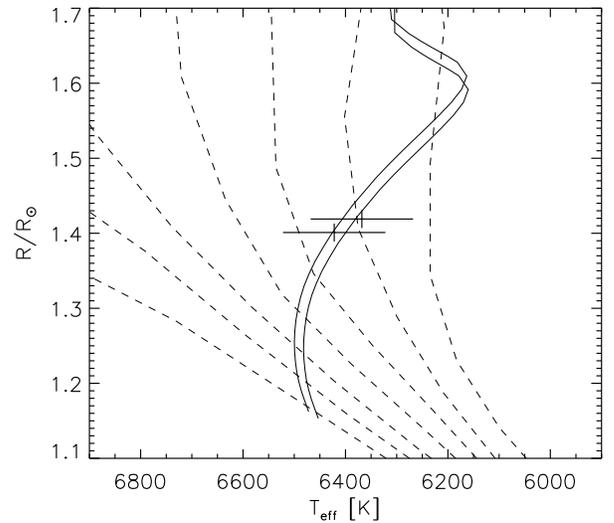}
\caption[]{\label{fig:wzoph_tr_alt}
Assuming E(b-y) = 0.066:
WZ Oph compared with $Y^2$ models for ($X$,$Y$,$Z$) = 
(0.72899,0.25734,0.013670),
equivalent to \feh\,$=-0.13$ for \afe\,$=0.0$.
Tracks for the component masses (full drawn) and isochrones from 0.5
to 3.5 Gyr (dashed; step 0.5 Gyr) are shown.
See text for details.
}
\end{figure}

As stated in Sect.~\ref{sec:eby}, we have adopted an
interstellar reddening of $E(b-y) = 0.033 \pm 0.012$ for \WZ,
equi\-va\-lent to $A_V = 0.14 \pm 0.05$ for a standard
extinction law. There were, however, some indications of higher
absorption from other sources.
If we therefore somewhat arbi\-tra\-rily double it and
assume $A_V = 0.28$ and $E(b-y) = 0.066$, the effective
temperatures derived from the $uvby$ indices increase by about 250~K,
i.e. to about 6400~K.
This has an effect on the abundances. From the Fe~I lines we would
derive \feh\,$\sim-0.13$ for both components, whereas the results 
from the Fe~II lines would be practically unchanged at $-0.27$. 
The photometric determinations from the calibration by Holmberg et al. 
(\cite{holmberg07}) would increase by about 0.12 dex for both components.
The discrepancy between neutral and ionized Fe lines argues against
the higher temperatures, but assuming them, together with the
higher abundance from the Fe~I lines, does on the other hand remove the
model discrepancies, as seen in Fig.~\ref{fig:wzoph_tr_alt}.
The age of \WZ\ then becomes about 3 Gyr. 

An independent check would be to derive intrinsic colours and/or
effective temperatures from line-depth ratios measured in the 
disentangled FEROS spectra, see e.g. Gray (\cite{gray94}), but
\WZ\ is unfortunately slightly too hot for the calibrations available.
Adopting a subset of those by Kovtyukh et al. (\cite{kovtyukh03}), 
valid to 6100 or 6150~K, we derive mean effective temperatures close to
6200~K for both components. The individual results from the 9 (primary) 
and 16 (secondary) measured line-depth ratios scatter by 100-200~K, 
and since we are at the edge of the calibrations, the method does 
not at the moment provide new firm information.

Although we cannot pinpoint the reason for the discrepancy between the 
observations and the models for \WZ, we suspect they are connected to 
composition (He) or convection anomalies in the stars and/or the 
determination of interstellar reddening, rather than to physical or 
numerical problems with the model calculations.
This is supported by the general good agreement seen for \AD\ and especially 
\VZ, but detailed studies of additional systems similar to these are needed 
for full clarification.
Work on several new F-G type binaries is in progress.

\section{Summary and conclusions}
\label{sec:sum}

We have presented precise absolute dimensions and abundances for the
three F-type double-lined eclipsing binaries \AD, \VZ, and \WZ. 
The results, which are based on state-of-the-art analyses of new 
photometric and spectroscopic observations, were used to 
test the recent $Y^2$, Victoria-Regina, and BaSTI stellar 
evolutionary models. 

Masses and radii precise to 0.5--0.7\% and 0.4--0.9\% have been established
for the binary components. This level can only be reached if excellent
observations are available and are analysed carefully using adequate tools.
Special care was taken to identify and avoid possible systematic errors. 
Several methods were applied in order to determine reliable and accurate 
(100--150~K) effective temperatures and interstellar extinctions, still
a challenging task whose importance can not be overestimated.

System \feh\ abundances were based on numerous lines and were derived relative
to the Sun. We have reached precisions between 0.07 and 0.15 dex. 
Even ha\-ving many Fe~I and Fe~II lines of different strength and excitation 
potential available, the strong correlation between abundance and effective
temperature is difficult to break. $T_{\rm eff}$ and microturbulence 
uncertainties contribute significantly to the final abundance precision. 
Abundances for other heavy elements, based on fewer lines, were also
derived; in one case (\WZ) indications of a slight $\alpha$-element
overabundance is seen.
Comparing the results for the three binaries clearly demonstrates that
analyses of disentangled spectra are far superior to those of
compo\-si\-te spectra, essentially because many more lines become available and
blending becomes less of an issue.

\AD\ has components with the greatest difference in mass and radius
among well-studied F-type binaries (1.41~$M_{\sun}$, 1.61~$R_{\sun}$; 
1.21~$M_{\sun}$, 1.22~$R_{\sun}$) and an abundance of \feh\,$=+0.10$.
Slight surface activity is present on the secondary component.
The system is fitted well by $Y^2$ and Victoria-Regina evolutionary 
models for identical component ages (as determined from 
masses and radii) of 1.75 and 1.5 Gyr, respectively. The fact that
the se\-con\-dary appears marginally cooler than the corresponding evolutionary 
track may be related to mild surface activity.
BaSTI models predict different ages of 2.25 and 1.75 Gyr 
for the components, and therefore do not fit \AD\ as well.

\VZ\ has somewhat different components 
(1.27~$M_{\sun}$, 1.31~$R_{\sun}$; 1.14~$M_{\sun}$, 1.11~$R_{\sun}$) 
and an abundance of \feh\,$=-0.20$. No signs of surface activity are seen.
$Y^2$ and Victoria-Regina evolutionary tracks fit \VZ\ well. Nearly identical
component ages (as determined from masses and radii) of 1.25 Gyr
are obtained from the $Y^2$ isochrones, whereas the Victoria-Regina 
calculations give 1.0 and 0.7 Gyr for the primary and secondary. 
An even larger difference, 1.25 versus 0.75 Gyr, is found when using the 
BaSTI models.

\WZ\ consists of two nearly identical components (1.22~$M_{\sun}$, 
1.41~$R_{\sun}$). The \feh\ abundance is $-0.27$, and as mentioned above 
a slight $\alpha$-element overabundance is possible. Photometry and 
spectroscopy suggest low-level surface activity on both components. 
Contrary to the case for \AD\ and \VZ, neither of the three model grids 
are able to fit the observations for \WZ; models are significantly hotter 
than observed. 
We tentatively conclude that this is caused either by \WZ\ anomalies
such as a low He content or decreased envelope convection,
or by underestimated interstellar reddening. Further observations will be
required to resolve the issue.

We believe our study is among the most detailed carried out to date for 
main-sequence eclipsing binaries, and that it sets new standards for 
critical tests of stellar evolutionary models based on accurate
binary data. With only three systems available, it is premature to
draw firm conclusions and suggest specific model shortcomings
and/or improvements. We find, however, better
agreement with the $Y^2$ and Victoria-Regina models than with the BaSTI models.
In contradiction to this, Tomasella et al. (\cite{asiago07}) recently
found that dedicated BaSTI models, including core overshoot and both
helium and heavy element diffusion, represent the components of
the young $\sim1.25 M_{\sun}$ system V505\,Per well. 
Their study is based on new absolute dimensions of extraordinary high formal
precision. 

As shown in Fig.~\ref{fig:debs}, several additional F-type binaries
are available, and we are presently working on re-analyses for
some of them, including abundance determinations. 
We are also conducting a larger program on new F-G type systems
exhibiting various levels of surface activity (Clausen et al.
\cite{jvcetal01}), and 
we expect some very useful insights to come from these studies, and from a 
parallel program on eclipsing binaries in clusters spanning a wide range 
in age and metallicity (e.g. Grundahl et al. \cite{fgj06}). 

\begin{acknowledgements}

The spectroscopic observations at the CfA facilities were obtained
with the able help of J.\ R.\ Caruso, R.\ J.\ Davis, E.\ Horine, R.\
D.\ Mathieu, A.\ A.\ E.\ Milone, J.\ Peters, and J.\ M.\ Zajac. We are
grateful to R.\ J.\ Davis for maintaining the echelle database at CfA.
We thank U.G. J{\o}rgensen for many inspiring discussions and for
allowing us to use his MARCS code.
A.\ Kaufer, O.\ Stahl, S.\ Tubbesing, and B.\ Wolf kindly obtained
a FEROS spectrum of WZ Oph spectra during Heidelberg/Copenhagen guaranteed 
time in 1999.
We thank H.\ Hensberge for making his improved
version of the MIDAS FEROS package available, and both him and
L.\ Freyhammer for valuable advise and help during the reduction
of the spectra.
We are grateful to E.\ Sturm for providing his original disentangling
code, and to him and J.\ D.\ Pritchard for modifying it for use
at Linux/Unix computer systems.
J.\ Knude kindly provided interstellar absorption results from
his $JHK_s$ calibration.
The projects "Stellar structure and evolution -- new challenges from
ground and space observations" and "Stars: Central engines of the evolution 
of the Universe", carried out at Copenhagen University and Aarhus University, 
are supported by the Danish National Science Research Council.
G.T.\ acknowledges partial support from NSF grants AST-0406183 and AST-0708229,
and JA and BN from the Carlsberg Foundation.
The following internet-based resources were used in research for
this paper: the NASA Astrophysics Data System; the SIMBAD database
and the VizieR service operated by CDS, Strasbourg, France; the
ar$\chi$iv scientific paper preprint service operated by Cornell University.
This publication makes use of data products from the Two Micron
All Sky Survey, which is a joint project of the University of
Massachusetts and the Infrared Processing and Analysis Center/
California Institute of Technology, funded by the National
Aeronautics and Space Administration and the National Science
Foundation.
\end{acknowledgements}

{}

\begin{appendix}
\section{Radial velocity tables}
\label{sec:rvtables}
 \onecolumn
\begin{longtable}{llrrrr}
\caption[]{\label{tab:adboo_rv}
Radial velocities (corrections applied) of AD\,Boo and residual from the final spectroscopic orbit.} \\
\hline\hline\noalign{\smallskip}
\multicolumn{1}{c}{HJD}&\multicolumn{1}{c}{Phase}      &\multicolumn{1}{c}{$RV_p$}&\multicolumn{1}{c}{$RV_s$}&\multicolumn{1}{c}{$(O-C)_p$}&\multicolumn{1}{c}{$(O-C)_s$} \\
$-$2\,400\,000&                         &\multicolumn{1}{c}{\kms}&\multicolumn{1}{c}{\kms}&\multicolumn{1}{c}{\kms}&\multicolumn{1}{c}{\kms} \\
\noalign{\smallskip}
\endfirsthead
\caption{continued.}\\
\hline\hline\noalign{\smallskip}
\multicolumn{1}{c}{HJD}&\multicolumn{1}{c}{Phase}      &\multicolumn{1}{c}{$RV_p$}&\multicolumn{1}{c}{$RV_s$}&\multicolumn{1}{c}{$(O-C)_p$}&\multicolumn{1}{c}{$(O-C)_s$} \\
$-$2\,400\,000&                         &\multicolumn{1}{c}{\kms}&\multicolumn{1}{c}{\kms}&\multicolumn{1}{c}{\kms}&\multicolumn{1}{c}{\kms} \\
\noalign{\smallskip}
\hline\noalign{\smallskip}
\endhead
\hline
\hline\noalign{\smallskip}
 47555.8293 &   0.3938 &   -83.75 &    58.17 &    -1.57 &    -2.09 \\
 47598.8459 &   0.1868 &  -114.44 &    97.36 &    -0.05 &    -0.57 \\
 47612.6969 &   0.8820 &    56.10 &  -103.17 &     0.92 &    -2.81 \\
 47613.8580 &   0.4432 &   -53.94 &    25.71 &    -0.34 &    -1.13 \\
 47628.7443 &   0.6388 &    66.83 &  -113.43 &     2.09 &    -1.89 \\
 47640.6739 &   0.4052 &   -77.17 &    57.34 &    -1.11 &     4.23 \\
 47642.7577 &   0.4125 &   -69.91 &    49.99 &     2.09 &     1.63 \\
 47643.6147 &   0.8267 &    75.39 &  -124.58 &    -2.14 &     1.91 \\
 47644.7409 &   0.3711 &   -93.70 &    76.77 &    -0.31 &     3.39 \\
 47675.5822 &   0.2788 &  -123.92 &   103.72 &    -3.01 &    -1.84 \\
 47676.5584 &   0.7507 &    88.38 &  -142.11 &    -1.23 &    -1.48 \\
 47693.7772 &   0.0738 &   -66.80 &    36.38 &    -2.83 &    -2.59 \\
 47700.6648 &   0.4030 &   -77.85 &    59.20 &    -0.59 &     4.69 \\
 47701.7853 &   0.9446 &    20.07 &   -53.80 &     0.42 &     5.02 \\
 47702.6453 &   0.3603 &   -97.54 &    79.78 &     0.61 &     0.84 \\
 47702.7425 &   0.4073 &   -74.63 &    52.97 &     0.25 &     1.24 \\
 47703.6426 &   0.8424 &    70.23 &  -120.29 &    -1.99 &     0.00 \\
 47703.7439 &   0.8914 &    51.04 &   -97.68 &     0.62 &    -2.88 \\
 47729.5844 &   0.3819 &   -86.88 &    72.30 &     1.36 &     4.95 \\
 47731.5528 &   0.3334 &  -106.61 &    91.39 &     1.81 &     0.44 \\
 47734.5463 &   0.7803 &    88.81 &  -141.47 &     1.12 &    -3.09 \\
 47957.8075 &   0.6982 &    81.97 &  -130.23 &    -2.07 &     3.88 \\
 47958.8221 &   0.1886 &  -115.20 &    94.94 &    -0.34 &    -3.53 \\
 47959.8666 &   0.6935 &    81.81 &  -127.99 &    -1.19 &     4.90 \\
 47960.9249 &   0.2050 &  -119.65 &   105.59 &    -1.20 &     2.92 \\
 48279.8560 &   0.3669 &   -96.11 &    72.02 &    -0.82 &    -3.58 \\
 48280.9257 &   0.8840 &    53.24 &  -104.72 &    -0.95 &    -5.51 \\
 48311.9322 &   0.8716 &    59.11 &  -101.51 &    -1.02 &     4.64 \\
 48499.5443 &   0.5577 &    19.69 &   -53.87 &    -1.43 &     6.67 \\
 48638.8910 &   0.9138 &    36.51 &   -79.95 &    -1.70 &     0.57 \\
 48639.9724 &   0.4365 &   -57.70 &    32.06 &     0.06 &     0.35 \\
 48699.9584 &   0.4319 &   -61.66 &    32.15 &    -1.12 &    -2.80 \\
 48701.8654 &   0.3537 &  -101.73 &    83.27 &    -0.83 &     1.12 \\
 48704.9191 &   0.8298 &    79.57 &  -125.73 &     3.02 &    -0.37 \\
 48705.8840 &   0.2962 &  -120.74 &   107.80 &    -2.53 &     5.40 \\
 48706.8287 &   0.7528 &    92.19 &  -139.10 &     2.59 &     1.51 \\
 48711.8041 &   0.1578 &  -106.84 &    83.82 &    -1.50 &    -3.52 \\
 48725.8331 &   0.9390 &    21.13 &   -60.51 &    -2.04 &     2.42 \\
 48728.8111 &   0.3785 &   -93.65 &    70.35 &    -3.74 &     1.05 \\
 48730.9634 &   0.4188 &   -69.17 &    44.35 &    -0.84 &     0.28 \\
 48754.5676 &   0.8284 &    77.25 &  -127.14 &     0.26 &    -1.27 \\
 48759.7935 &   0.3545 &   -99.14 &    87.17 &     1.47 &     5.36 \\
 48783.6614 &   0.8915 &    52.54 &   -97.21 &     2.17 &    -2.48 \\
 48783.7736 &   0.9457 &    18.44 &   -58.86 &    -0.54 &    -0.83 \\
 48784.7101 &   0.3984 &   -78.44 &    62.33 &     1.33 &     4.89 \\
 48810.6101 &   0.9177 &    34.75 &   -77.67 &    -1.21 &     0.21 \\
 48824.6164 &   0.6879 &    79.52 &  -130.74 &    -2.12 &     0.57 \\
 48828.5744 &   0.6011 &    44.50 &   -88.00 &    -1.96 &     2.16 \\
 48840.5419 &   0.3858 &   -87.71 &    69.26 &    -1.42 &     4.19 \\
 49106.8924 &   0.1318 &   -91.82 &    76.57 &     2.87 &     1.68 \\
 49111.6651 &   0.4388 &   -57.14 &    30.24 &    -0.78 &     0.17 \\
 49112.6802 &   0.9294 &    27.39 &   -73.06 &    -1.63 &    -3.29 \\
 49114.7754 &   0.9422 &    22.66 &   -59.87 &     1.46 &     0.75 \\
 49115.7720 &   0.4239 &   -65.94 &    41.63 &    -0.59 &     1.05 \\
 49117.7719 &   0.3906 &   -84.69 &    61.93 &    -0.83 &    -0.30 \\
 49136.7641 &   0.5709 &    28.22 &   -72.23 &    -0.97 &    -2.26 \\
 49138.8295 &   0.5692 &    24.31 &   -67.62 &    -3.89 &     1.19 \\
 49140.6382 &   0.4435 &   -54.56 &    19.20 &    -1.14 &    -7.43 \\
 49141.6541 &   0.9345 &    28.08 &   -65.19 &     2.17 &     0.94 \\
 49142.6823 &   0.4315 &   -59.39 &    41.82 &     1.38 &     6.59 \\
 49146.7972 &   0.4206 &   -66.61 &    40.58 &     0.71 &    -2.30 \\
 49173.6118 &   0.3820 &   -85.69 &    65.83 &     2.53 &    -1.49 \\
 49174.5980 &   0.8587 &    65.86 &  -114.47 &     0.03 &    -1.66 \\
 49174.7590 &   0.9365 &    25.50 &   -66.53 &     0.77 &    -1.78 \\
 49175.6609 &   0.3724 &   -93.04 &    71.89 &    -0.27 &    -0.76 \\
 49401.7803 &   0.6718 &    76.53 &  -127.27 &    -0.54 &    -1.31 \\
 49447.7007 &   0.8684 &    63.10 &  -116.37 &     1.52 &    -8.53 \\
 49564.6149 &   0.3813 &   -86.88 &    69.00 &     1.68 &     1.28 \\
 49761.8372 &   0.7127 &    85.16 &  -131.47 &    -1.55 &     5.76 \\
 49765.8501 &   0.6524 &    69.42 &  -116.29 &    -0.85 &     1.72 \\
 49767.8090 &   0.5993 &    45.62 &   -89.81 &     0.15 &    -0.81 \\
 49825.8712 &   0.6648 &    74.12 &  -122.47 &    -0.65 &     0.80 \\
 49843.8216 &   0.3415 &  -104.74 &    86.98 &     0.85 &    -0.66 \\
 49873.7442 &   0.8052 &    83.09 &  -125.64 &    -0.21 &     7.60 \\
 49891.6211 &   0.4464 &   -52.53 &    26.95 &    -0.92 &     2.43 \\
 49901.5946 &   0.2673 &  -123.33 &   105.21 &    -1.30 &    -1.65 \\
 49909.5921 &   0.1330 &   -96.19 &    77.05 &    -0.94 &     1.51 \\
 49920.5892 &   0.4487 &   -48.66 &    24.62 &     1.49 &     1.81 \\
 49939.5702 &   0.6235 &    60.30 &   -99.67 &     2.46 &     3.80 \\
 49965.5149 &   0.1644 &  -105.60 &    96.18 &     2.08 &     6.10 \\
 49979.5013 &   0.9250 &    36.07 &   -70.40 &     4.43 &     2.43 \\
 50119.8669 &   0.7736 &    87.83 &  -134.44 &    -0.62 &     4.82 \\
 50126.9140 &   0.1800 &  -113.41 &    96.07 &    -0.86 &     0.30 \\
 50140.7929 &   0.8886 &    51.90 &   -99.18 &     0.06 &    -2.73 \\
 50146.8822 &   0.8320 &    77.74 &  -125.27 &     1.91 &    -0.76 \\
 50154.8316 &   0.6745 &    76.19 &  -127.41 &    -1.71 &    -0.48 \\
 50170.8382 &   0.4116 &   -73.04 &    54.00 &    -0.57 &     5.10 \\
 50507.8470 &   0.3117 &  -113.80 &    96.01 &     0.98 &    -2.37 \\
 50514.8317 &   0.6879 &    80.94 &  -130.93 &    -0.70 &     0.37 \\
 50516.9061 &   0.6906 &    85.91 &  -131.11 &     3.60 &     0.97 \\
 50534.8342 &   0.3565 &  -100.21 &    76.31 &    -0.45 &    -4.51 \\
 50535.6409 &   0.7464 &    89.82 &  -139.25 &     0.23 &     1.35 \\
 50535.6560 &   0.7537 &    91.99 &  -146.77 &     2.40 &    -6.18 \\
 50535.6710 &   0.7610 &    90.27 &  -135.81 &     0.91 &     4.52 \\
 50540.8206 &   0.2502 &  -124.08 &   109.00 &    -1.43 &     1.41 \\
 50541.7939 &   0.7206 &    86.82 &  -140.28 &    -0.99 &    -1.76 \\
 50543.8365 &   0.7080 &    83.48 &  -131.91 &    -2.45 &     4.41 \\
 50561.7089 &   0.3469 &  -105.72 &    80.35 &    -2.15 &    -4.92 \\
 50563.7115 &   0.3149 &  -113.48 &    96.44 &     0.46 &    -0.96 \\
 50564.6929 &   0.7893 &    86.35 &  -139.03 &    -0.04 &    -2.17 \\
 50565.6508 &   0.2523 &  -120.05 &   101.11 &     2.59 &    -6.47 \\
 50567.7134 &   0.2493 &  -123.07 &   104.91 &    -0.42 &    -2.68 \\
 50568.6959 &   0.7243 &    90.61 &  -138.11 &     2.38 &     0.90 \\
 50569.7779 &   0.2473 &  -121.60 &   109.17 &     1.04 &     1.60 \\
 50571.6641 &   0.1590 &  -105.86 &    88.54 &    -0.09 &     0.69 \\
 50573.7839 &   0.1836 &  -116.28 &    98.59 &    -2.72 &     1.63 \\
 50589.5959 &   0.8267 &    77.20 &  -127.53 &    -0.33 &    -1.03 \\
 50590.6390 &   0.3309 &  -107.69 &    93.36 &     1.55 &     1.46 \\
 50591.6685 &   0.8285 &    75.92 &  -124.18 &    -1.04 &     1.65 \\
 50592.6487 &   0.3023 &  -116.76 &   100.03 &     0.21 &    -0.91 \\
 50593.5959 &   0.7602 &    87.43 &  -140.95 &    -1.97 &    -0.57 \\
 50595.6617 &   0.7587 &    86.84 &  -135.87 &    -2.62 &     4.57 \\
 50596.5967 &   0.2107 &  -117.38 &   102.96 &     2.05 &    -0.86 \\
 50597.6074 &   0.6992 &    82.68 &  -135.60 &    -1.58 &    -1.24 \\
 50619.6064 &   0.3329 &  -106.61 &    94.13 &     1.98 &     2.99 \\
 50619.6215 &   0.3402 &  -106.01 &    91.36 &     0.06 &     3.16 \\
 50620.6194 &   0.8225 &    79.11 &  -126.68 &     0.33 &     1.28 \\
 50621.5810 &   0.2873 &  -120.35 &   107.51 &    -0.60 &     3.32 \\
 50622.6012 &   0.7805 &    89.09 &  -133.20 &     1.41 &     5.16 \\
 50623.6005 &   0.2635 &  -122.58 &   108.27 &    -0.31 &     1.13 \\
 50624.6042 &   0.7487 &    87.14 &  -139.68 &    -2.47 &     0.94 \\
 50625.6138 &   0.2367 &  -122.61 &   108.53 &    -0.33 &     1.37 \\
 50627.6490 &   0.2204 &  -123.42 &   109.14 &    -2.59 &     3.68 \\
 50629.5948 &   0.1610 &  -106.47 &    90.60 &     0.01 &     1.92 \\
\noalign{\smallskip}
\hline
\end{longtable}

\twocolumn
\begin{table}
\caption[]{\label{tab:vzhya_rv}
Radial velocities (corrections applied) of VZ\,Hya and residuals from the final spectroscopic orbit.}
\scriptsize{
\begin{flushleft}
\begin{tabular}{llrrrr} \hline
\hline\noalign{\smallskip}
\multicolumn{1}{c}{HJD}&\multicolumn{1}{c}{Phase}      &\multicolumn{1}{c}{$RV_p$}&\multicolumn{1}{c}{$RV_s$}&\multicolumn{1}{c}{$(O-C)_p$}&\multicolumn{1}{c}{$(O-C)_s$} \\
$-$2\,400\,000&                         &\multicolumn{1}{c}{\kms}&\multicolumn{1}{c}{\kms}&\multicolumn{1}{c}{\kms}&\multicolumn{1}{c}{\kms} \\
\hline\noalign{\smallskip}
 47555.7226 &   0.8107 &    82.13 &  -101.37 &    -1.40 &     0.94 \\
 47558.7158 &   0.8413 &    75.81 &   -89.61 &     0.66 &     3.40 \\
 47568.6242 &   0.2530 &   -99.07 &   101.23 &     0.40 &     0.51 \\
 47570.7110 &   0.9715 &    13.91 &   -20.42 &     1.56 &     2.92 \\
 47574.6105 &   0.3141 &   -93.26 &    92.10 &    -1.38 &    -0.20 \\
 47575.6910 &   0.6862 &    82.32 &  -100.78 &    -0.50 &     0.74 \\
 47577.6268 &   0.3527 &   -79.70 &    81.21 &     0.70 &     1.64 \\
 47583.6110 &   0.4132 &   -54.33 &    49.75 &    -0.51 &    -0.33 \\
 47587.6242 &   0.7950 &    86.63 &  -105.73 &     0.05 &    -0.03 \\
 47603.5188 &   0.2678 &   -98.83 &   100.09 &     0.06 &     0.00 \\
 47609.5595 &   0.3477 &   -80.89 &    83.10 &     1.27 &     1.58 \\
 47613.5894 &   0.7352 &    90.37 &  -110.66 &     0.43 &    -1.24 \\
 47628.5692 &   0.8930 &    54.76 &   -71.57 &     0.23 &    -1.44 \\
 47642.5540 &   0.7082 &    85.75 &  -106.46 &    -1.35 &    -0.19 \\
 47645.5312 &   0.7333 &    89.68 &  -108.31 &    -0.15 &     0.99 \\
 47664.5534 &   0.2830 &   -97.70 &   100.29 &    -0.25 &     1.81 \\
 47670.6171 &   0.3708 &   -74.23 &    70.09 &    -0.82 &    -1.72 \\
 47671.6217 &   0.7167 &    87.91 &  -106.94 &    -0.37 &     0.64 \\
 47672.6262 &   0.0626 &   -39.66 &    38.82 &     1.29 &     3.02 \\
 47899.7202 &   0.2549 &  -101.14 &   100.79 &    -1.70 &     0.10 \\
 47905.7028 &   0.3149 &   -92.07 &    93.39 &    -0.36 &     1.27 \\
 47928.7245 &   0.2416 &   -98.52 &   102.25 &     0.83 &     1.66 \\
 47941.6387 &   0.6882 &    83.84 &  -100.78 &     0.56 &     1.25 \\
 47956.6250 &   0.8482 &    72.18 &   -91.00 &    -0.66 &    -0.55 \\
 47957.6126 &   0.1883 &   -91.30 &    96.46 &     1.14 &     3.54 \\
 47959.5793 &   0.8654 &    65.89 &   -83.00 &    -0.56 &     0.36 \\
 47960.5813 &   0.2105 &   -95.81 &    98.29 &     0.76 &     0.78 \\
 48191.8651 &   0.8454 &    74.21 &   -89.36 &     0.41 &     2.16 \\
 48191.8800 &   0.8505 &    72.28 &   -87.06 &     0.25 &     2.50 \\
 48226.7915 &   0.8712 &    64.13 &   -83.69 &    -0.02 &    -2.88 \\
 48230.9468 &   0.3019 &   -96.99 &    89.83 &    -2.51 &    -5.36 \\
 48236.8621 &   0.3386 &   -87.41 &    83.89 &    -2.27 &    -0.94 \\
 48252.8201 &   0.8332 &    77.88 &   -94.81 &     0.22 &     0.99 \\
 48256.8706 &   0.2279 &   -98.67 &    99.73 &    -0.10 &     0.00 \\
 48258.8791 &   0.9195 &    42.00 &   -56.01 &     0.56 &    -0.40 \\
 48259.7319 &   0.2131 &   -96.77 &    97.61 &     0.17 &    -0.31 \\
 48264.7826 &   0.9521 &    22.39 &   -34.46 &    -1.16 &     1.30 \\
 48280.7728 &   0.4578 &   -30.24 &    23.96 &    -0.82 &     0.95 \\
 48281.7752 &   0.8030 &    82.86 &  -103.45 &    -2.28 &     0.64 \\
 48290.6911 &   0.8729 &    63.96 &   -84.32 &     0.52 &    -4.30 \\
 48315.6410 &   0.4636 &   -24.29 &    18.58 &     1.82 &    -0.76 \\
 48354.5562 &   0.8627 &    68.11 &   -84.77 &     0.59 &    -0.22 \\
\noalign{\smallskip}
\hline
\end{tabular}
\end{flushleft}
}
\end{table}

\begin{table}
\caption[]{\label{tab:wzoph_rv}
Radial velocities (corrections applied) of WZ\,Oph and residuals from the final spectroscopic orbit.}
\scriptsize{
\begin{flushleft}
\begin{tabular}{llrrrr} \hline
\hline\noalign{\smallskip}
\multicolumn{1}{c}{HJD}&\multicolumn{1}{c}{Phase}      &\multicolumn{1}{c}{$RV_p$}&\multicolumn{1}{c}{$RV_s$}&\multicolumn{1}{c}{$(O-C)_p$}&\multicolumn{1}{c}{$(O-C)_s$} \\
$-$2\,400\,000&         &\multicolumn{1}{c}{\kms}&\multicolumn{1}{c}{\kms}&\multicolumn{1}{c}{\kms}&\multicolumn{1}{c}{\kms} \\
\hline\noalign{\smallskip}
 47583.9253 &   0.4058 &   -78.81 &    22.61 &    -0.94 &     1.19 \\
 47605.8830 &   0.6545 &    45.26 &  -101.95 &     0.37 &     0.06 \\
 47608.8786 &   0.3705 &   -94.35 &    35.97 &    -1.48 &    -0.54 \\
 47612.8755 &   0.3259 &  -107.37 &    49.40 &    -0.15 &    -1.54 \\
 47628.8294 &   0.1394 &   -95.42 &    41.35 &     1.13 &     1.14 \\
 47639.8265 &   0.7681 &    58.80 &  -117.67 &    -1.03 &    -0.63 \\
 47641.8367 &   0.2486 &  -116.31 &    59.25 &     0.81 &    -1.64 \\
 47643.7878 &   0.7150 &    60.26 &  -113.94 &     1.99 &     1.53 \\
 47644.8166 &   0.9609 &    -5.58 &   -47.03 &     1.20 &     3.03 \\
 47645.8002 &   0.1960 &  -111.58 &    55.62 &     0.49 &    -0.20 \\
 47662.6849 &   0.2320 &  -117.86 &    60.36 &    -1.30 &     0.03 \\
 47664.7592 &   0.7279 &    58.48 &  -117.41 &    -1.07 &    -0.66 \\
 47674.8228 &   0.1334 &   -93.47 &    39.12 &     0.89 &     1.12 \\
 47675.7352 &   0.3515 &   -99.81 &    43.29 &    -0.14 &    -0.06 \\
 47689.7883 &   0.7107 &    56.82 &  -117.76 &    -0.89 &    -2.85 \\
 47691.7534 &   0.1804 &  -110.29 &    53.39 &    -1.52 &     0.89 \\
 47691.9235 &   0.2211 &  -116.21 &    59.69 &    -0.55 &     0.26 \\
 47693.8686 &   0.6860 &    54.15 &  -110.29 &     0.82 &     0.21 \\
 47693.9378 &   0.7025 &    56.23 &  -112.40 &    -0.26 &     1.28 \\
 47700.6733 &   0.3126 &  -109.29 &    55.41 &     1.06 &     1.32 \\
 47723.6572 &   0.8065 &    56.53 &  -111.37 &     1.66 &     0.68 \\
 47725.6208 &   0.2759 &  -117.06 &    61.70 &    -1.10 &     1.98 \\
 47727.6324 &   0.7567 &    58.66 &  -118.05 &    -1.67 &    -0.51 \\
 47729.6085 &   0.2291 &  -116.36 &    60.66 &     0.00 &     0.53 \\
 47731.6271 &   0.7116 &    57.97 &  -113.55 &     0.14 &     1.48 \\
 47775.5008 &   0.1989 &  -112.54 &    57.69 &     0.04 &     1.36 \\
 47777.5180 &   0.6811 &    51.47 &  -110.36 &    -0.74 &    -0.99 \\
 47779.5228 &   0.1603 &  -103.69 &    47.89 &    -0.30 &     0.81 \\
 47788.5003 &   0.3062 &  -111.36 &    56.91 &     0.29 &     1.52 \\
 48194.4589 &   0.3441 &   -99.98 &    43.14 &     2.09 &    -2.62 \\
 48405.7132 &   0.8410 &    46.85 &  -104.96 &     0.56 &    -1.54 \\
 48409.7815 &   0.8135 &    54.04 &  -111.28 &     0.60 &    -0.67 \\
 48429.7546 &   0.5877 &    17.51 &   -73.70 &    -0.61 &     1.40 \\
 48432.7522 &   0.3042 &  -113.49 &    54.28 &    -1.47 &    -1.48 \\
 48462.5915 &   0.4369 &   -63.20 &     5.35 &    -0.54 &    -0.78 \\
 48466.5983 &   0.3946 &   -81.02 &    25.90 &     1.92 &    -0.62 \\
 48485.5754 &   0.9308 &     8.77 &   -65.98 &    -0.27 &    -0.01 \\
 48491.5446 &   0.3576 &   -98.40 &    39.30 &    -0.81 &    -1.96 \\
 48499.5903 &   0.2808 &  -114.65 &    57.64 &     0.82 &    -1.59 \\
 48501.5840 &   0.7574 &    59.46 &  -117.06 &    -0.85 &     0.46 \\
\noalign{\smallskip}
\hline
\end{tabular}
\end{flushleft}
}
\end{table}

\end{appendix}

\listofobjects
\end{document}